\documentclass[preprint]{aastex}
\usepackage[usenames]{color}
\usepackage{epsfig}
\usepackage{natbib}
\usepackage{longtable}
\usepackage{multirow}
\usepackage{amsmath}

\newcommand{\myemail}{mtpatter@nmsu.edu}

\defcitealias{Garnett1987}{GS87}
\defcitealias{Stanghellini2010}{S10} 	
\defcitealias{Croxall2009}{C09}
\defcitealias{KK2004}{KK04}
\defcitealias{PT2005}{PT05}
\defcitealias{KD2002}{KD02}
\defcitealias{B07}{B07}

\slugcomment{}

\shorttitle{Oxygen Abundances of M81 HII Regions}
\shortauthors{M. T. Patterson, et al.}

\begin{document}

\title{An Oxygen Abundance Gradient into the Outer Disk of M81\footnote{Observations reported here were 
obtained at the MMT Observatory, a joint facility of the Smithsonian Institution and the 
University of Arizona.} \footnote{Based on observations obtained with the Apache Point Observatory 
3.5-meter telescope, which is owned and operated by the Astrophysical Research Consortium}
 \footnote{Observations made with the Burrell Schmidt of the Warner and Swasey Observatory,
Case Western Reserve University.} }

\author{Maria T. Patterson$^1$ \footnote{E-mail: \myemail} , Rene A.M. Walterbos$^{1}$ \footnote{Visiting Astronomer,
Kitt Peak National Observatory, National Optical Astronomy Observatory, which is operated by the
Association of the Universities for Research in Astronomy (AURA) under cooperative agreement with
the National Science Foundation} , Robert C. Kennicutt$^2$, 
Cristina Chiappini$^3$, and David A. Thilker$^4$}
\affil{$^1$Department of Astronomy, New Mexico State University, P.O. Box 30001, MSC 4500, Las Cruces, NM 88003, USA}
\affil{$^2$Institute of Astronomy, Madingley Road, Cambridge, CB3 0HA, United Kingdom}
\affil{$^3$Leibniz-Institut f\"ur Astrophysik Potsdam (AIP), An der Sternwarte 16 D - 14482, Potsdam, Germany}
\affil{$^4$Department of Physics \& Astronomy, Johns Hopkins University, 3400 N. Charles St., Baltimore, MD 21218, USA}

\begin{abstract}
The extended HI disk and tidal tails of M81 present an interesting environment to study the effects 
of galaxy interaction on star formation and chemical evolution of the outer disk of a large spiral galaxy. We present 
H$\alpha$ imaging of the outer disk of M81 and luminosities for 40 HII regions out to $\sim$3$\times$R$_{25}$.  
We have also obtained MMT spectra for 21 HII regions out to more than twice R$_{25}$.  
We derive strong line oxygen abundances for all HII regions using $R_{23}$ based and [NII]/[OII] based calibrations 
and electron temperature abundances for seven regions spanning a galactocentric distance between 5.7 and 32 kpc.  
We also comment on the abundances of HII regions near KDG 61 and the ``tidal dwarf'' candidate HoIX.  
Our results constitute the most radially extended metallicity study for M81 to date.  
With this extended data set, we find an overall oxygen abundance gradient of
$\Delta(\log$(O/H))$/\Delta$R$_G\sim-$0.013 dex kpc$^{-1}$ over the entire radial range.  This is significantly flatter 
than what has been found in previous studies which were limited to the optical disk.  
From our temperature based abundances, we find  $\Delta(\log$(O/H))$/\Delta$R$_G\sim-$0.020 dex kpc$^{-1}$ and 
present the possibility of a broken gradient from these data, but note the need to obtain more temperature
based abundances at intermediate galactocentric distances ($\sim$10-20 kpc) to verify whether or not this may be the case.  
We discuss our main result of a rather flat gradient for M81 in the context of 
simulations and observations of abundance gradients in other galaxies. 
We find that the shallow abundance gradient of M81 is likely a result of the interaction history of this galaxy.

\end{abstract}

\keywords{galaxies: individual (M81, NGC 3031) -- galaxies: abundances -- galaxies: ISM -- ISM: HII regions}


\section{Introduction}

M81 is a moderately-inclined Sa galaxy with remarkably well-defined spiral arms at a distance of 
3.63$\pm$0.34 Mpc \citep{Freedman2001}.   At this distance, 1$^{\prime\prime}$= 17.5 pc.  It has a
total mass of 2.6$\times$10$^{11}$ M$_{\odot}$ \citep[corrected for distance]{Appleton1981}, 
similar to the Milky Way galaxy.  M81 is an interesting object in light of its tidal
interactions with surrounding companion galaxies.  It has a large outer disk of HI gas and large HI 
tidal tails over a wide area, caused by the interactions with M82 and NGC 3077. 
However, there is debate over the origin of some of the features in the area surrounding M81, 
such as ``Arp's loop'' \citep{Arp1965}, due to the possible confusion with foreground galactic cirrus
\citep[see][and references therein]{Sollima2010,Davies2010}.
M81 has extended HI arms, filaments, and clouds \citep{Yun1994,Walter2002,Chynoweth2008}, within which 
can be found distant HII regions \citep{Munch1959} and dwarf galaxies, including KDG 61 and ``tidal dwarf'' 
candidate HoIX \citep[see e.g.][hereafter, \citetalias{Croxall2009}]{Makarova2002,Croxall2009}, 
supporting the extra-galactic origin of many observed outer disk features. 
As such, it provides a fertile ground for exploring star formation in low density
environments, and much attention has focused in past years on the evidence for star formation 
and on the properties of the young stellar populations detected in these HI features, especially from recent
GALEX and HST observations \citep{Durrell2004,Thilker2007,deMello2008,Sabbi2008,Weisz2008,
Chiboucas2009,Davidge2009,Gogarten2009,Mouhcine2009,Durrell2010}.

HII regions are of particular interest since their emission line spectra trace the temperature and 
metallicity of the gas in each region, unveiling the current abundance radial profile in spiral galaxies
\citep[recently,][]{Kennicutt2003,Bresolin2004,Bresolin2009,Stanghellini2010,
Garcia2010,Goddard2011,Werk2011}, which constitutes an important constraint to chemical evolution 
models \citep[e.g.,][]{Prantzos2000,Chiappini2001,Chiappini2003,Molla2005}.  The chemical abundances may 
help constrain the origin of outer disk and inter-galaxy HI gas and the formation of tidal dwarf galaxies.  
Previous metallicity studies of HII regions in M81 found a steeper than expected radial abundance gradient for this galaxy, 
given that more massive galaxies, like M81, tend to have shallower gradients than smaller galaxies \citep{Zaritsky1994}.  
These studies, however, were limited to HII regions mainly within the disk between 3-12 kpc, with only one outer disk object 
past R$_{25}$ \citep[][(hereafter \citetalias{Garnett1987} and \citetalias{Stanghellini2010})]{Garnett1987,Stanghellini2010}.
Only a handful of galaxies have well-characterized gradients traced by HII regions past $R_{25}$
(M101 \citep{Kennicutt2003}; M83 \citep{Bresolin2009}; NGC 4625 \citep{Goddard2011}).

In this paper, we present a combined imaging and spectroscopic study of HII regions in the outer disk
of M81.  We present the sizes and H$\alpha$ luminosities of newly discovered HII regions from
a survey of the M81-NGC 3077-M82 complex with the Burrell Schmidt telescope at the Kitt Peak National 
Observatory.  For several of these regions, we describe their morphological features based on separate 
high resolution H$\alpha$ imaging obtained with the APO ARC 3.5-meter telescope.  We also present optical 
spectra for 21 HII regions obtained with the MMT.  We use the MMT spectra to derive strong line oxygen 
abundances using the metallicity-sensitive $R_{23}$ parameter and the [NII]/[OII] ratio for all regions and 
temperature derived oxygen abundances for seven regions with detectable temperature lines.  From the 
oxygen abundances, we derive a metallicity gradient into the outskirts of M81 and comment on the possible 
effects of the tidal interaction on the abundance gradient.

In \S\ref{observationssec} we describe our H$\alpha$ imaging and spectroscopic observations and 
our data reduction process.  In \S\ref{oxygenabundancessec} we derive oxygen abundances from our 
HII region spectra and describe and compare our results from the direct method of abundance determination
via electron temperature lines versus indirect metallicity-sensitive strong line calibrations.
In \S\ref{dwarfsec} we discuss the metallicities of HII regions near two dwarf galaxies, HoIX and KDG 61.  
We discuss our abundance gradient and implications for the evolution of M81 in \S\ref{abundancegradientsec} 
and \S\ref{discussionsec}, and conclude with a summary of our results in \S\ref{summarysec}.  


\section{Observations and data reduction}
\label{observationssec}


\subsection{Observations}

The outer disk HII regions were found using existing deep H$\alpha$ + [NII] imaging of M81 obtained 
with the Burrell-Schmidt telescope at KPNO.  The images were obtained through a 75 \AA~filter for a total
exposure time of 3.5 hours.  Details of these observations are described in \citet{Greenawalt1998}.  The 
1.5 degree field observed encompasses the entire M81-M82-NGC 3077 triplet.

\begin{figure}
  \centering
  \epsfig{width=0.95\linewidth,file=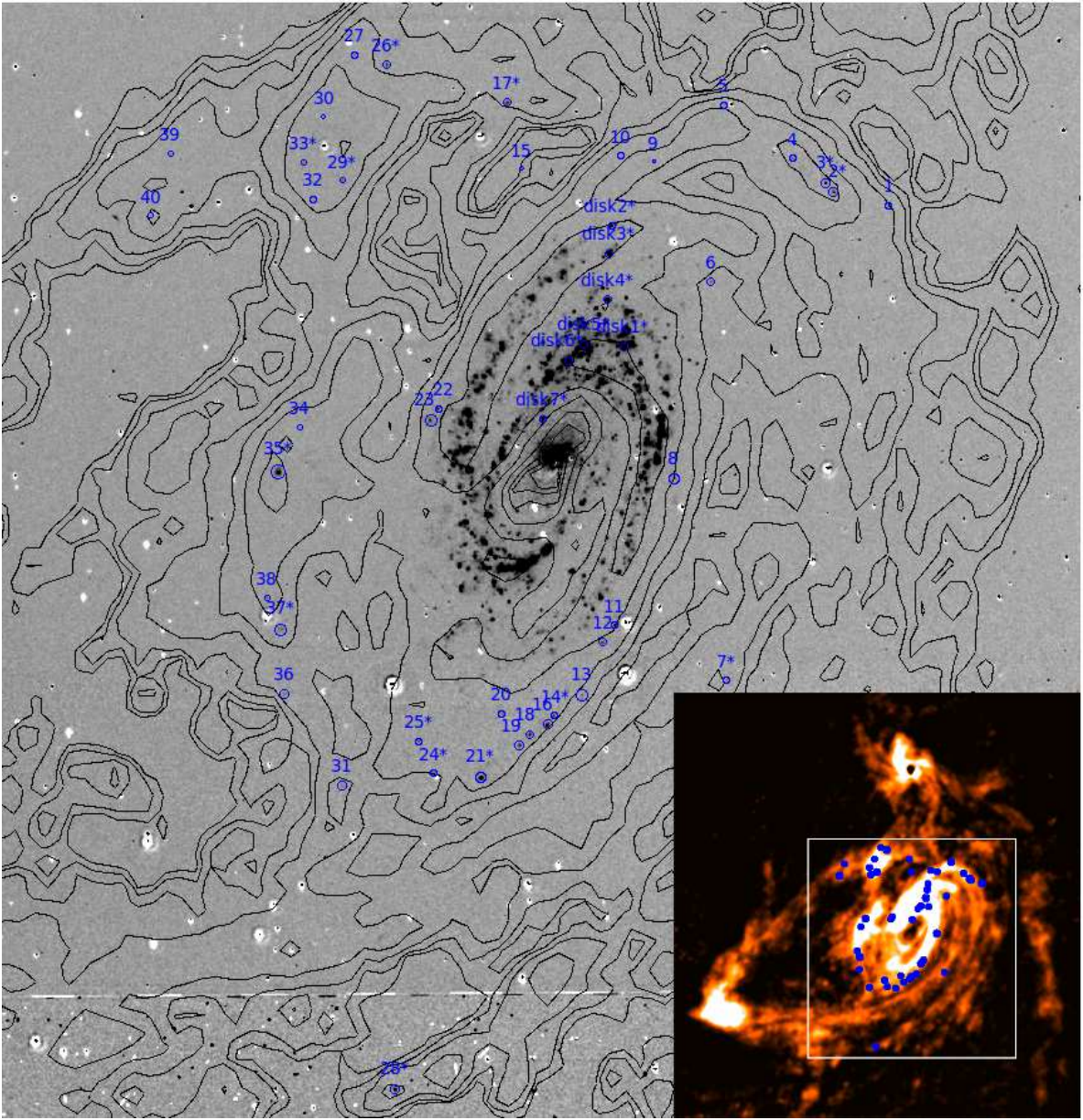}
  \caption{Burrell-Schmidt image of M81 in H$\alpha$ with the HII regions in our sample marked.  The contours correspond 
    to HI column densities of 1.8$\times$10$^{20}$ cm$^{-2}$ times 1, 2, 4, 8, 16, 32, and 64 from \citet{Yun1994}.    
    HII regions with spectra are marked with asterisks.  A list of all HII regions and their properties are provided in Table
    \ref{tab:sample}.  In this paper, we discuss in particular region 21 (M\"unch 1), region 28 (the HII region near KDG 61),
    and region 35 (a bright ionized nebula near tidal dwarf candidate HoIX).  In the bottom right corner, we show the HII regions
    marked on the full HI column density map of the M81-M82-NGC 3077 triplet, with the white box marking the size of the 
    section of the Burrell-Schmidt H$\alpha$ image shown here.}
  \label{fig:HIIregs}
\end{figure}

\begin{figure}
  \centering
  \epsfig{width=0.95\linewidth,file=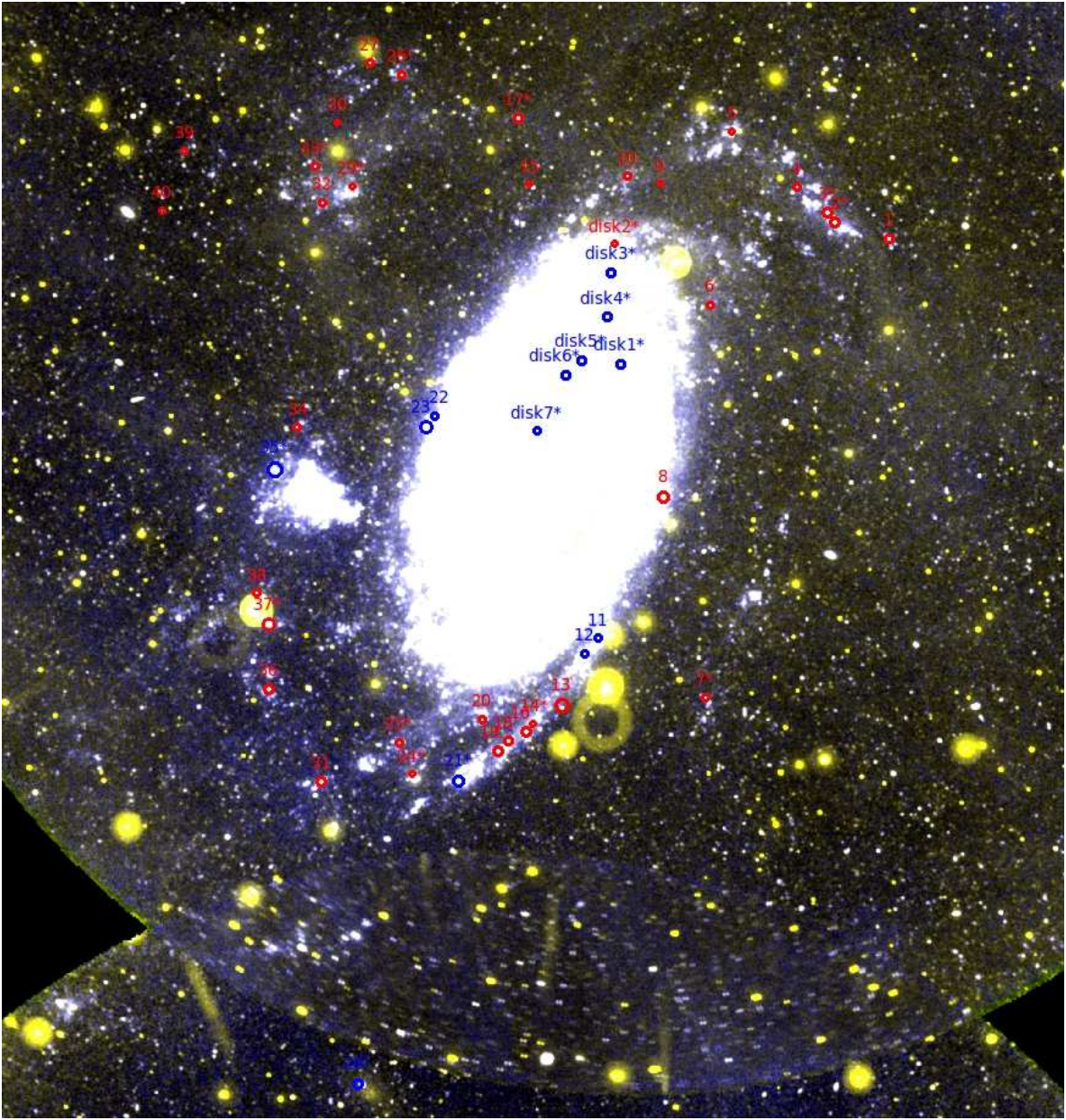}
  \caption{GALEX color-composite image shown on the same scale as the previous figure and with HII regions
    marked for comparison.  We show GALEX FUV in blue, GALEX NUV in red, and an average of the two in green.  The image
    is stretched to a square root scale in order to enhance the faint outer disk star forming features.    
    The HII regions marked in blue signify previously catalogued
    objects, whereas the red regions denote newly found objects.  Of particular interest is region 35, which is offset from the
    UV emission of the dwarf galaxy HoIX.  We show a larger image of this object in Fig. \ref{fig:colorHoIX}.   }
  \label{fig:HIIregs2}
\end{figure}

Candidate HII regions were selected by visual comparison of the H$\alpha$ + [NII] image to the continuum 
image.  We limited our search to a 42\arcmin$\times$63\arcmin~box centered on the galaxy center.
We obtained fluxes for the HII regions within an appropriately sized circular aperture chosen
individually for each region.  The aperture size for each HII region was chosen to have a radius just large
enough so that the enclosed flux profile flattened sufficiently, meaning that the level of the background had been
reached.  We subtracted the background level as determined from a small annulus around each aperture.  
The average aperture size of our HII regions is $\sim$9.5$^{\prime\prime}$, which corresponds to a physical size of
170 pc.  This is approximately the same size as the faintest ($\log(L_{H\alpha})$$\sim$37.3) outer disk HII regions
described by \citet{Ferguson1998}.
Only candidate regions with signal to noise greater than 4 were kept,
setting our detection limit to $\log(L_{H\alpha})$$\sim$36.6.  Note that this is brighter than the expected
H$\alpha$ luminosity for the faintest single star HII regions ($\log(L_{H\alpha})$$\sim$36.15-36.30 for
a B0.5 star \citep{Vacca1996,Sternberg2003}).  The faintest HII regions detected here are the 
equivalent of what is predicted for a single ionizing O9 star \citep{Sternberg2003}.
We found 40 HII regions outside of the main disk at a 4$\sigma$ confidence level within a distance of
$\sim$3$\times$$R_{25}$.  The locations of our HII regions are shown in Figs. \ref{fig:HIIregs} and \ref{fig:HIIregs2},
in which we show the H$\alpha$ image as compared to HI \citep{Yun1994} and GALEX UV data \citep{Thilker2007}.
We verified our flux calibration by comparison with the fluxes of several HII regions in the main disk
given by \citet{Lin2003} and \citet{Perez2006}, using the aperture sizes quoted in each paper. 
We estimate our absolute fluxes to be accurate to $\sim$9\%, based on the uncertainties of \citet{Lin2003} 
and \citet{Perez2006} and the uncertainty in our calibration.  To convert to H$\alpha$ luminosities, we
assumed the [NII]/H$\alpha$ ratio to be 0.4 and a distance to M81 of 3.63 Mpc.  In Table \ref{tab:sample},
we list the HII regions in our sample, noting their locations, galactocentric distances, sizes, and H$\alpha$ luminosities.
For previously catalogued objects, we provide alternative names from \citet{Hodge1983}, \citet{Miller1994}, 
\citet{PSK1988}, \citet{Munch1959}, and \citetalias{Croxall2009}.  

Higher resolution H$\alpha$ plus continuum images of several HII regions are included in Fig. \ref{fig:spicam} 
to show the morphologies of several interesting regions.  We chose to show a few of the brightest
HII regions located in various areas of the outskirts. These images were obtained with the SPICAM instrument 
on the Apache Point Observatory ARC 3.5-meter telescope.  The full images have a 4.8$\times$4.8
arcmin$^2$ field of view and were obtained in 2$\times$2 pixel binning mode, giving 0.28 arcsec pixel$^{-1}$.  
The coverage of the SPICAM images was such that they overlapped with 36 of the Schmidt HII region detections; all 
those were confirmed.  For each field, two 420 second H$\alpha$ exposures were taken with a 25\AA~(FWHM) filter.
Two 60-90 second exposures in R band were also obtained for continuum imaging.  In Fig. \ref{fig:colorHoIX},
we show an RGB image of region 35 and the dwarf galaxy HoIX, using continuum subtracted H$\alpha$ (red), GALEX NUV
(green), and GALEX FUV (blue).  This HII region is noticeably offset from the main body of the dwarf galaxy, which shows little
H$\alpha$ emission.  However, there is UV emission within this region, and, as shown in Fig. \ref{fig:HIIregs}, 
the region lies in a peak of the HI.  We will discuss this region extensively in a later section of this paper.     

We obtained 21 spectra for HII regions in the disk and outskirts of M81 identified in the Burrell-Schmidt image 
using the Blue Channel spectrograph on the 6.5-meter MMT telescope over four nights in January 2002.  
The selected regions lie at various galactocentric distances from $\sim$3 to 33 kpc and range in log($L_{H\alpha}$)
luminosity from $\sim$36.9 to 38.9 ergs s$^{-1}$.  HII regions for which we have spectra are marked in 
Fig. \ref{fig:HIIregs} with asterisks.  We observed with a 500 mm$^{-1}$ grating, covering the
wavelength range from 3650-7150 \AA~with a dispersion of 1.19 \AA~pixel$^{-1}$.  We used a 
2\arcsec$\times$180\arcsec~slit aligned along the average parallactic angle for each observation. The final
spectral resolution is 7 \AA~FWHM. The CCD camera produced 3072 pixel (wavelength) $\times$ 220 pixel 
(spatial) images with 0.56\arcsec~per pixel spatial scale (after two pixel binning only in the spatial direction).  
Each night we observed a quartz lamp for flatfielding, twilight exposures for illumination correction, a Helium 
Neon Argon lamp for wavelength calibration, and 4-5 standard stars for flux calibration.  For objects with spectra,
the exposure times are noted in the last column.


\subsection{Spectroscopic Data Reduction}

We used the \textsc{iraf}\footnote{IRAF is distributed by the National Optical Astronomy Observatory, 
which is operated by the Association of Universities for Research in Astronomy (AURA) under cooperative agreement 
with the National Science Foundation.} task \textsc{ccdproc} to remove overscan, trim, and flatfield all 
the MMT spectra \citep{Tody1993}.  We applied a slit illumination correction to the data derived from twilight 
exposures frames.  A $\sim$4\% gradient along the spatial axis was typically present before the twilight correction.
To remove cosmic rays, we averaged frames for each object using \textsc{imcombine} with the cosmic ray reject 
option (\textsc{crreject}) enabled.  All 2-D spectra were wavelength calibrated using a Helium Neon Argon image 
taken near the rotation angle of each science exposure. The spectrum of each object was extracted using an 
appropriate size aperture based on a visual inspection of the spatial extent of each target.  A standard star
spectrum was used as a reference for the trace along the dispersion axis if the HII region continuum was too weak 
to use as a trace.  We flux calibrated all spectra using a separate sensitivity function for each of the four nights made 
from at least four standard star observations.  The 1$\sigma$ uncertainty introduced in the absolute flux calibration 
for the first three nights is $\sim$3\%, across the wavelength range.  The uncertainty in our absolute flux calibration 
for the fourth night of observations is $\sim$6\%, slightly higher due to varying weather conditions.

To correct for interstellar reddening, we used the \textsc{iraf} task \textsc{deredden}, assuming R=3.1 and 
using the reddening law from \citet{Cardelli1989}.  We derived the logarithmic extinction at H$\beta$, $c(H\beta)$,
for each spectrum using the observed H$\alpha$ to H$\beta$ ratio and taking an intrinsic value of 
H$\alpha$/H$\beta$= 2.86, which assumes photoionization and Case B recombination \citep{Osterbrock2006}.
Note, however, that this assumption may be not be valid for region 35, which harbors the ultraluminous x-ray source M81 X-9.  
We will address the effect of shock ionization on the abundance for this object specifically in \S\ref{dwarfsec}.   
The uncertainty in our extinction correction was calculated by propagating the errors in 
the H$\alpha$ and H$\beta$ lines.  Because the underlying continuum was very weak in most of our objects, 
the Balmer line fluxes were not corrected for stellar absorption.
We measured the flux in each detectable line of our dereddened spectra 
with a Gaussian fit to the line profile using the task \textsc{splot}.  The final errors in our dereddened line fluxes are 
due to a combination of uncertainties from our flux calibration, the line flux measurements with \textsc{splot}, and 
the extinction correction.  Our dereddened line fluxes relative to H$\beta$ and extinction coefficients for a subset 
of emission lines are listed in Tables \ref{tab:fluxes1} and \ref{tab:fluxes2}.  In the last line of each table,
we also mark the regions which have Wolf-Rayet features in their spectra, such as the 4660\AA~blue bump or emission
at HeII $\lambda$4686.


\section{Oxygen Abundances}
\label{oxygenabundancessec}

We derive oxygen abundances for HII regions in our sample using two different methods.  The first method is 
a ``direct'' method, using temperature sensitive emission lines to constrain the oxygen abundance.  Because this 
method requires measuring very weak lines, we were able to calculate an oxygen abundance for only seven HII 
regions in our sample.  We describe our determination of oxygen abundances from electron temperature lines in 
\S\ref{electrontempsec}. 

The second method, which we describe in \S\ref{stronglineabundancessec}, uses strong lines to constrain the 
oxygen abundance without directly measuring the electron temperature.  Although this is an indirect method for
abundance determination, the strong line calibrations allow us to derive an oxygen abundance for all HII regions 
in our data set.  We use two $R_{23}$ based metallicity calibrations and two metallicity calibrations based on the 
[NII]/[OII] ratio.  Because the $R_{23}$ parameter is a non-monotonic function of metallicity, we must first decide 
whether each region lies on the upper or lower branch of the $R_{23}$ vs. metallicity relation.  We describe the 
upper vs. lower branch determination in \S\ref{upperlowerbranchsec}.  Having determined each region's branch
placement, we then use two separate strong line $R_{23}$ calibrations to derive oxygen abundances in 
\S\ref{empthesec}.  In \S\ref{NIIOIIsec}, we derive strong line abundances using the 
[NII]$\lambda$6584/[OII]$\lambda$3727 ratio and compare these results to our $R_{23}$ based abundances.  
We mark abundances determined from electron temperatures distinctly from our strong line abundances throughout 
plots in the next sections for easy comparison.  For a thorough discussion of a comparison of methods of abundances 
determination see \citet{Kennicutt2003}. 


\subsection{Electron Temperature Abundances}
\label{electrontempsec}

For seven of our HII regions, we were able to derive electron temperatures using either the [NII] 
($\lambda$6583 and $\lambda$5755) or [OIII] ($\lambda$5007 and $\lambda$4363) lines.  Note that there may be 
some bias as to the metallicities of the regions with detectable electron temperature lines.  HII regions with higher 
abundances will have lower electron temperatures and therefore more difficult to detect weak electron temperature 
lines.  Yet, we are still able to derive electron temperature lines for regions between 5.7 to 32 kpc, which covers a 
larger radial range than any previous HII region metallicity study for M81.

We calculate the electron temperatures using the \textsc{iraf} task \textit{temden} which calculates either temperature 
or density as part of the \textsc{nebular} package \citep{Shaw1995}.  The tasks in the \textsc{nebular} package 
are based on a 5-level atom program that approximates the physical conditions in a nebula, 
originally described by \citet{DeRobertis1987}.
Since the ratio of the [SII] lines $\lambda$6716/$\lambda$6731 $\sim$1.4 for all seven
regions, we assume that $n_e$ is in the low density limit and is approximately 100 cm$^{-3}$. Table \ref{tab:temps}
gives the measured temperatures derived from [NII] and [OIII] lines in cases where these lines were detectable.  Where
these lines were not detected, we adopt temperatures derived using the following equations from \citet{Garnett1992}:
\begin{align}
T(OII) &= T(NII)\\
T(OII) &= 0.7T(OIII) + 3000 K 
\end{align}
We then calculate oxygen ion abundances using the \textsc{iraf} task \textit{nebular.ionic}.  We used the flux 
from [OII]$\lambda$3727 for the O$^+$/H$^+$ calculation, and [OIII]$\lambda$5007 for the O$^{+2}$/H$^+$
calculation.  Table \ref{tab:ionic} gives the derived oxygen ion and element abundances for the seven HII regions 
with detectable electron temperature lines.  The errors in the abundances are dominated by uncertainty in the 
adopted temperatures.


\subsection{Strong Line Abundances}
\label{stronglineabundancessec}

Most of the HII region spectra do not have strong enough [OIII]$\lambda$4363 or [NII]$\lambda$5755 lines to 
derive reliable temperatures for abundance determinations.  We use both $R_{23}$ based and [NII]/[OII]  ratio based 
metallicity calibrations to derive oxygen abundances for all HII regions in our sample. 

First, we discuss abundances based on the metallicity-sensitive parameter $R_{23}$ \citep{Pagel1979}, 
which is defined as the flux ratio of lines as follows:

\begin{equation}
  R_{23} \equiv \frac{[OII]\lambda3727 + [OIII]\lambda\lambda4959,5007}  
  {H\beta\lambda4861 }
\end{equation}

The main advantage of using the $R_{23}$ parameter as an indication of oxygen abundance is that it is a 
direct function of the strength of the lines for the first and second ionization states of oxygen, rather than
depending on line ratios of other elements.  To use the $R_{23}$ parameter, we must decide a priori whether
the HII region is on the upper or lower branch of the $R_{23}$ to O/H relation. 


\subsubsection{Upper or lower $R_{23}$ branch?}
\label{upperlowerbranchsec}

\begin{figure}
  \centering
  \epsfig{width=0.5\linewidth,file=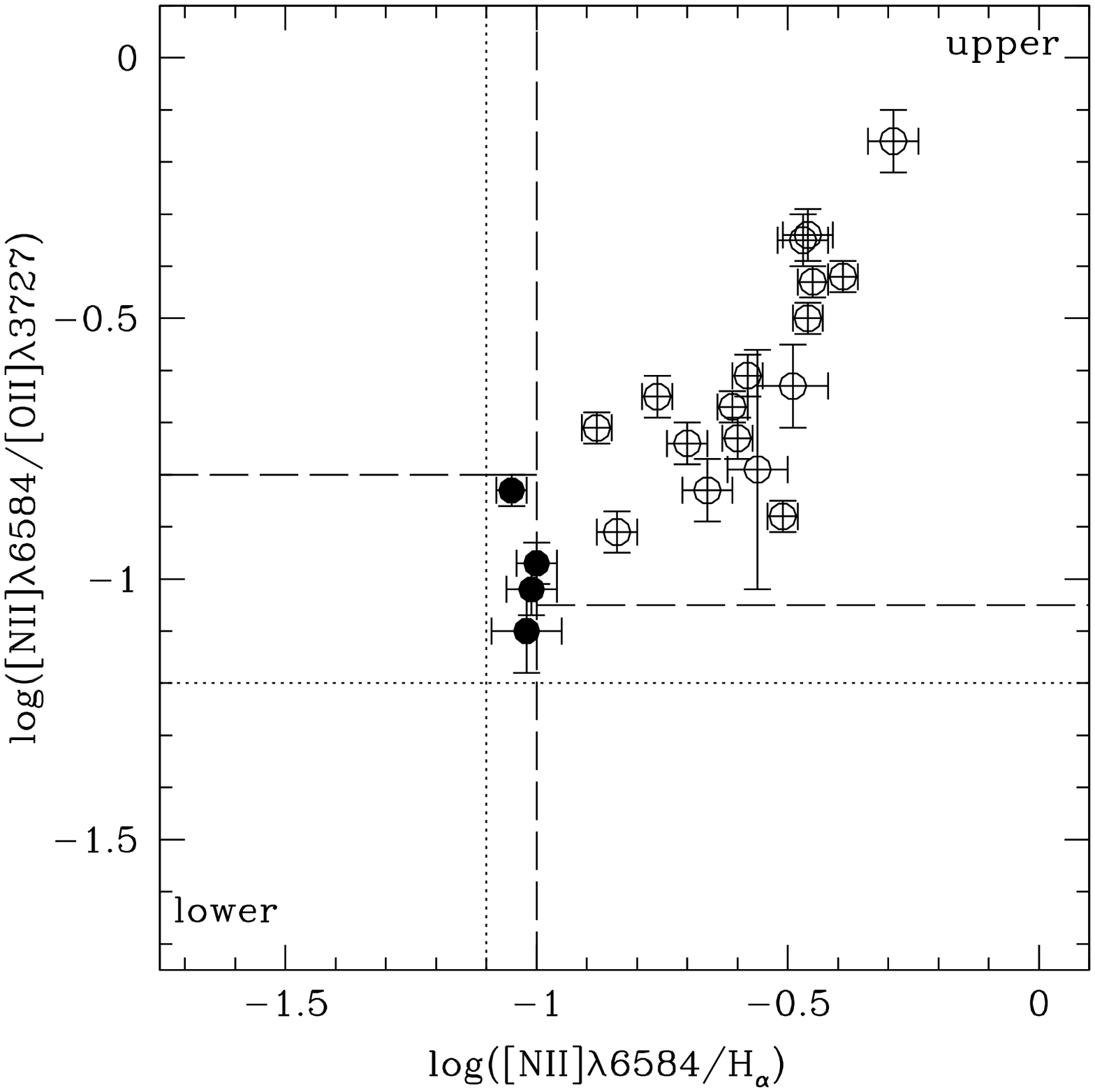}
  \caption{The relation between $R_{23}$ and 12+log(O/H) is a non-monotonic function.  Here, we plot 
    log([NII]$\lambda$6584/[OII]$\lambda$3727) vs. log([NII]$\lambda$6584/H$\alpha$).  The dashed lines mark
    the division between upper and lower branch regions described by \citet{Contini2002}.  The dotted lines mark the division
    between the upper and lower branches adopted by \citet{Kewley2008}.  As described in the text,
    we use this diagram to infer that most of the HII regions, marked with open circles, lie on the upper branch of the 
    $R_{23}$ vs. metallicity relation.  It is immediately obvious, however, that there is an ambiguous region in which an object
    may lie on the upper branch of the relation according to one method and the lower branch according to the other.
    The filled circles mark the regions which lie in the ambiguous area and are not clearly on the upper or lower branch 
    according to both line ratio cuts.  We label these as ``turnaround'' regions.}
  \label{fig:branches}
\end{figure}

We use both the [NII]$\lambda$6584/[OII]$\lambda$3727 and [NII]$\lambda$6584/H$\alpha$ ratios to break this 
upper vs. lower branch degeneracy, comparing the methods of \citet{Contini2002} and \citet{Kewley2008}.
From \citet{Contini2002},  HII regions with 
\begin{equation}
  \log\left( \frac{\mbox{[NII]}}{\mbox{[OII]}}\right) > -1.05
  \mbox{~~~and~~~}
  \log\left( \frac{\mbox{[NII]}}{\mbox{H}\alpha}\right) > -1 
\end{equation}
lie on the upper branch, and HII regions with  
\begin{equation}
  \log\left( \frac{\mbox{[NII]}}{\mbox{[OII]}}\right) < -0.8
  \mbox{~~~and~~~}
  \log\left( \frac{\mbox{[NII]}}{\mbox{H}\alpha}\right) < -1
\end{equation}
lie on the lower branch.
\citet{Kewley2008}, however, choose a division between the upper and lower branches at
the following values:
\begin{equation}
  \log\left( \frac{\mbox{[NII]}}{\mbox{[OII]}}\right) \sim -1.2
   \mbox{~~~and~~~}
   -1.3 < \log\left( \frac{\mbox{[NII]}}{\mbox{H}\alpha}\right) \lesssim -1.1 
\end{equation}
We plot the values of these line ratios for all regions and overlay the branch divisions
of \citet{Contini2002} as dashed lines and \citet{Kewley2008} as dotted lines in Fig. \ref{fig:branches}.  
Using these line ratio cuts, most of the HII regions in our sample lie unambiguously on the upper branch of 
the $R_{23}$ vs. O/H relation, according to both methods.  None of the regions in our sample lie on the lower branch. 
Four of the regions, however, lie on the lower branch according to \citet{Contini2002} and on the upper
branch according to \citet{Kewley2008}.  We mark these ambiguous 
regions as ``turnaround'' (T) regions and keep them separately visible in all figures.  We note that two of these 
T regions do have an electron temperature derived abundance.  We will address our treatment of these T regions 
in the next section.


\subsubsection{$R_{23}$ Metallicity Calibrations}
\label{empthesec}

In order to derive oxygen abundances from the $R_{23}$ parameter, we calculated 12+log(O/H) using two methods, 
an ``empirical'' and a ``theoretical'' calibrations.  Because theoretical calibrations generally overestimate oxygen 
abundances and empirical calibrations underestimate them \citep{Kennicutt2003}, we derive abundances using 
two methods, one theoretical calibration and one empirical calibration, following the procedure described in 
\citet{Moustakas2010}.  The empirical calibration used is based on the oxygen abundances of observed HII regions 
directly determined from temperature sensitive lines from \citet{PT2005}, hereafter \citetalias{PT2005}.
The theoretical calibration used is based on the relationship between line strengths and metallicity from 
photoionization models of HII regions from \citet{KK2004}, hereafter \citetalias{KK2004}.

For the empirical abundance calibration, we have the following equations from \citetalias{PT2005} for the upper 
and lower branches of the $R_{23}$ vs. O/H relation:
\begin{equation}
  12 + \log{\mbox{(O/H)}}_{\mbox{lower}} = 
  \frac{R_{23}+106.4+106.8P-3.40P^2}{17.72+6.60P+6.95P^2-0.302R_{23}} 
\end{equation}
and
\begin{equation}
  12 + \log{\mbox{(O/H)}}_{\mbox{upper}} = 
  \frac{R_{23}+726.1+842.2P+337.5P^2}{85.96+82.76P+43.98P^2+1.793R_{23}},
\end{equation}
where 
\begin{equation}
  P \equiv \frac{\mbox{[OIII]}\lambda\lambda4959,5007}{\mbox{[OII]}\lambda3727
    +\mbox{[OIII]}\lambda\lambda4959,5007}.
\end{equation}

In Table \ref{tab:R23abund}, we list our calculated values of 12+log(O/H) via this method as well as the 
values of $P$ for each region.  

For the theoretical abundance calibration, we have the following equations from \citetalias{KK2004}:
\begin{equation}
  12 + \log{\mbox{(O/H)}}_{\mbox{lower}} = 
  9.40+4.65x-3.17x^2-\log(q)(0.272+0.547x-0.513x^2)
\end{equation}
and
\begin{multline}
  12 + \log{\mbox{(O/H)}}_{\mbox{upper}}=9.72-0.777x-0.951x^2-0.072x^3-0.811x^4 \\
  -\log(q)(0.0737-0.0713x-0.141x^2+0.0373x^3-0.058x^4),
\end{multline}
where
$x \equiv \log(R_{23})$ and $q$ is the ionization parameter given by
\begin{multline}
  \log(q)=[32.81-1.153y^2+z(-3.396-0.025y+0.1444y^2)] \\
  \times[4.603-0.3119y-0.163y^2+z(-0.48+0.0271y+0.02037y^2)]^{-1}
\end{multline}
in units of cm s$^{-1}$.  In these equations $z \equiv 12+\log$(O/H), $y \equiv \log$(O$_{32}$), and 
\begin{equation}
  \mbox{O}_{32} \equiv \frac{\mbox{[OIII]}\lambda\lambda4959,5007}
       {\mbox{[OII]}\lambda3727}.
\end{equation}
The 12+log(O/H) formulas for both the upper and lower branches of this calibration are a function of $\log(q)$, 
which is dependent upon 12+log(O/H), requiring an iterative calculation to converge upon a solution.  We 
were able to derive an $R_{23}$ oxygen abundance from this calibration for all HII regions in our sample.
Our values of 12+log(O/H) and $O_{32}$ from this calibration are listed in Table \ref{tab:R23abund}.  

\begin{figure}
  \centering
  \vspace{.5truein}
  \epsfig{width=0.5\linewidth,file=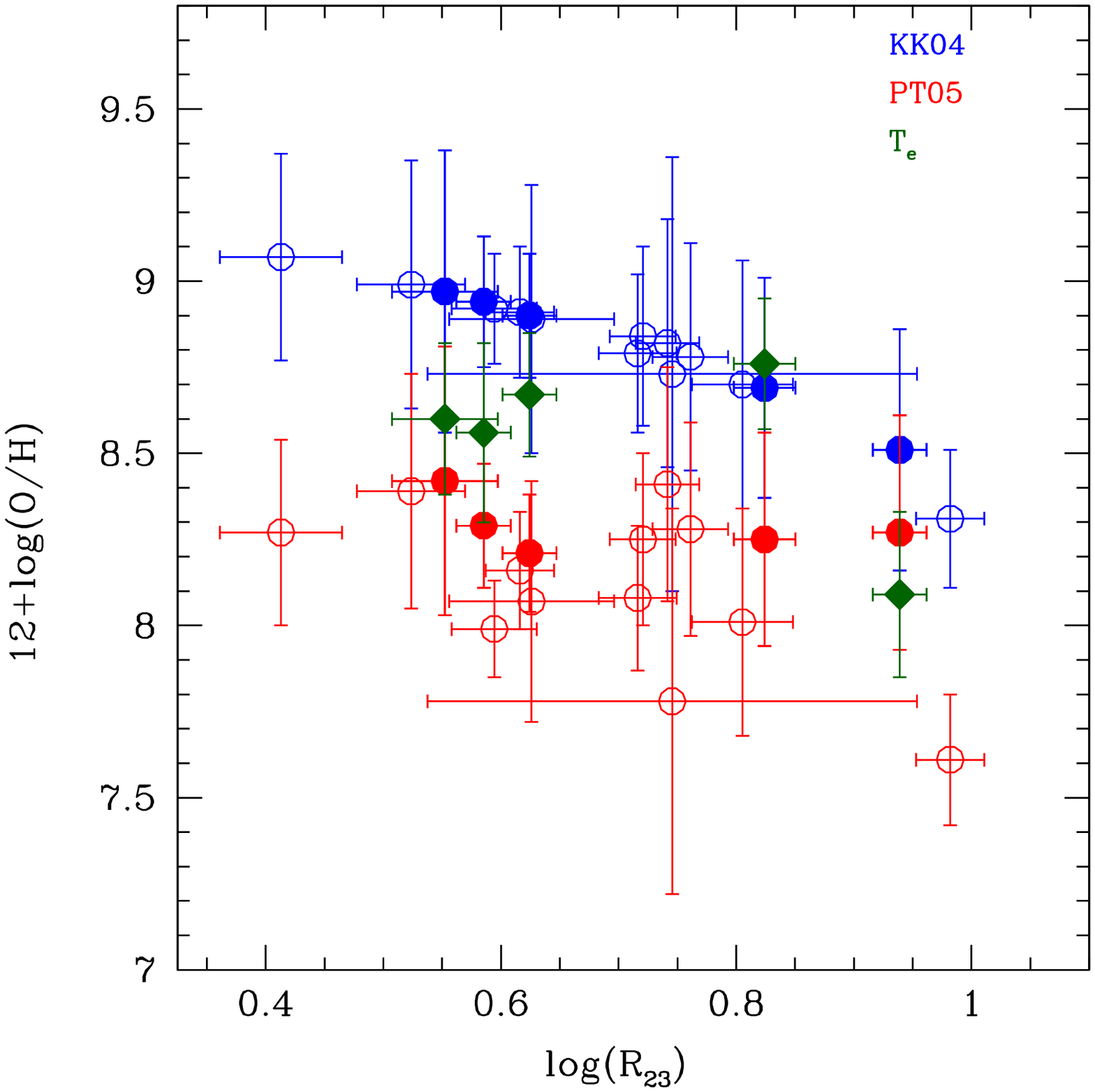} \\
  \vspace{.75truein}
  \epsfig{width=0.5\linewidth,file=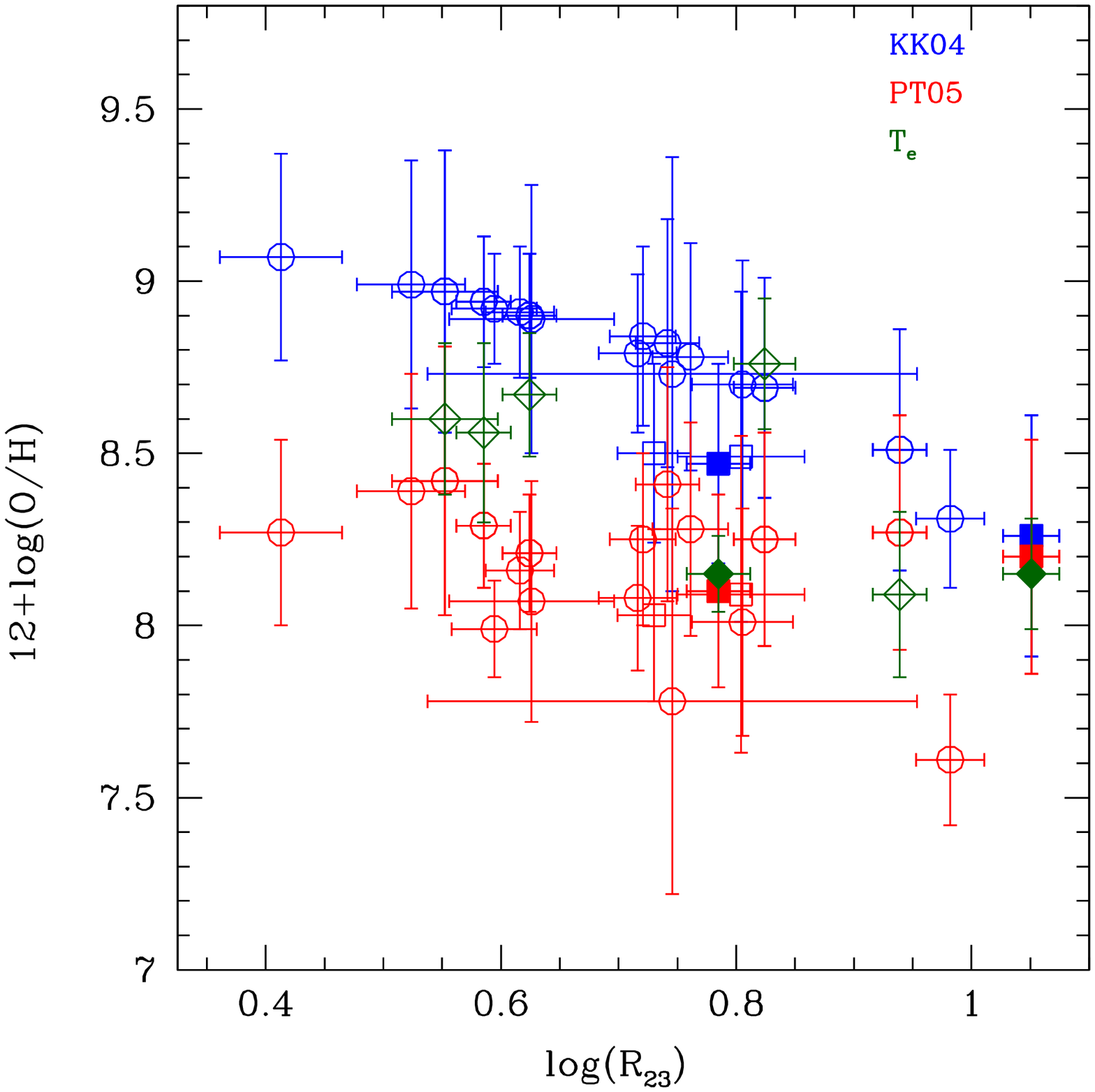}
  \caption{In the top graph, we show $R_{23}$ vs. 12+log(O/H) for the HII regions in our sample that lie on the upper branch. 
    We show the values for 12+log(O/H) derived with the \citetalias{KK2004} \citep{KK2004} method (blue) vs. the 
    \citetalias{PT2005} \citep{PT2005} method (red).  The filled green diamonds and the filled circles show the temperature derived 
    abundances and the strong line abundances, respectively, for the same HII region.  The \citetalias{KK2004} 
    calibration gives a higher upper branch than the \citetalias{PT2005} calibration.
    In the bottom graph, we now include the  turnaround regions here as squares.   Two 
    turnaround HII regions also have temperature derived abundances, and we mark all derived metallicities 
    for these two objects with filled points.  } 
  \label{fig:R23vsOH}
\end{figure}

In Fig. \ref{fig:R23vsOH}, we plot the $R_{23}$ vs. metallicity relation for our HII regions for both the 
theoretical \citetalias{KK2004} and empirical \citetalias{PT2005} metallicity calibrations.  
In the top graph, we are showing only the HII regions on the upper branch, to compare the two calibrations.
The \citetalias{KK2004} 
strong line metallicity calibration gives an upper branch on the $R_{23}$ vs. metallicity relation that lies higher 
than the \citetalias{PT2005} strong line metallicity calibration.  The solid points mark upper branch HII regions that 
have both a temperature derived metallicity (filled diamonds) and a strong line metallicity (filled circles).  The
temperature derived metallicities are in agreement with the \citetalias{KK2004} and 
\citetalias{PT2005} metallicities, within errors, though are, on average, between the two calibrations.

For each turnaround (T) region, we average the upper and lower branch solutions of the \citetalias{KK2004} 
calibration and average the upper and lower solutions of the \citetalias{PT2005} calibration to get a turnaround 
metallicity estimate from each calibration.  In the bottom graph of Fig. \ref{fig:R23vsOH}, we plot the final abundances
for all regions, with the T regions as squares on the $R_{23}$ vs. O/H relation.   The filled points again mark two T HII 
regions with both a strong line metallicity and a temperature metallicity.  Here, the temperature metallicities agree with
both strong line abundances, within the errors, but may be in closer agreement with the \citetalias{PT2005} derived 
abundances.  In Table \ref{tab:R23abund}, we list the values of $R_{23}$ for each region and mark which branch 
we assume each region lies on. 


\subsubsection{[NII]/[OII] Metallicity Calibrations}
\label{NIIOIIsec}

In addition to the two $R_{23}$ based strong line metallicity calibrations, we also use two calibrations based on the
[NII]$\lambda$6584/[OII]$\lambda$3727 lines flux ratio.  While we used the [NII]/[OII] ratio to determine the upper and 
lower branches of the $R_{23}$ vs. metallicity relation, the [NII]/[OII] ratio is itself also a function of metallicity.  
Additionally, it does not suffer from the upper and lower branch ambiguity of the $R_{23}$ calibrations, since it is
a monotonic function.

The first calibration we use is a theoretical calibration based on photoionization and stellar population synthesis 
models of \citet{KD2002}, hereafter \citetalias{KD2002}, as follows:
\begin{equation}
  \log{\mbox{([NII]/[OII])}} = 1106.87 - 532.154Z + 96.3733Z^2 - 7.81061Z^3 + 0.239282Z^4,
\end{equation}
where \begin{math} Z = 12 + \log{\mbox{(O/H)}} \end{math}.
This equation assumes ionization parameter $q$ = 2 $\times$ 10$^7$ cm s$^{-1}$, which is appropriate given 
the range of ionization parameters for our sample ($\sim$ 1-7 $\times$ 10$^7$ cm s$^{-1}$).  To find
values of $Z$, we use the IDL task \textsc{fz\_roots}, based on the numerical recipe \textsc{zroots} \citep{Press1992}
which finds the roots of an $n$-order polynomial.  We list the values of log([NII]/[OII]) and oxygen abundances 
derived using this calibration in Table \ref{tab:R23abund}.  Since the two lines are strong, the errors are dominated by 
a systematic error of $\sim$0.1 dex given by the rms of the line fit defining the calibration.  The abundances we derive 
from this calibration agree most closely with the \citetalias{KK2004} $R_{23}$ abundances, which is not surprising 
since both calibrations are based on the same models.  This calibration is simpler than that of \citetalias{KK2004}, 
however, since we do not explicitly calculate an ionization parameter for each HII region, but assume one appropriately 
chosen value for all.

We also use an empirical calibration from \citet{B07}, hereafter \citetalias{B07}, based on data from a number of 
HII regions with electron temperature abundances:
\begin{equation}
  12 + \log(\mbox{O/H}) = 8.66 + 0.36x - 0.17x^2, 
\end{equation}
where \begin{math} x = \log(\mbox{[NII]/[OII]}) \end{math}.  Here the errors again are dominated by the systematic
error of $\sim$0.2 dex from the rms of the fit to the log([NII]/[OII]) metallicity relation.  We list the values of 
12+log(O/H) from this calibration in Table \ref{tab:R23abund}.  

The various calibrations all seem to have different metallicity scales (i.e. they are offset from each other).
We will therefore focus in our discussion on the metallicity {\it gradients} implied by each of them 
in \S\ref{abundancegradientsec}.


\section{Holmberg IX and KDG 61}
\label{dwarfsec}

Several objects in our sample deserve special discussion.  These include the brightest ionized nebula near the claimed
tidal dwarf candidate Holmberg IX (region 35 in our labeling), a bright HII region near KDG 61 (region 28), and M\"unch 1
(region 21).  We will discuss M\"unch 1 in the next section.

\begin{figure}
  \centering
  \begin{tabular}{cc}
    \epsfig{width=0.3\linewidth,file=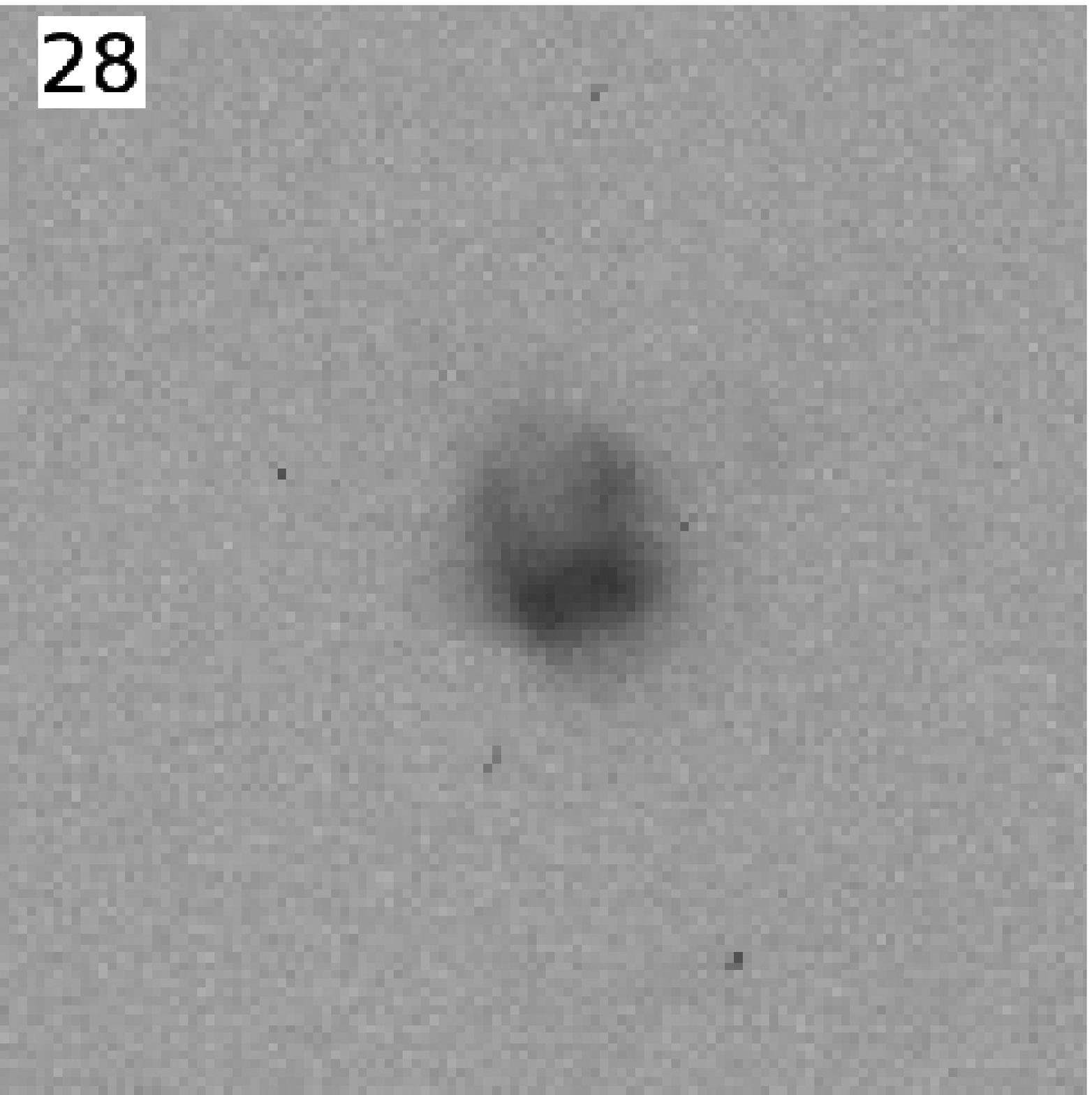} &
    \epsfig{width=0.3\linewidth,file=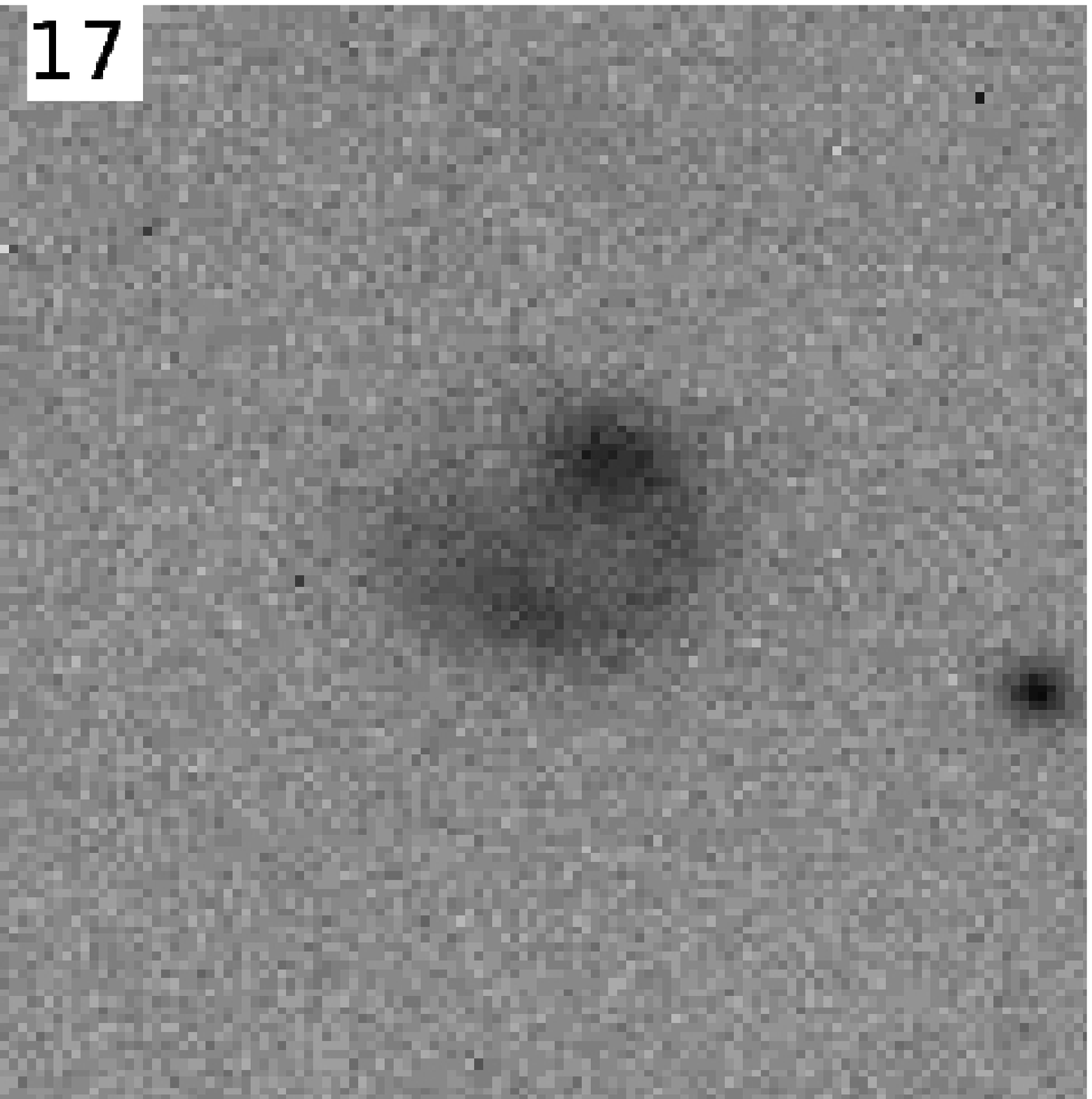} \\ \\
    \epsfig{width=0.3\linewidth,file=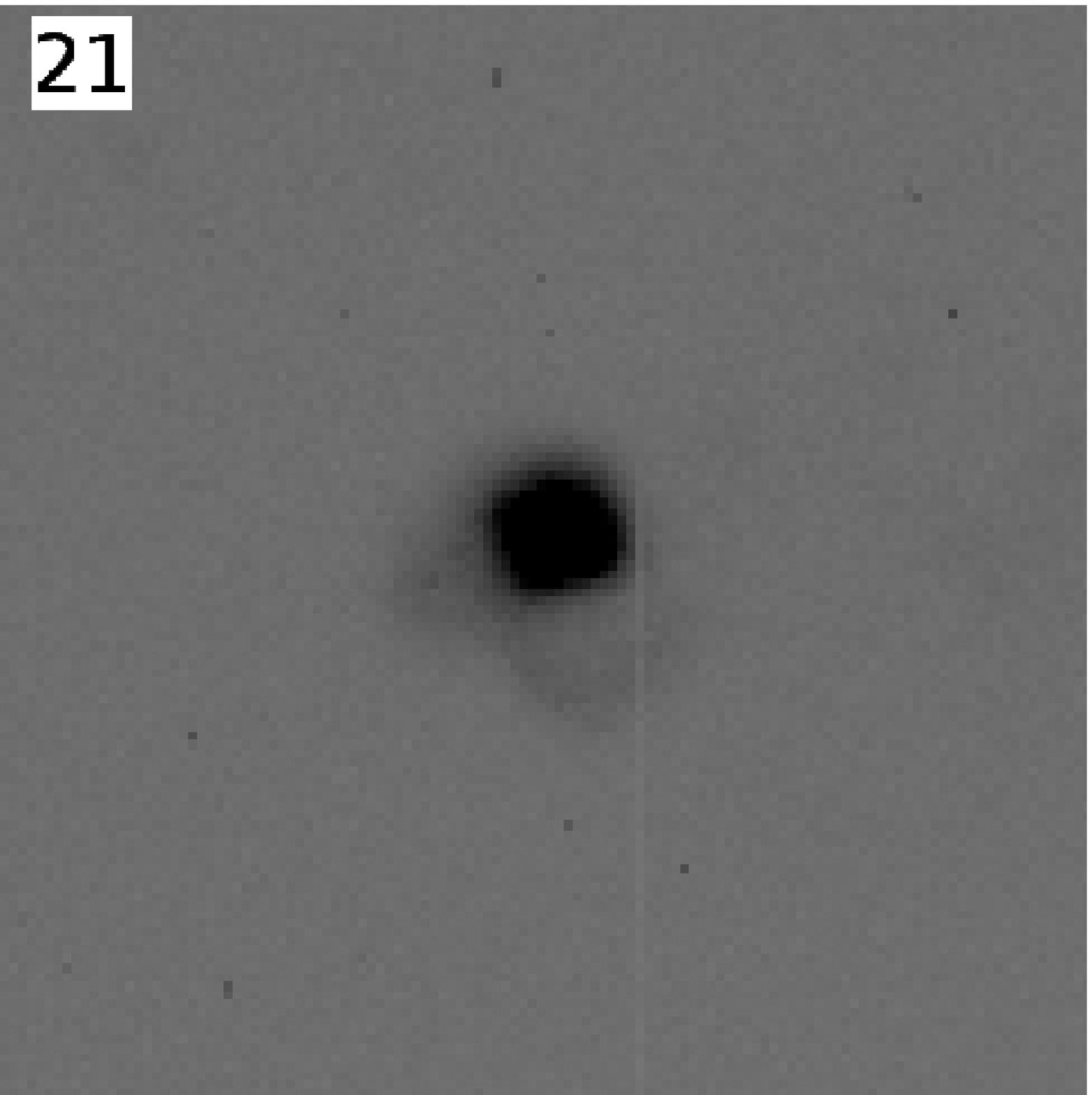} &
    \epsfig{width=0.3\linewidth,file=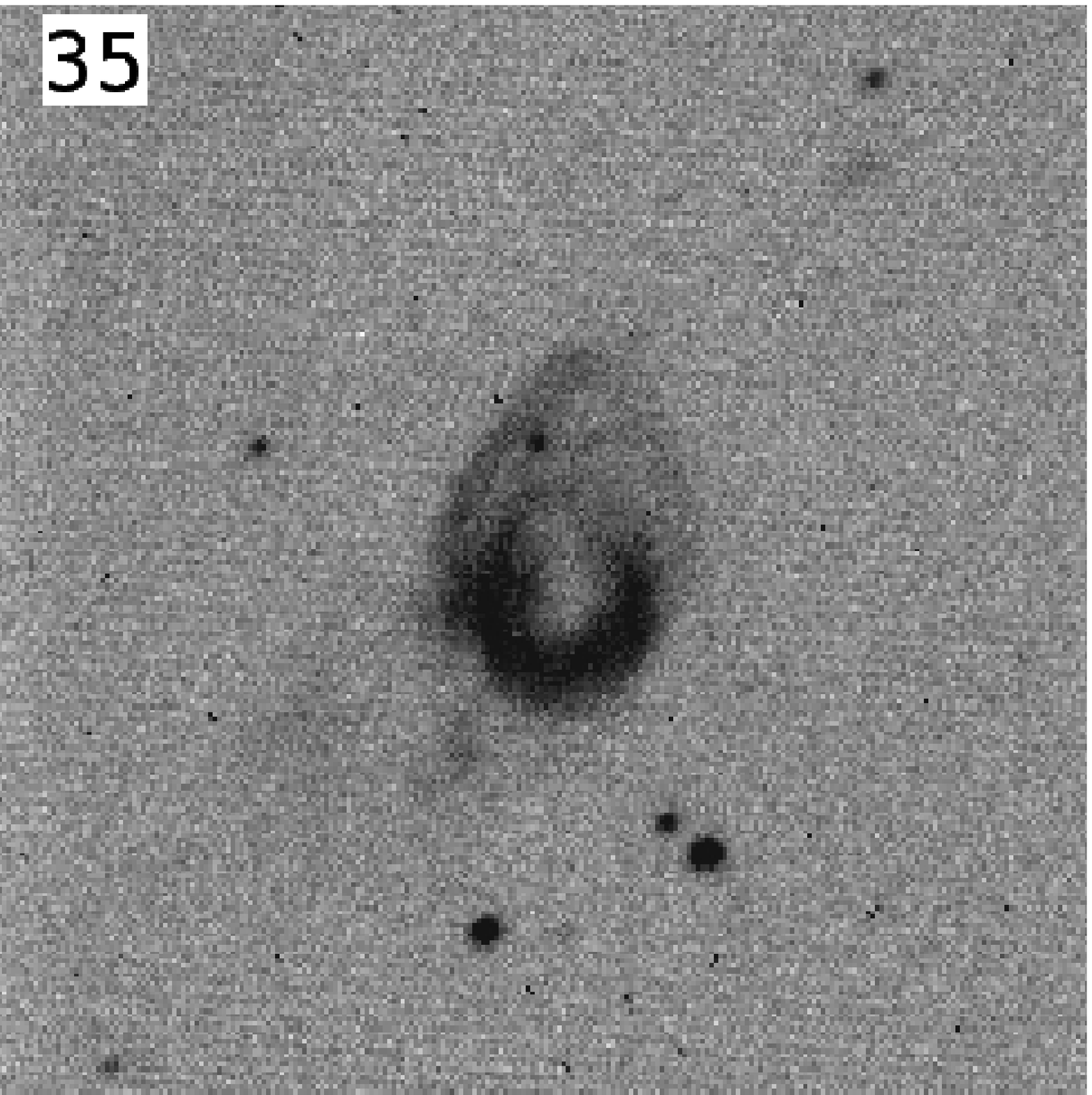} \\ \\
    \epsfig{width=0.35\linewidth,file=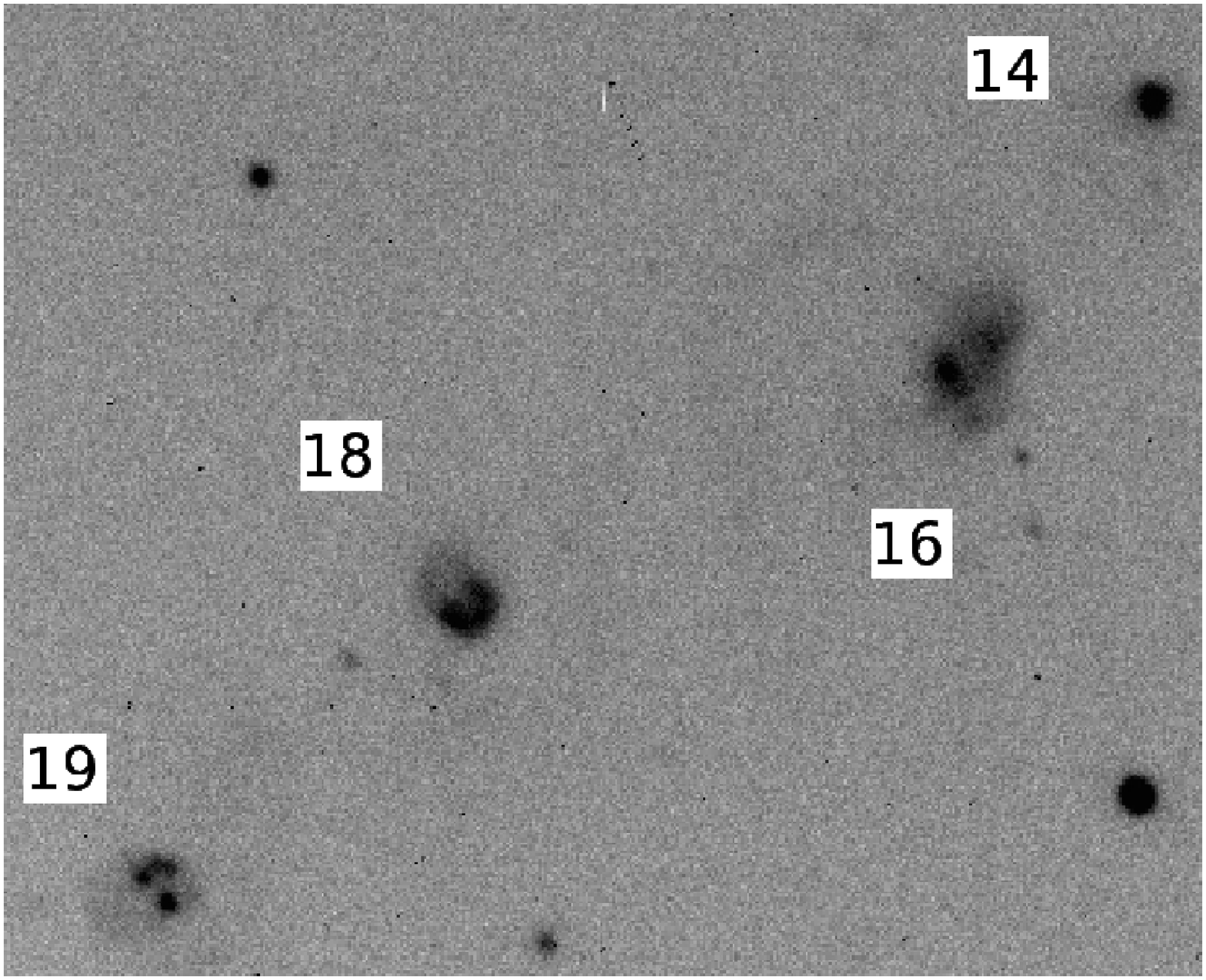} &
    \epsfig{width=0.30\linewidth,file=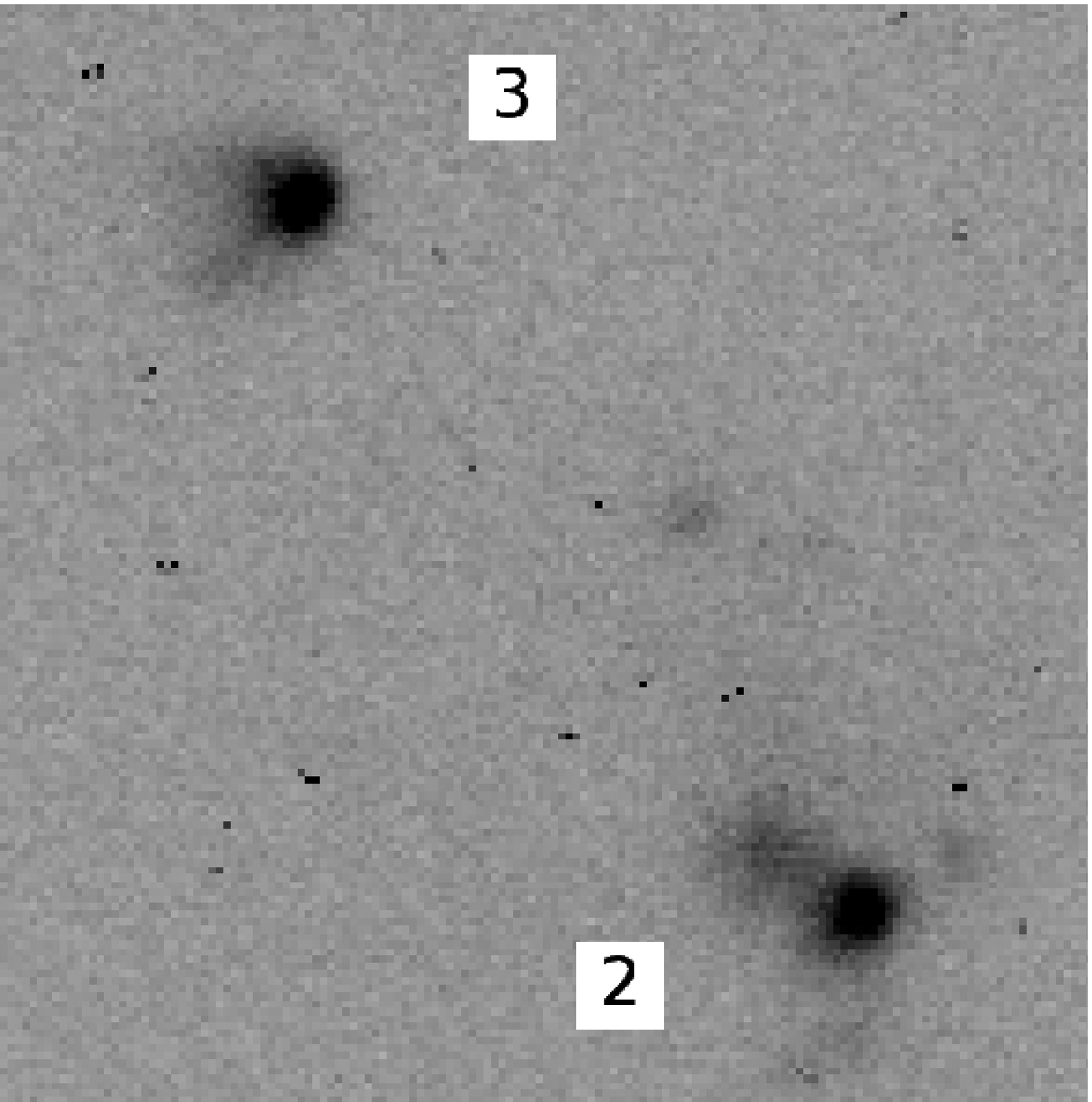} 
  \end{tabular}
  \caption{
    {\footnotesize H$\alpha$ SPICAM images of several interesting HII regions.  The red continuum
        light has not been removed in these images to show stars for reference.
      {\bf Top Left:} HII region 28, associated with the dwarf spheroidal galaxy KDG 61 (not shown to the lower right.)  
      It has a horseshoe shape morphology with a cavity-like feature to the north.
      {\bf Top Right:} HII region 17, located to the north of M81 with a comma shape morphology.
      {\bf Middle Left:} HII region 21 (M\"unch 1), a bright (1.5 $\times 10^{38}$ ergs s$^{-1}$) region
      in the Southern tidal arm of M81.  It has two prominent loops of gas protruding from a bright central region.
      {\bf Middle Right:} HII region 35, associated with the tidal dwarf HoIX, whose main body is outside 
      this image to the lower right shown in full in Fig \ref{fig:colorHoIX}).  
      It has a similar horseshoe morphology to KDG 61.  The source within the shell just to the lower
        left of center is a star.
      {\bf Bottom Left:} String of HII regions in the Southern tidal arm of M81.  From left to right, region 
      19 has four visible knots of gas.  Region 18 has a horseshoe shape similar to KDG 61 and HoIX.  
      Region 16 has a very flocculent morphology.  Region 14 appears smooth and round.
      {\bf Bottom Right:} Two HII regions in the Northern tidal arm of M81.  On the left, region 3 has a 
      bright concentration with extended features.  On the right, region 2 has a bright region with several 
      flocculent tufts of gas.}   
  }
  \label{fig:spicam}
\end{figure}

\begin{figure}
  \centering
  \epsfig{width=0.9\linewidth,file=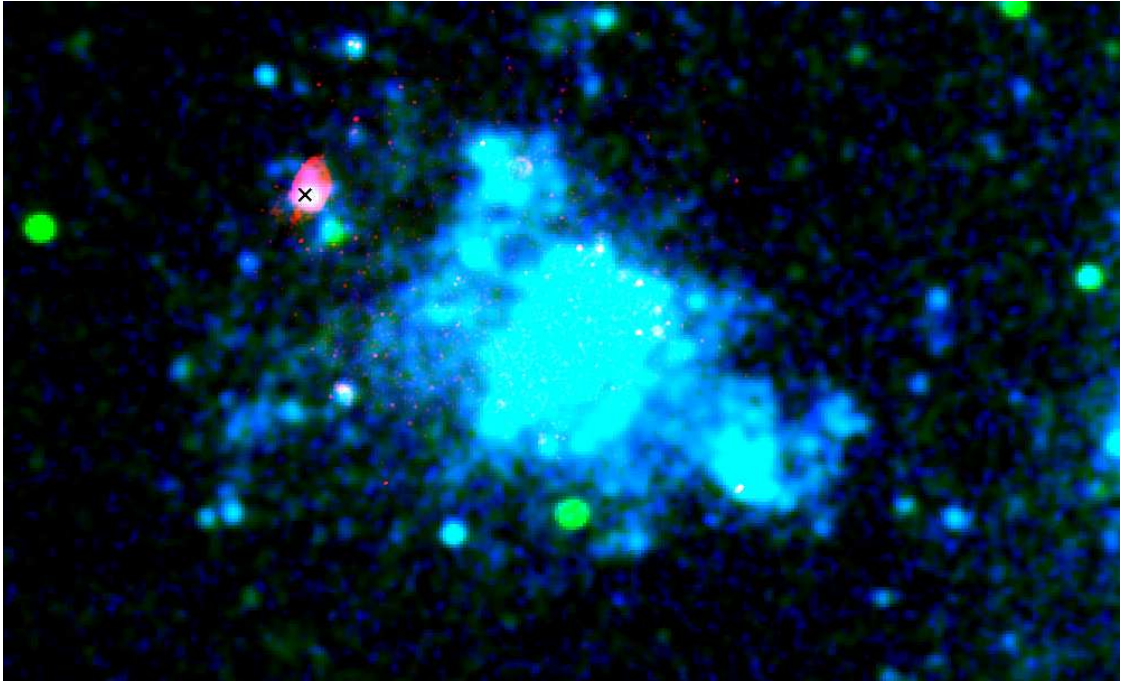}
  \caption{Three color image using the SPICAM continuum subtracted H$\alpha$ image (red) combined with GALEX FUV 
    (blue) and NUV (green) of Holmberg IX and the HII region referred to here as region 35.  Note that the bright HII region is
    offset from the main body of the dwarf galaxy, though there is a peak in the UV emission within the lower central part of the 
    HII region.  The HII region does, however, lie within a peak in the HI column density map, as can be seen in Fig. \ref{fig:HIIregs}.
    The 'x' marks the location of the x-ray source HoIX X-1 (or M81 X-9), which is contained within the extent of the 
    H$\alpha$ emission \citep{Immler2001,Wang2002}.}
  \label{fig:colorHoIX}
\end{figure}

\begin{figure}
  \centering
  \epsfig{width=0.5\linewidth,file=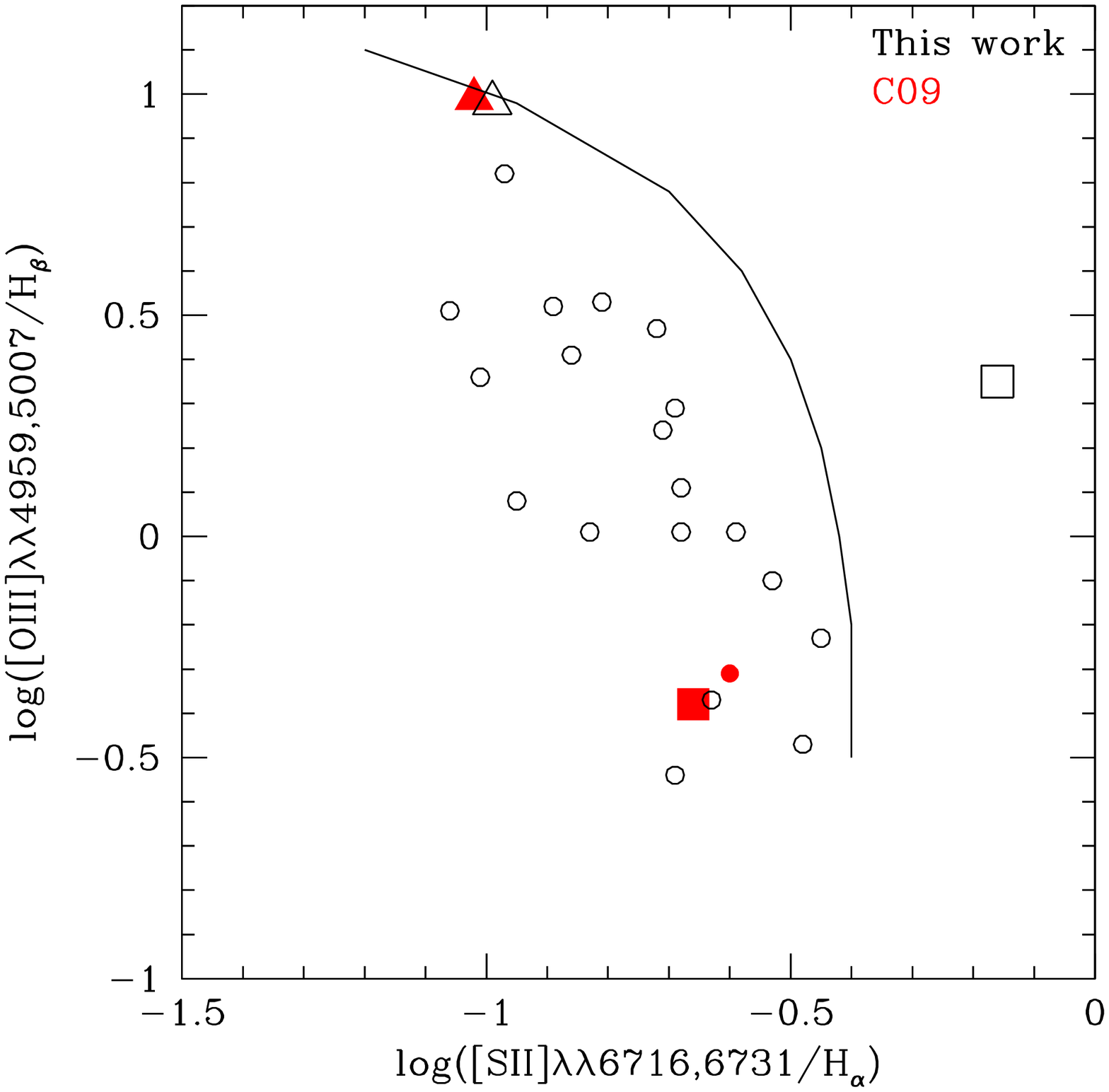}
  \caption{BPT diagram \citep{Baldwin1981} of our HII region data (open points) compared with three data points (filled) from 
    \citetalias{Croxall2009} \citep{Croxall2009}.  The curved line marks the empirical division between HII region-like 
    objects (to the lower left) and AGN (upper right), taken from \citet{Osterbrock2006}.  The two triangles mark our 
    region 28 (or KDG 61-9).  Both our data and that of \citetalias{Croxall2009} show a highly excited HII region.  The 
    square points mark the bright ionized gas region associated with HoIX (MH9/MH10).  Our data for this region 
    shows evidence of shock ionization, whereas the data of \citetalias{Croxall2009} does not.  We attribute this to the 
    spatial offset of our long-slit spectrum as compared to the GMOS spectrum of \citetalias{Croxall2009}, and the 
    complex ionization conditions in this large object. } 
  \label{fig:BPT}
\end{figure}

The large object near the dwarf galaxy Holmberg IX was first identified in H$\alpha$ imaging by \citet{Miller1994} as three 
HII regions- MH9, MH10, and MH11.  We show an H$\alpha$ plus continuum image of the entire object (our region 35)
in Fig. \ref{fig:spicam} and an RGB image combining H$\alpha$ (red), GALEX NUV (green), and 
GALEX FUV (blue) of the HII region and the dwarf galaxy HoIX in Fig. \ref{fig:colorHoIX}.  From the image in Fig. \ref{fig:colorHoIX},
it is obvious that the bright H$\alpha$ emission is offset from the main body of the dwarf galaxy, but coincides with a peak
in the HI distribution (see Fig. \ref{fig:HIIregs}).  Studies of Holmberg IX suggest that the dwarf galaxy itself has a tidal origin, given its location in the tidal
HI streams and young stellar population dominated by stars less than $\sim$200 Myr old, which is consistent with star formation triggered by the
past interaction of M81 and M82 \citep{Sabbi2008,Weisz2008,Hoversten2011}.   
The nature of the large ionized object is discussed in \citet{Miller1995} as a U-shaped ``supershell'' approximately 250 pc 
wide and 475 pc in the north-south dimension, with strong [SII], [NII], [OI], and [OII] emission indicative of shock 
heating.  The supershell is the optical counterpart of the x-ray source ULX HoIX X-1 (M81 X-9) \citep{Fabbiano1988}.
In Fig. \ref{fig:colorHoIX}, we mark the location of the x-ray source, which is nested in the lower eastern portion of 
the ``U'' shape and appears to be confined within the spatial extent of the H$\alpha$ emission \citep{Immler2001,Wang2002}.
A recent HST/ACS study by \citet{Grise2011} finds an OB association of young stars ($\lesssim$20 Myrs) in MH10. 
\citetalias{Croxall2009} find an oxygen abundance of 12+log(O/H)=8.91$\pm$0.20 for this HII region (called UGC 
5336-3 in that paper), using a strong line calibration from \citet{McGaugh1991}.  The data comes from the Gemini 
Multiple Object Spectrograph (GMOS) and is focused on the lower eastern section of the ``U''.  The slit of our spectrum 
for this object was aligned so as to include both MH9 and MH10, both halves of the ``U''.  We find lower abundances 
of 12+log(O/H) = 7.61$\pm$0.19 \citepalias{PT2005}, 8.31$\pm$0.20 \citepalias{KK2004}, 8.68$\pm$0.10 
\citepalias{KD2002}, and 8.21$\pm$0.20 \citepalias{B07}.  
These theoretical model based calibrations assume that the emission is solely due to photoionization,
but the strong [SII] emission in our data suggests the presence of some shock ionization.  In Fig. \ref{fig:BPT}, 
we plot a BPT emission line diagnostic diagram \citep{Baldwin1981} showing the [OIII] and [SII] emission lines
normalized to Balmer lines for our HII regions, as well as the data of \citetalias{Croxall2009}.  Our line ratios 
indeed place the section of the object we observed in the shock excited range, but note that the 
\citetalias{Croxall2009} data are not in that same section.  Clearly, this object appears to be a complex mixture
of H$\alpha$ emission from both photoionization and shock ionization.  We compare our line ratios to the shock 
models of \citet{Allen2008} and find that our data is consistent with a low velocity shock ionized region with a 
12+log(O/H) metallicity between that of the LMC (8.35) and solar (8.93), which is in agreement with the abundances 
we derive using the \citetalias{KK2004} and \citetalias{KD2002} methods.  The data of \citetalias{Croxall2009} 
for this region does not show evidence of shock ionization.  We take the emission line fluxes from the 
\citetalias{Croxall2009} data and re-calculate strong line abundances of 8.21$\pm$0.16 \citepalias{PT2005}, 
9.02$\pm$0.18 \citepalias{KK2004}, 8.76$\pm$0.10 \citepalias{KD2002}, and 8.30$\pm$0.20 \citepalias{B07}.  
The \citepalias{KD2002} calibration of the \citetalias{Croxall2009} data for this object is most consistent 
with the metallicity of our data assuming some shock ionization for this region.  

We also observe an HII region near the dwarf galaxy KDG 61, region 28 in our data (see Fig. \ref{fig:spicam}) 
and KDG 61-9 in \citetalias{Croxall2009}.  This object is a highly ionized object with strong [OIII] emission
as shown in Fig. \ref{fig:BPT}.  Like \citetalias{Croxall2009}, we also detect the emission line HeII $\lambda$4686 
which indicates that this object may have a central Wolf-Rayet star.  We derive an electron temperature oxygen 
abundance of 8.15$\pm$0.11 and strong line abundances of 8.20$\pm$0.34 \citepalias{PT2005}, 8.26$\pm$0.35 
\citepalias{KK2004}, 8.71$\pm$0.10 \citepalias{KD2002}, and 8.24$\pm$0.20 \citepalias{B07} for this object.
\citetalias{Croxall2009} report an electron temperature abundance of 8.35$\pm$0.05, which is slightly higher 
than our value, since we measure T[OIII] to be $\sim$1800 K higher.  \citet{Makarova2010} find a radial 
velocity for the stellar light of KDG 61 of +221$\pm$3 km s$^{-1}$, whereas for this HII region they find a velocity of 
$-$123$\pm$6 km s$^{-1}$.  From the velocities, the authors conclude that this HII region is not bound to the dwarf spheroidal galaxy and is likely
a chance projection.  This region marks one of the most distant points in our abundance gradient, and its oxygen 
abundance is comparable to that of other regions at its radial distance.  If we compare our abundance results for this region
to the expected abundance for the dwarf galaxy KDG 61, we find higher abundances than what would be estimated
from the metallicity-luminosity relation for dwarf galaxies \citep[see e.g.,][]{Skillman1989}.  For a low luminosity
dwarf like KDG 61, with $M_B$=-12.83$\pm$0.30 \citep{Karachentsev2004}, \citetalias{Croxall2009} shows that 
the metallicity-luminosity relation for dwarf galaxies in the M81 group predicts an oxygen abundance of $\sim$7.6.   
We agree with \citetalias{Croxall2009} that the abundance of this region is more consistent with 
an origin in enriched gas from the tidal interactions of M81 rather than from the dwarf galaxy KDG 61.  We will discuss in detail the
possible effects of M81's interaction history on the abundances of the outer disk and the abundance gradient in \S\ref{discussionsec}.


\section{Abundance Gradient}
\label{abundancegradientsec}

To obtain deprojected distances for all HII regions in our sample, we assume a flat planar geometry, with a 
rotation angle of 157$^{\circ}$ and an inclination angle of 59$^{\circ}$ for M81 \citep{Kong2000}.  
Clearly, the assumption of a planar geometry with the same orientations as the M81 disk may be incorrect for the
outermost HII regions, given the role of tidal interaction in creating the HI tails.  However, our regions do not probe
all of the tidal tails and do not stray very far from M81 proper.  For regions along the southern- and northern- most
spiral arms, the deviation from the M81 disk is probably minor.  The southern arm extends into the HoIX region
(see Fig. \ref{fig:HIIregs}, our region 35.)  The regions where the radial distance to M81 is most uncertain then likely
include 26, 29, and 33 (all in Arp's loop), and region 28 (near KDG 61).  The latter is near the projected major axis of
M81, so its actual radial distance from M81 cannot be less than the one we use.  In the subsequent presentation
of the radial abundance gradient, it is good to keep in mind that these regions are the four outermost points in
radial distance.  We will use our electron temperature and $R_{23}$ abundance determinations to discuss the metallicity 
gradient out to $\sim$2.25$\times$$R_{25}$, or $\sim$33 kpc from the center.

\begin{figure}
  \centering
  \vspace{.75truein}
  \epsfig{width=0.5\linewidth,file=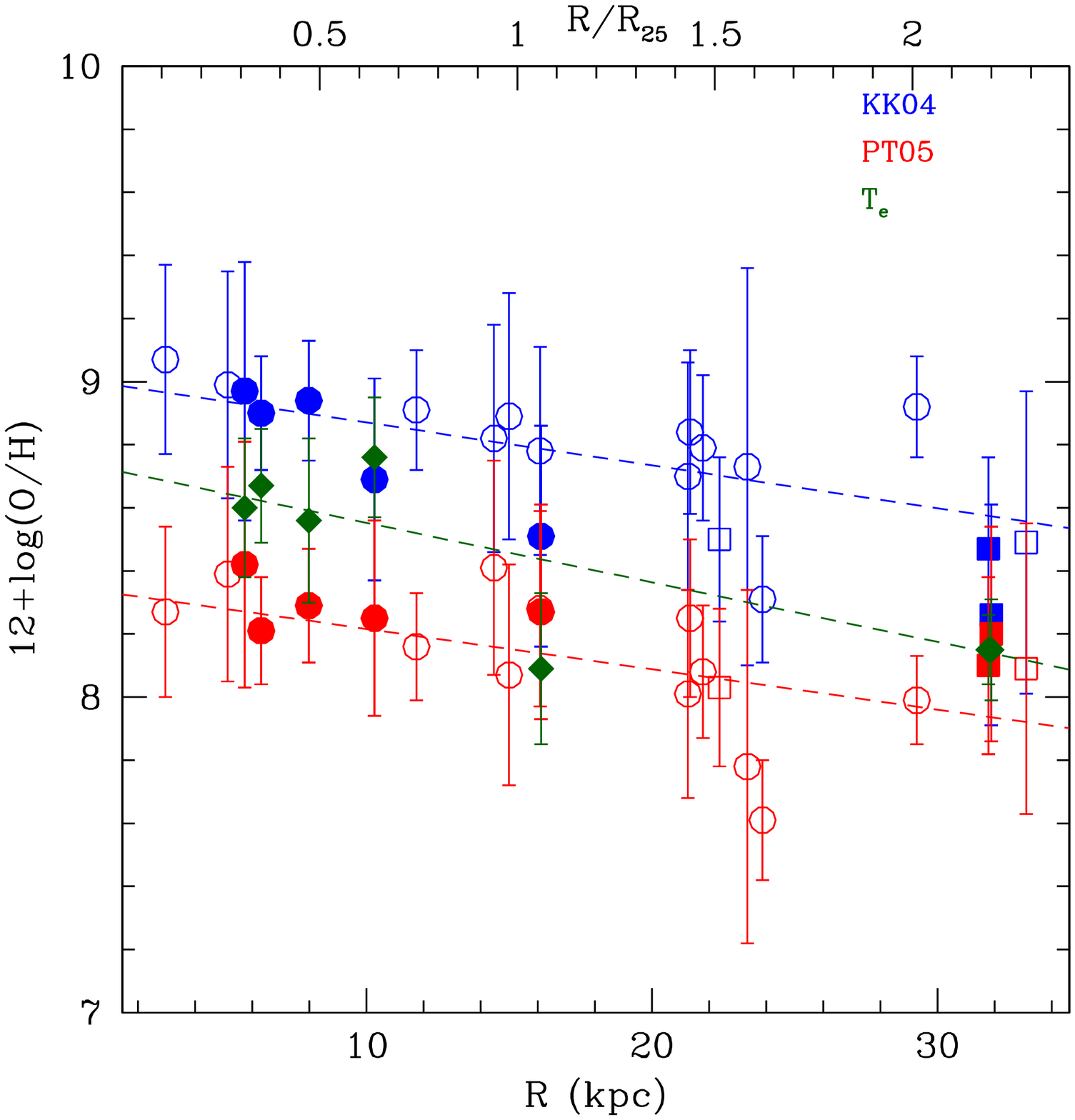} \\
  \vspace{1.0truein}
  \epsfig{width=0.5\linewidth,file=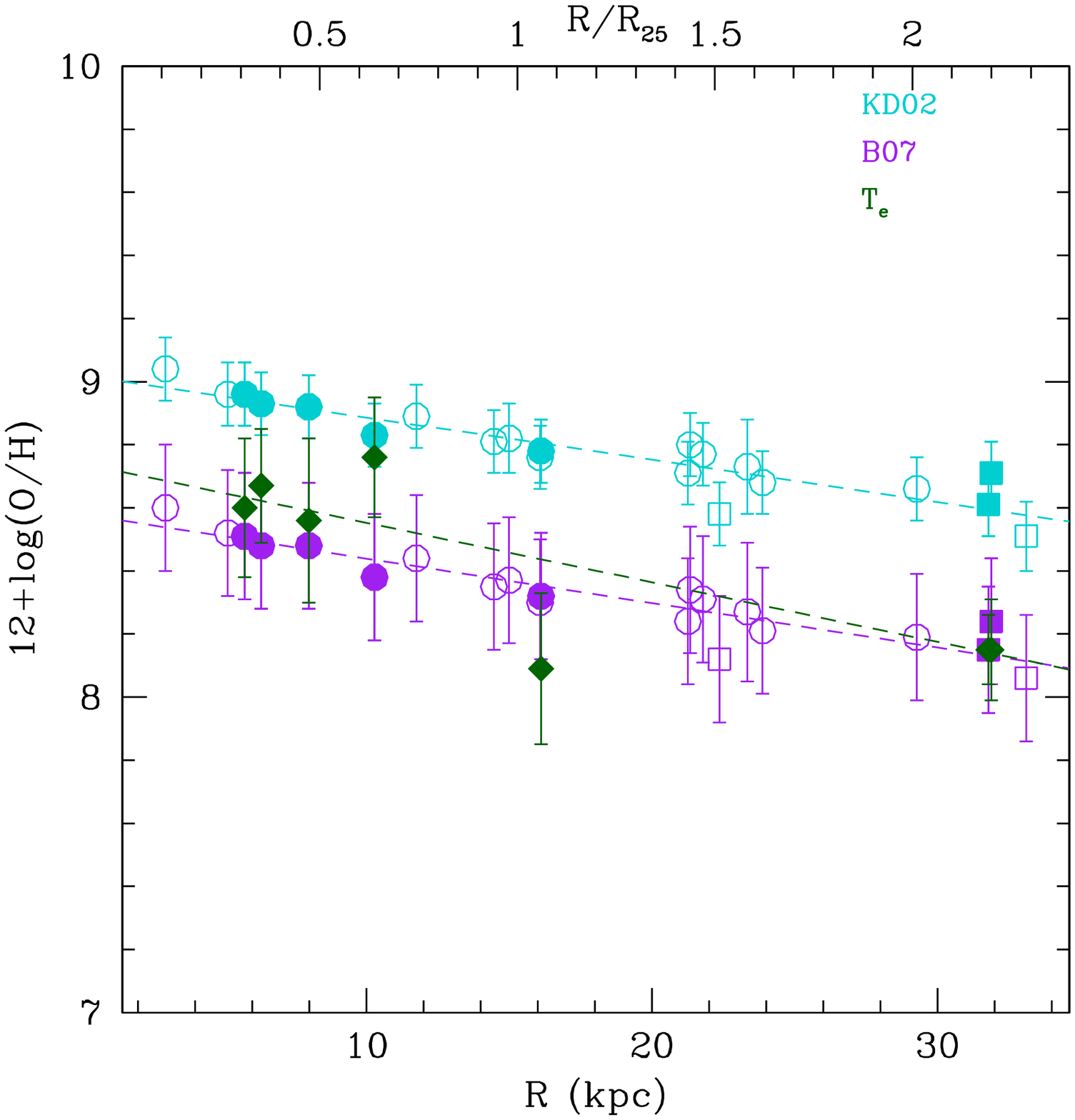}
  \caption{12+log(O/H) vs. galactocentric distance for the HII regions in our data set.  In the top graph, we show 
    the metallicities from both $R_{23}$ calibrations as compared to the electron temperature metallicities.  
    Blue points mark the abundances from the \citetalias{KK2004} calibration, and red points mark the abundances
    from the \citetalias{PT2005} calibration.  Turnaround regions are marked by squares.  Green diamonds show 
    the temperature derived abundances.  The strong line abundances that also have a temperature derived 
    abundance are marked with filled circles. M\"unch 1 is marked by the green diamond at $\sim$16 kpc and the corresponding 
    filled circles of the strong line abundances.  The dashed lines show our weighted least-squares fit to the $R_{23}$ 
    abundances and temperature abundances.  In the bottom graph, we compare the abundances from the
    \citetalias{KD2002} \citep{KD2002} and \citetalias{B07} \citep{B07} [NII]/[OII] metallicity calibrations 
    to the temperature derived abundances.  The slopes of the abundance gradients derived from all four strong 
    line methods are roughly the same. The temperature derived abundance gradient is slightly steeper. }
  \label{fig:OHvsDistance}
\end{figure}

In the top graph of Fig. \ref{fig:OHvsDistance}, we show abundance gradients from both the \citetalias{KK2004}
and \citetalias{PT2005} strong line $R_{23}$ metallicity calibrations as well as the electron temperature abundances 
for the HII regions in our sample.   The strong line abundances derived from the \citetalias{KK2004}
calibration are higher but show a gradient with a slope similar to the abundance gradient derived using 
the \citetalias{PT2005} calibration.  We use a weighted least-squares fitting routine
with uncertainties in both error and distance and derive a gradient of 
$\Delta(\log$(O/H))$/\Delta$R$_G=-$0.014$\pm$0.006 dex kpc$^{-1}$ from the \citetalias{KK2004} abundances and
$\Delta(\log$(O/H))$/\Delta$R$_G=-$0.013$\pm$0.006 dex kpc$^{-1}$ from the \citetalias{PT2005} abundances.  
The abundance gradient from electron temperature derived abundances is slightly steeper than those of the 
$R_{23}$ based calibrations, with a gradient of 
$\Delta(\log$(O/H))$/\Delta$R$_G=-$0.020$\pm$0.006 dex kpc$^{-1}$.
In the bottom graph of Fig. \ref{fig:OHvsDistance}, we plot our abundance gradient derived from the [NII]/[OII] 
based calibrations of \citetalias{KD2002} and \citetalias{B07} compared again to our electron temperature abundance gradient.
We find a metallicity gradient of
$\Delta(\log$(O/H))$/\Delta$R$_G=-$0.013$\pm$0.002 dex kpc$^{-1}$ from the \citetalias{KD2002} abundances and
 $\Delta(\log$(O/H))$/\Delta$R$_G=-$0.014$\pm$0.005 dex kpc$^{-1}$ from the \citetalias{B07} abundances,
similar to the gradients given by both $R_{23}$ metallicity calibrations.  The four strong line abundance calibrations
all give a consistently shallow negative metallicity gradient.  Of the two $R_{23}$ metallicity calibrations we use,
the \citetalias{PT2005} abundances appear to be in closer agreement to our electron temperature abundances.  
The \citetalias{B07} [NII]/[OII] calibration gives the strong line abundances closest to the electron temperature abundances 
of our data set.

We compare our results 
with previously published HII region abundance results from \citetalias{Garnett1987}, \citetalias{Stanghellini2010}, 
and three regions from \citetalias{Croxall2009}, which studied the abundances of HII regions in M81 dwarfs. 
Two of these HII regions are located near Holmberg IX (UGC5336-3 and UGC5336-12), and the third is located 
near KDG 61 (KDG 61-9, also our region 28).  The abundances quoted in \citetalias{Stanghellini2010} are derived 
from the electron temperatures from one or both of the [NII] and [OIII] lines.  Where only one of the lines is detected, 
the authors assume the same temperature for all ions.  In \citetalias{Garnett1987}, the authors use a theoretical 
calibration of the $R_{23}$ parameter from photoionization models.  For the HII regions we include here from the data 
set of \citetalias{Croxall2009}, the authors derive an electron temperature abundance for KDG 61-9 and use an 
$R_{23}$ calibration from \citet{McGaugh1991} for the two regions near HoIX.  

\begin{figure}
  \centering
  \vspace{.75truein}
  \epsfig{width=0.5\linewidth,file=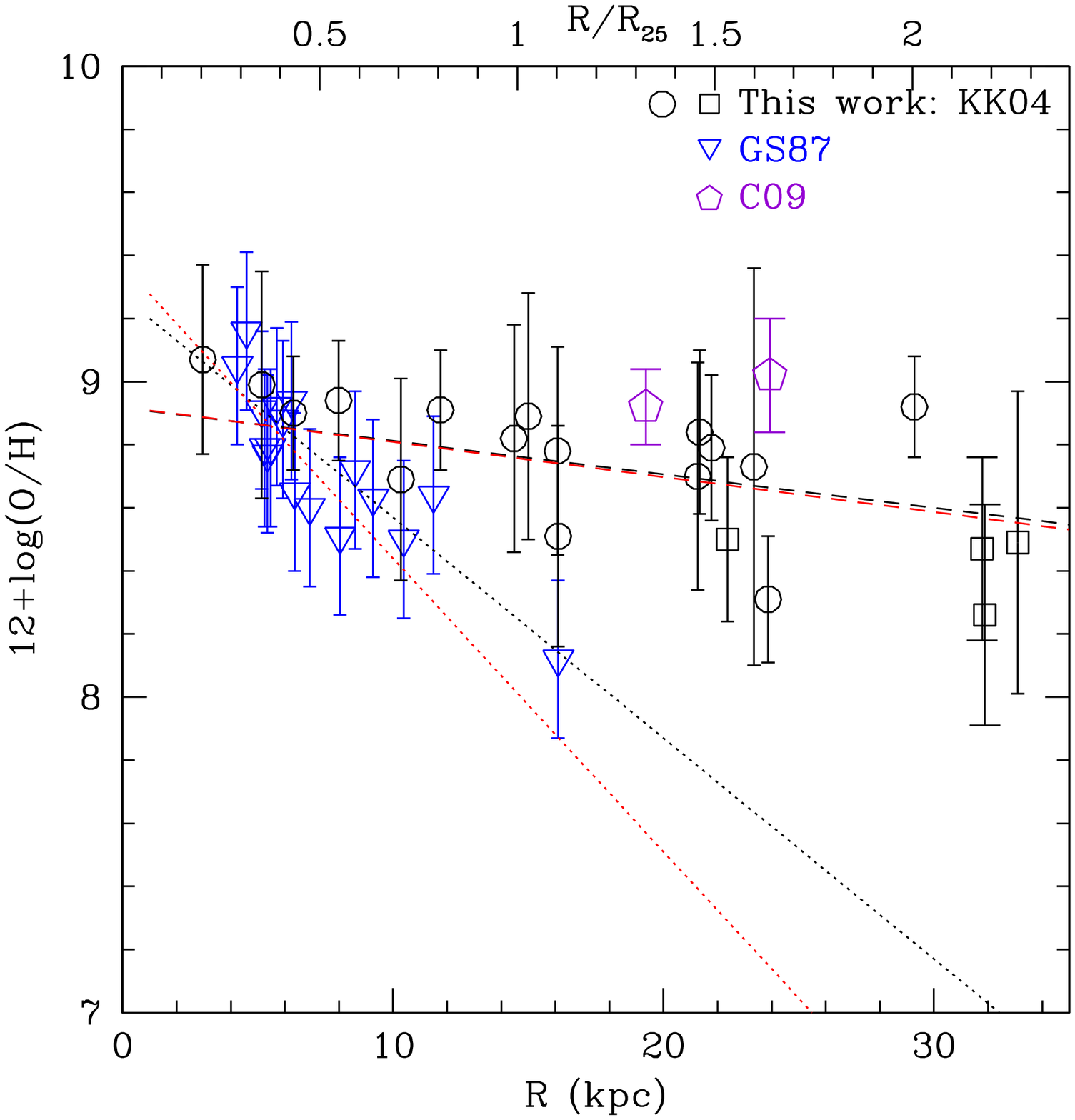} \\
  \vspace{1.0truein}
  \epsfig{width=0.5\linewidth,file=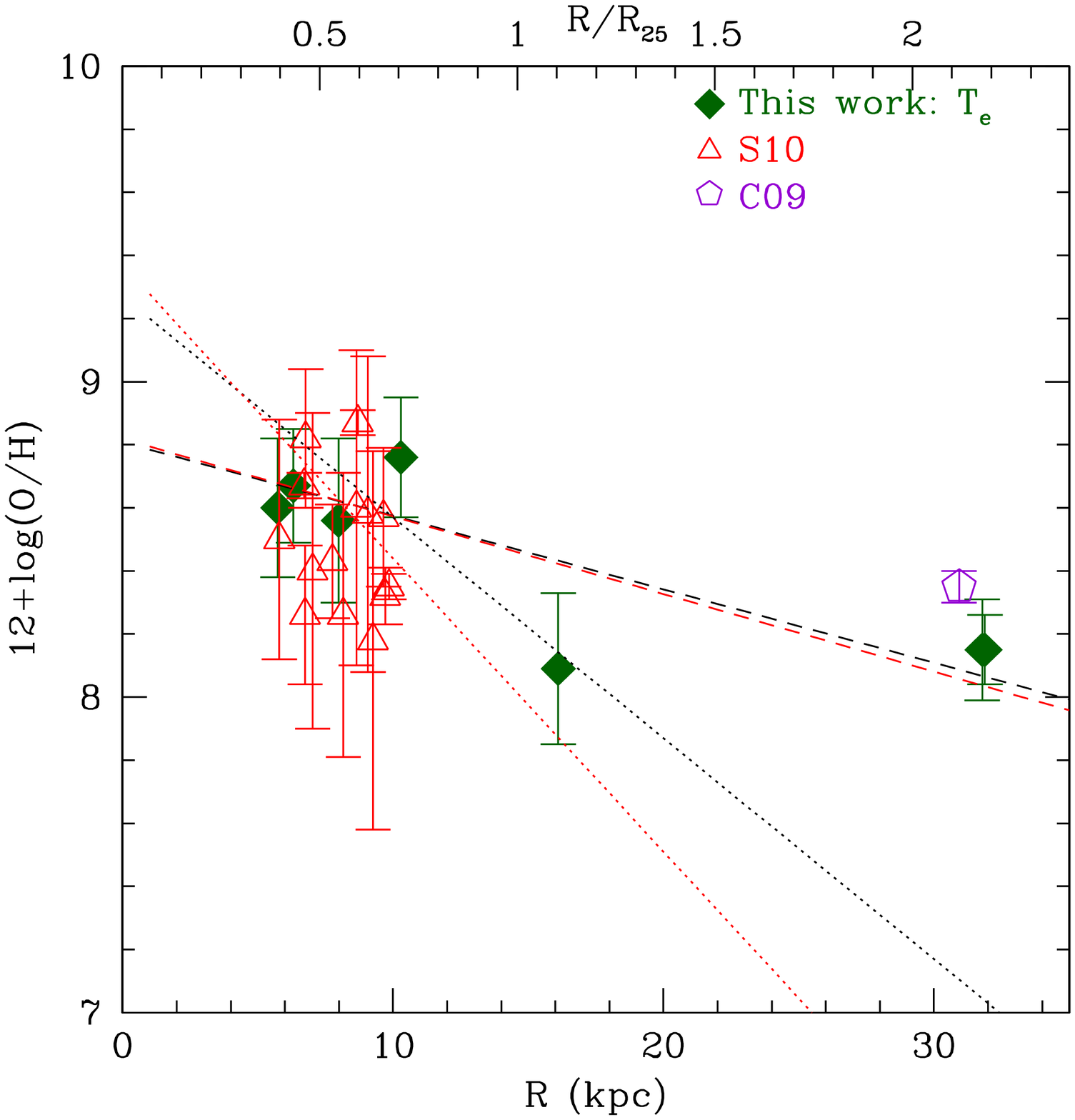}
  \caption{12+log(O/H) vs. galactocentric distance for all HII regions including previous abundance work
    using the published values in those papers.  
    We plot the abundances for our HII regions derived from the method most similar to the methods used in previous
    studies.  In the top graph, we compare abundances published by \citetalias{Croxall2009} and \citetalias{Garnett1987} 
    \citep{Garnett1987}, which both use $R_{23}$ based photoionization models, with the abundances for our HII regions
    derived using the \citetalias{KK2004} calibration.  M\"unch 1 is marked by the blue triangle at $\sim$16 kpc, and the
    lower abundance black circle at the same radius.  In the bottom graph, we compare electron temperature based 
    abundances from \citetalias{Stanghellini2010} \citep{Stanghellini2010} and \citetalias{Croxall2009}, with the 
    electron temperature based abundances from our HII regions.  The dashed lines are our fits to all points shown, 
    taking only our data if the HII region overlaps with another data set.  The dotted lines shows the gradient derived by
    \citetalias{Stanghellini2010}, which fit HII regions from both \citetalias{Stanghellini2010} and \citetalias{Garnett1987} 
    in the inner 17 kpc. (Red represents the fit from the routine {\it fitexy} \citep{Press1992}, and black represents a 
    weighted least-squares fit.)  } 
  \label{fig:zgradallcompare}
\end{figure}

Because different methods of 
abundance calculations may yield different absolute abundances, in order to compare our abundances with 
previously published abundances, we separate the results into groups derived by similar methods.
In the top graph of Fig. \ref{fig:zgradallcompare}, we compare the published abundances from 
\citetalias{Garnett1987} and \citetalias{Croxall2009} with the \citetalias{KK2004} strong line abundances from 
our data set, since these abundances were all derived using the $R_{23}$ parameter and a theoretical calibration 
based on photoionization models.  If we perform a weighted least-squares fit to this group of data, using only our 
abundance result if a region is in multiple data sets, we find a gradient of
$\Delta(\log$(O/H))$/\Delta$R$_G=-$0.011$\pm$0.005 dex kpc$^{-1}$.
In the bottom graph of Fig. \ref{fig:zgradallcompare}, we plot only HII regions with electron temperature derived 
abundances, using our data and the abundances published in \citetalias{Stanghellini2010} and 
\citetalias{Croxall2009}.  A weighted least-squares fit to these points gives a gradient of 
$\Delta(\log$(O/H))$/\Delta$R$_G=-$0.023$\pm$0.004 dex kpc$^{-1}$,
but there is an obvious lack of radial coverage for these results, with only one object between 12 and 30 kpc.
If these electron temperature based abundances are closer to true abundances than the $R_{23}$ derived ones,
it may be the case that the abundances can be described by a broken gradient, with a drop in metallicity from 
10 to 15 kpc and a flatter outer gradient.  However, we have no temperature derived abundances in the
10 to 15 kpc range, and the HII region at 16 kpc, M\"unch 1, may be an anomalous point, which we will discuss 
later in this paper.  
\citetalias{Stanghellini2010} derive an abundance gradient using a composite data set of HII regions from their 
paper and HII regions published by \citetalias{Garnett1987}, over a radial range from 3 to 17 kpc.  They find a 
noticeably steeper gradient of 
$\Delta(\log$(O/H))$/\Delta$R$_G=-$0.093$\pm$0.02 dex kpc$^{-1}$
using the {\it fitexy} routine and a slightly less steep gradient of 
$\Delta(\log$(O/H))$/\Delta$R$_G=-$0.07 dex kpc$^{-1}$
using a least-squares fit.  We plot these lines for comparison in this figure.  The steep gradient that 
\citetalias{Stanghellini2010} derive is based on the steeper inner gradient found by \citetalias{Garnett1987}
from $\sim$3 to 10 kpc and the low abundance of M\"unch 1 at 16 kpc.
Note that the \citetalias{Stanghellini2010} regions are too limited in radial range to derive 
an inner gradient.

\begin{figure}
  \centering
  \vspace{.75truein}
  \epsfig{width=0.5\linewidth,file=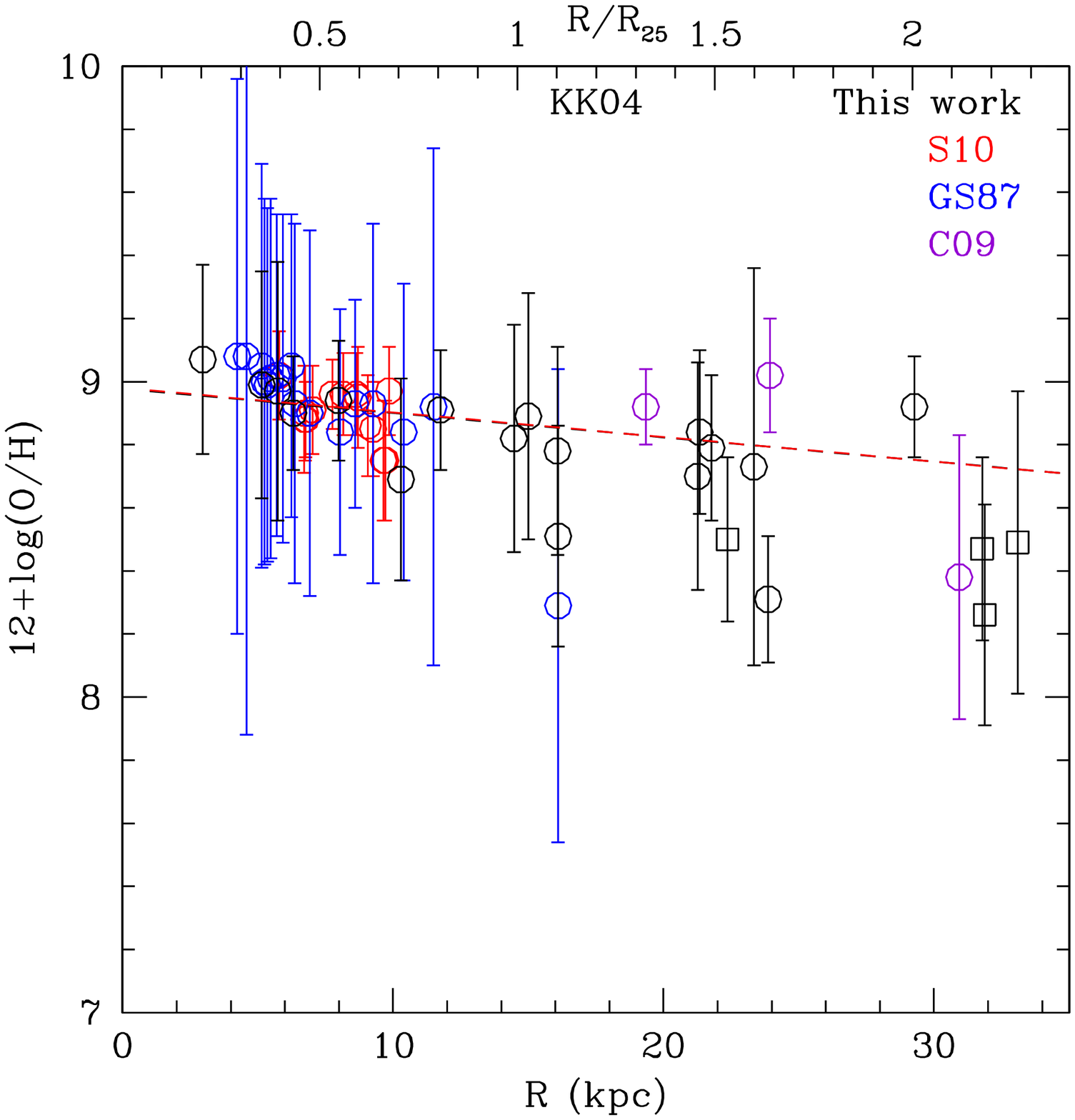} \\
  \vspace{1.0truein}
  \epsfig{width=0.5\linewidth,file=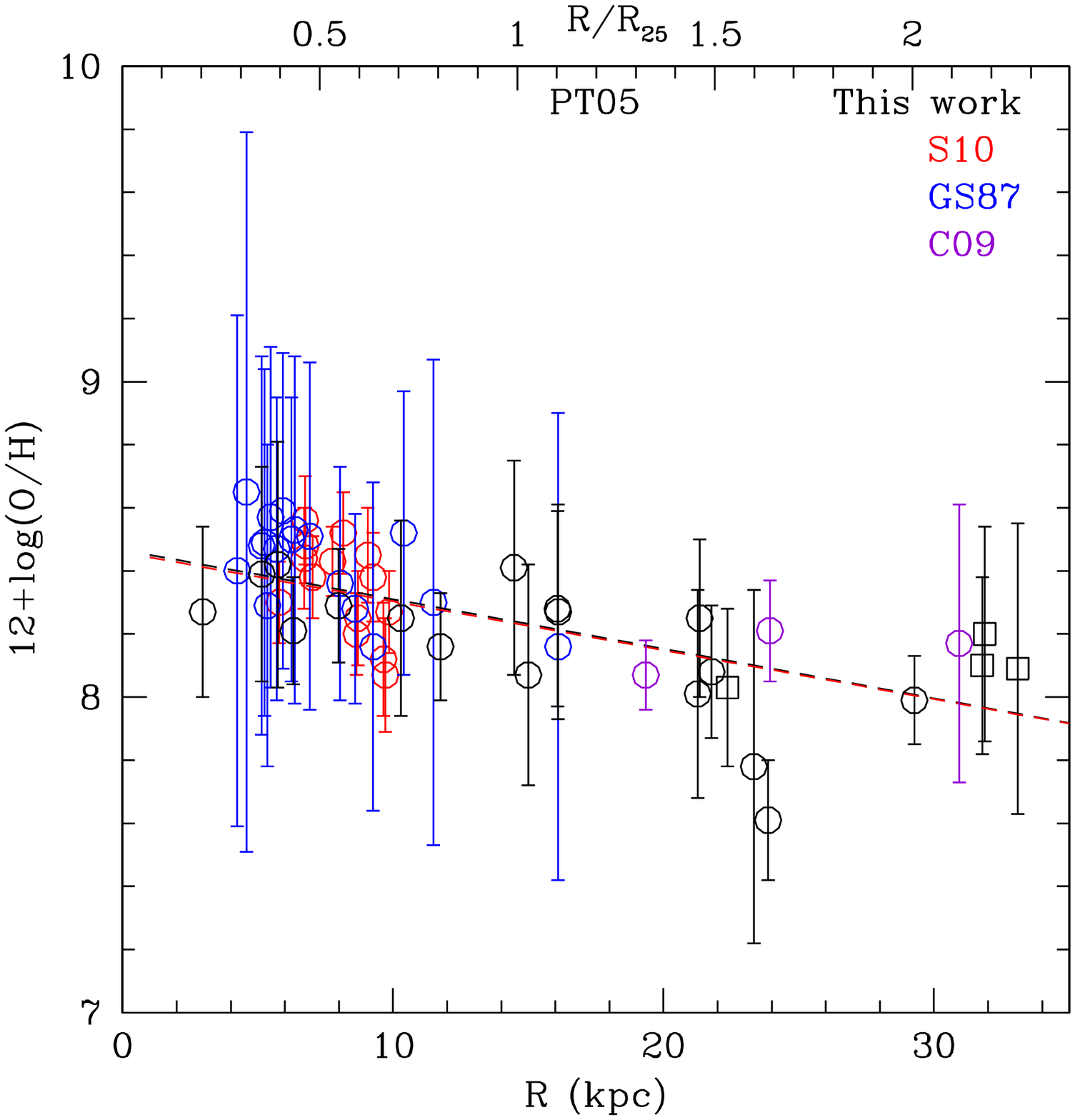}
  \caption{12+log(O/H) from each strong line calibration vs. galactocentric distance for all HII regions including 
    the reanalysis of previous data sets.  Here, we rederive 
    abundances for the HII regions in previous data sets using the same strong line $R_{23}$ based calibration for all 
    regions.  In the top graph, we calculate abundances for each HII region using the \citetalias{KK2004} calibration.  
    In the bottom graph, we derive abundances using the \citetalias{PT2005} calibration.  The dashed lines represent
    our weighted least-squares fits to all HII regions, taking only our data if the HII region overlaps with another data 
    set.    } 
  \label{fig:zgradallnewR23}
\end{figure}

\begin{figure}
  \centering
  \vspace{.75truein}
  \epsfig{width=0.5\linewidth,file=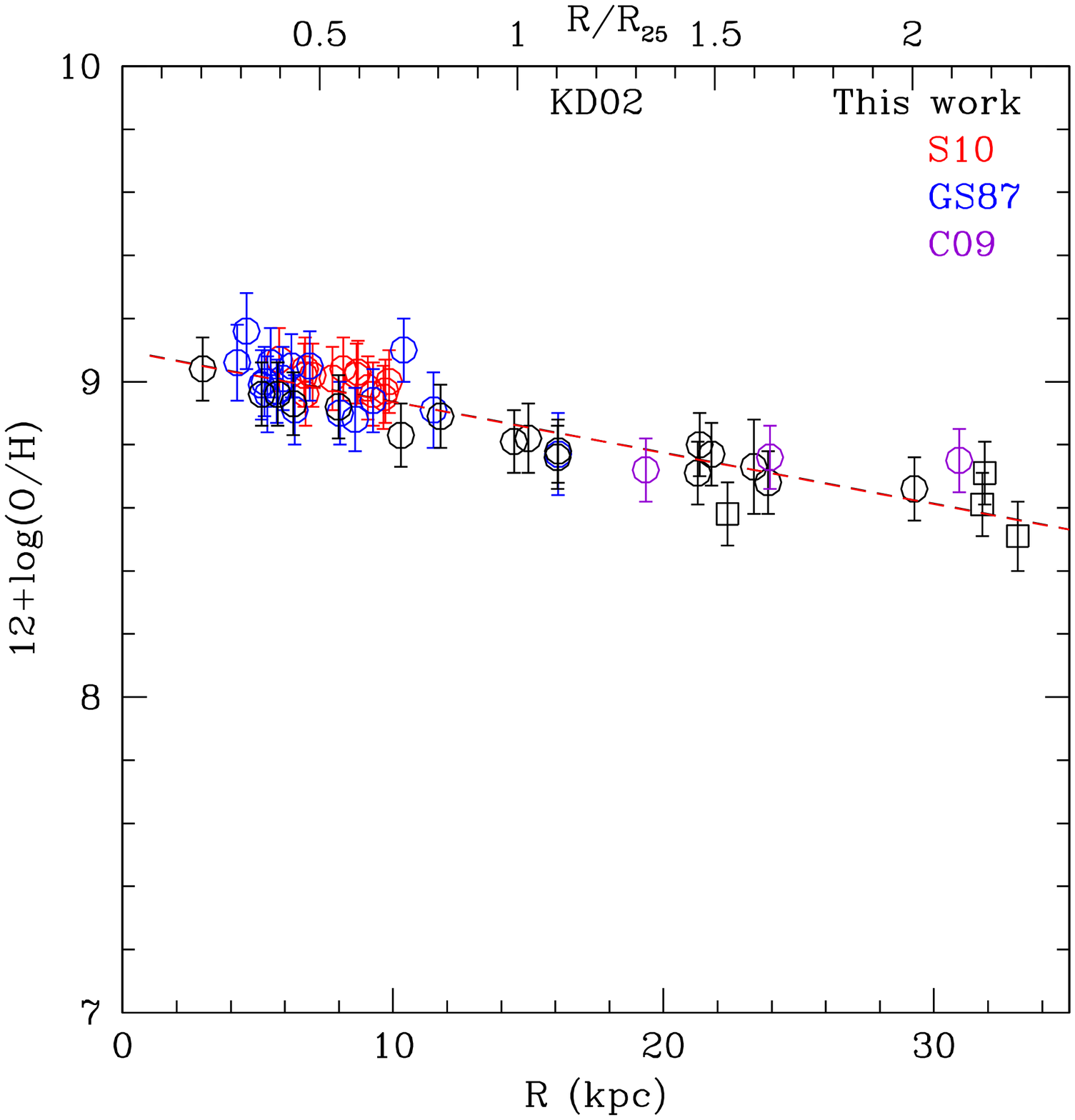}\\
  \vspace{1.0truein}
  \epsfig{width=0.5\linewidth,file=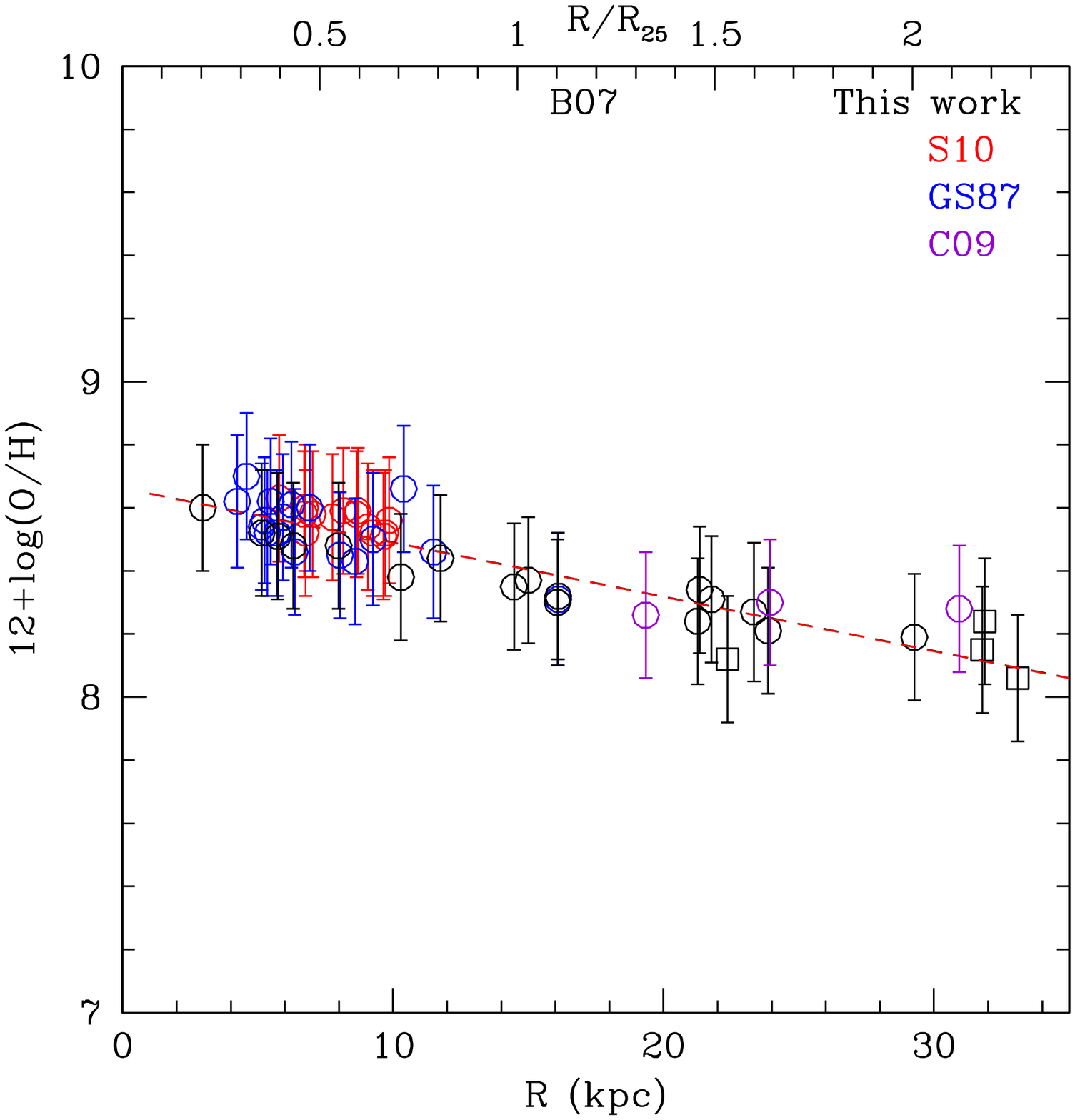}
  \caption{12+log(O/H) vs. galactocentric distance derived from the [NII]/[OII] based calibrations of \citetalias{KD2002}
    (top) and \citetalias{B07} (bottom) using the data in this paper and data from \citetalias{Croxall2009}, 
    \citetalias{Stanghellini2010}, and \citetalias{Garnett1987} reanalysed in the same method.  The dashed lines 
    are our fits to all points regions shown, taking only our data if the region falls in multiple data sets.   } 
  \label{fig:zgradallNII}
\end{figure}

One important point to emphasize is that not all strong line calibration methods are alike.  In the past decade, 
various refinements have been introduced.  We therefore considered it useful to recalculate strong line 
abundances from previous data using the methods adopted in this paper.  In Fig. \ref{fig:zgradallnewR23}, 
we have taken the emission line fluxes for the HII regions published by \citetalias{Stanghellini2010}, 
\citetalias{Garnett1987}, and \citetalias{Croxall2009} and recalculated new strong line $R_{23}$ based 
abundances in the same way as for our data set, using again the two calibrations of \citetalias{KK2004} and 
\citetalias{PT2005}.  For the combined four data sets, using only our data if an HII region is also included in another 
data set, we find abundance gradients of
$\Delta(\log$(O/H))$/\Delta$R$_G=-$0.008$\pm$0.005 dex kpc$^{-1}$
from the \citetalias{KK2004} calibration and 
$\Delta(\log$(O/H))$/\Delta$R$_G=-$0.016$\pm$0.004 dex kpc$^{-1}$
from the \citetalias{PT2005} calibration.
The abundances given by the \citetalias{KK2004} calibration for the HII regions in the data set from
\citetalias{Garnett1987} are closest to the published values in that paper.  The \citetalias{Garnett1987}
data for M\"unch 1 also give a relatively low abundance for this calibration, as shown by the open blue
circle near 16 kpc in Fig. \ref{fig:zgradallnewR23}.  Our data for this object, marked by a black open
circle at the same radius, shows a relatively low abundance derived from the \citetalias{KK2004} calibration
as compared to other objects near this distance.
In Fig. \ref{fig:zgradallNII}, we have recalculated strong line abundances for all previous data using the 
[NII]/[OII] ratio based calibrations used for our data set.  We find an abundance gradient of 
$\Delta(\log$(O/H))$/\Delta$R$_G=-$0.016$\pm$0.002 dex kpc$^{-1}$ from the \citetalias{KD2002} calibration and
$\Delta(\log$(O/H))$/\Delta$R$_G=-$0.017$\pm$0.004 dex kpc$^{-1}$ from the \citetalias{B07} calibration.
The location of the points for the \citetalias{KD2002} and \citetalias{B07} calibrations suggest that the two calibrations
are essentially the same with only a zero point offset. 

The new strong line abundances that we calculate for these data sets are consistent with the gradients derived
for only our data.  The reanalysis of these previous data with the four strong line abundance calibrations that we use 
yields oxygen abundances with relatively shallow gradients for M81.  The flatter gradient that we derive is not a consequence
of disagreements between our data and previous data, but a result of our reanalysis of previous data sets
using recent strong line calibration methods.  It is worth noting that in a recent abundance study of blue 
supergiants, which, like HII regions, trace the current metallicity of the interstellar medium, \citet{Kudritzki2011} also find a 
relatively shallow gradient of $-$0.034 dex kpc$^{-1}$ across the main disk of M81. 
This is only slightly steeper than the gradients we find here.
The coefficients to our least-squares fits to the metallicity gradient 
for different methods and sets of data are listed in Table \ref{tab:grad}.  
 
Our oxygen abundance gradient is noticeably less steep than the previously published gradient of M81 derived 
from HII regions.  We attribute this partially to the limited radial range of HII regions of previous studies
and the high electron temperature and low abundance derived for M\"unch 1, which has the largest galactocentric
radius for previous M81 metallicity gradient studies.  Further inspection of this HII region's spectrum
shows the presence of HeII $\lambda$4686, meaning that the region is being heated by a very hot central star.  
The spatial extent of the HeII $\lambda$4686 is centrally confined to not much further beyond the continuum of the
central star.  We had hypothesized that the temperature lines may also be restricted in spatial extent to the very 
core of the HII region, thus giving a very high temperature in the core of the region which would not reflect the 
temperature across the region in its entirety.  However, we find that the temperature lines have the same spatial
extent as H$\beta$, and therefore the high electron temperature we derive for this region is representative of the 
whole region.  It is worth noting that the [OIII] line at $\lambda$4363 lies under a mercury sky line, but we believe 
our background subtraction to be reliable.  This HII region compared to the more distant regions in our data set has 
an anomalously low oxygen abundance as derived from the electron temperature, as well as from the 
\citetalias{KK2004} calibration on this object.   

It is worth noting that the low abundance derived from electron temperature for M\"unch 1 may not be anomalous
to this HII region, since we do not currently have spectra deep enough to derive oxygen abundances based on 
temperature sensitive lines for other HII regions at this radius ($\sim$10-20 kpc).  More data are needed,
particularly for the nearby string of HII regions in the outer southern tidal arm, to conclude whether this HII
region is atypical in its low temperature derived oxygen abundance and to explore the possibility of a broken
gradient.  Though a continuously sloped gradient is a good fit to our strong line abundance gradients, they may
be equally well-described by a slight brokent gradient, with a flattening past $\sim$16 kpc, such as the broken gradients
seen in M83 \citep{GildePaz2007,Bresolin2009} and NGC 4625 \citep{Goddard2011}.

The slope of the oxygen abundance gradient may also be affected by our averaging of the upper and lower branch
metallicities for the HII regions which do not clearly lie on the upper or lower branch of the $R_{23}$ vs.
metallicity calibration.  These regions all lie at distances greater than $\sim$22 kpc from the galactic center.  
If we assume that these ambiguous regions lie on either the upper or lower branch, the values of 12+log(O/H) 
that we calculate are $\sim$0.25 higher or lower than our average metallicity, which would slightly change the 
slope of the abundance gradient.  In particular, if we assume that these regions lie on the lower branch, the gradient 
could be steeper, or could be represented by a broken gradient, shallow in the main disk and dropping off to a flatter
low abundance past $\sim$22 kpc.  However, our favored interpretation is that the metallicities of these ambiguous 
regions are best represented by the average of the upper and lower branch values.  These values appear to agree 
nicely with our final $R_{23}$ vs. (O/H) relation of both strong line abundances for the upper branch regions and 
our temperature derived abundances.  Additionally, the gradient we derive from the [NII]/[OII] calibrations, which 
does not suffer from this metallicity degeneracy, are similar in slope to both $R_{23}$ based abundance gradients 
with these four turnaround regions.


\section{Discussion}
\label{discussionsec}

We have used four strong line calibrations in addition to electron temperatures to derive abundances for our HII regions,
and while we find similar slopes to the abundance gradients from these different methods, the absolute abundance 
for a region may vary significantly depending on the method used.  This study demonstrates the need for adopting
robust and consistent abundance diagnostics to reliably determine abundance distributions in galaxies.  Using the
abundance gradients presented in the previous section,  here we discuss the possibilities of a broken gradient,
a single gradient as a result of chemical evolution, and a gradient as a result of M81's interaction history. 

On one hand, our data show good agreement between the slopes of abundance gradients derived from both strong
line abundances and temperature derived abundances, with the exception of M\"unch 1.  On the other hand, the
reanalysis of data from \citetalias{Stanghellini2010}, \citetalias{Garnett1987}, and \citetalias{Croxall2009}
to derive new strong line abundances leads to a significantly flatter abundance gradient in the inner part, consistent
with the gradient we find for the outer part.  It is important to confirm this through more electron temperature based 
abundances.  Because low metallicity regions have the strongest and most easily detectable electron temperature lines,
our temperature based abundances may reflect a bias towards these low abundances.  In the inner regions, this could
amount to underestimating the slope of the abundance gradient.  Additionally, if our temperature based abundance of 
M\"unch 1 is an accurate representation of the true abundance
at its radius, it is possible that M81 may have a broken metallicity gradient, with the break just outside of the optical
radius $R_{25}$, similar to the profile seen in M83.  M83 shows a negative oxygen abundance gradient in the inner disk 
and a flattening just outside of the optical radius out to 2.6 times $R_{25}$  \citep{GildePaz2007,Bresolin2009}.
It also has an extended UV (XUV) disk with recent star formation far into the outer disk \citep{Thilker2005,Thilker2007}.
\citet{Bresolin2009} attribute the abundance profile flattening in the XUV disk to its relatively unevolved state.
The authors draw the analogy of the low gas surface density and low star formation environment in the galaxy outskirts to low 
surface brightness galaxies, which have a nearly constant oxygen abundance as a function of radius \citep{deBlok1998}. 
M81 also has an XUV disk so it would not be unlikely for this galaxy to show a broken abundance profile that flattens
outside of the optical radius. 

While M81 may have a broken gradient, as discussed above, the present data are more consistent with a single shallow gradient.
The shallow negative metallicity gradient derived from the oxygen abundances of M81 imply only a slightly higher 
metallicity at small galactocentric radii than into the outskirts of the galaxy far beyond the optical radius of the disk.  
The flatter gradient reported here for M81 is now consistent previous observational findings showing that flatter gradients 
are observed in early-type spirals \citep{Zaritsky1994} and that this trend disappears when gradients are normalized to disk size. 
If we compare the metallicity gradient of M81,
$\Delta(\log$(O/H))$/\Delta$R$_G\sim-$0.013 to $-$0.020 dex kpc$^{-1}$ (or $\sim-$0.19 to $-$0.29 dex/$R_{25}$),
 to the gradient of the Milky Way as similarly derived from HII regions,
$\Delta(\log$(O/H))$/\Delta$R$_G\sim-$0.04 to $-$0.06 dex kpc$^{-1}$ (or $\sim-$0.45 to $-$0.68 dex/$R_{25}$)
\citep{Deharveng2000,Esteban2005,Rudolph2006,Rood2007}, both galaxies show a similarly shallow negative sloping trend, with a slightly 
steeper gradient for the Milky Way.  

Chemical evolution models of galaxy disks usually rely on four basic constraints: the exponential stellar profile, the gaseous profile, 
the star formation rate profile, and the abundance profiles of different elements (in external galaxies, most of the time, oxygen).
From these observational constraints, chemical evolution models can infer the star formation histories of disk galaxies. In the 90's, 
it was shown both for the Milky Way \citep[e.g.,][]{Matteucci1989,Chiappini1997,Boissier1999}, and for other spiral 
galaxies \citep{Molla1997}, that one way to reproduce these constraints is to assume a star formation history that is a function of 
galactic radius (i.e., more peaked in the inner parts and less efficient in the outer parts). Furthermore, these models explain the observed 
shallower abundance gradient in more luminous galaxies due to a faster formation of large spiral galaxies (see \citealt{Molla1997} for a 
comparative study of seven nearby spirals; \citealt{Chiappini2003} for M101; \citealt{Renda2005} and \citealt{Yin2009} for M31, 
\citealt{Marcon2010} for M33).  A common interpretation is that the variation of the accretion timescale of gas onto the disk 
should also be a function of the galaxy mass in the sense that the most massive galaxies formed in the shortest timescale 
\citep{Prantzos2000,Molla2005}.  The flatter oxygen gradient reported in this work for M81 is more consistent with this scenario 
than was reported in previous works.
For M81, an Sa galaxy with rotation velocity $V_{max}$ = 260 km s$^{-1}$ \citep{Rohlfs1980}, we would expect
an oxygen abundance gradient of $\Delta(\log$(O/H))$/\Delta$R$_G\sim-$0.04 \citep{Molla2005} to 
$-$0.05 dex kpc$^{-1}$ \citep{Prantzos2000}.  In comparison, the gradient
presented in \citetalias{Stanghellini2010} of $-$0.093 dex kpc$^{-1}$ is significantly steeper than these predictions, as the authors note.  
The gradient that we find in this paper, in the range of about $-$0.01 to $-$0.02 dex kpc$^{-1}$ depending on method, is shallower
than that of the Milky Way, as expected, but even flatter than the chemical evolution models of \citeauthor{Prantzos2000}
and \citeauthor{Molla2005} predict.  Note, however, that these models do not consider a gas threshold for star formation,
and models computed including this would predict shallower gradients in the outer parts of galaxies \citep[see e.g.,][]{Chiappini2001}. 

It is likely that the shallow gradient is partially a result of the interaction history of M81 with its companions.  
\citet{Yun1999} describes a simulation of the tidal interaction between M81, M82, and 
NGC 3077, in which much of the gas presently in the near outskirts of M81 originated in its disk and was pulled out 
by the interactions with these two galaxies.  In this scenario, the interactions between these galaxies may have 
served to flatten the metallicity gradient of M81.  Recent numerical simulations of chemical evolution in galaxy interactions
predict the flattening of radial metallicity gradients soon after the first passage of interacting galaxies, due to the radial 
redistribution of gas over the disk \citep{Kewley2006, Rupke2010a}.  
Moreover, the slope of the metallicity gradient of an interacting galaxy does not seem to be a strong function of the initial
gradient \citep{Rupke2010a}.  Though these authors did not include star formation in their interaction simulations, the models
of \citet{Perez2011}, which include star formation and subsequent supernova feedback, confirm the result of a flattened
metallicity gradient for interacting galaxies.  In the central regions, they find interaction-induced dilution, lowering the metallicity.  
In the outer regions, both in situ star formation activity and metal-rich gas transported from the inner disk to the outer disk
increase the metallicity.  These processes would produce a flattened gradient in an interacting galaxy.

Recent observational studies have also explored the connection between interacting galaxies and abundance gradients.
The possibility of a past interaction is invoked as a feasible explanation for the flat metallicity gradient observed in the blue compact dwarf galaxy 
NGC 2915 \citep{Werk2010b}.  A survey of galaxies in an early stage of strong interaction 
by \citet{Rupke2010b} provide observational confirmation of the flattening of radial metallicity gradients in interacting galaxies.  
The authors find that interacting galaxies, on average, have metallicity gradients that are less than half as steep as non-interacting 
galaxies.  Specifically, they find a median oxygen abundance slope of $-$0.041$\pm$0.009 dex kpc$^{-1}$,
or $-$0.57$\pm$0.05 dex/$R_{25}$, for the 11 isolated galaxies in the control sample (including NGC 300 and M101) and a shallower 
median slope of $-$0.017$\pm$0.002 dex kpc$^{-1}$, or $-$0.23$\pm$0.03 dex/$R_{25}$, for the 16 interacting galaxies studied.  
If we compare our results for M81, we find that our derived oxygen gradient has the shallow characteristic of a typical interacting galaxy.  
Though these studies find flat gradients for interacting galaxies, it is important to note that isolated galaxies
also display flat gradients.  For example, in a study of a sample of 13 galaxies in the HI Rogues catalog, \citet{Werk2011} find flat oxygen abundance 
gradients out to large radii, and most, though notably not all, of the galaxies are interacting.  Additionally, deep CMD studies of the isolated galaxies 
NGC 300 \citep{Vlajic2009,Gogarten2010} and NGC 7793 \citep{Vlajic2011} find flattened metallicity gradients into the outer disks of these galaxies. 
Inefficient star formation in the galaxy outskirts could also produce a flattened abundance profile.
It may be the case that the gas in the outskirts of M81 was pre-enriched by inefficient outer disk star formation and would have also displayed a shallow 
abundance gradient prior to its interaction with M82 and NGC 3077.  However, given the interaction history of M81 as traced by the HI gas and the recent
observational studies and simulations of flattened gradients in interacting galaxies, it is probable that the shallow abundance gradient into the outskirts
of M81 is at least partially a result of interaction with the companion galaxies.




\section{Summary}
\label{summarysec}

We have presented H$\alpha$ luminosities for 40 HII regions outside the main disk of M81.  We have also presented 
emission line fluxes for 21 HII regions ranging from $\sim$3-33 kpc from the galaxy center, corresponding to 
$\sim$2.3$\times$R$_{25}$.   We calculated strong line oxygen abundances from two $R_{23}$ based metallicity 
calibrations and two [NII]/[OII] based calibration.  We find that M81 has a shallow negative gradient of 
$\Delta(\log$(O/H))$/\Delta$R$_G\sim$$-$0.013 to $-$0.016 dex kpc$^{-1}$.  
For seven HII regions, we were able to derive oxygen abundances using electron temperature sensitive [NII] or [OIII] 
emission lines.  We find the temperature derived abundances gradient to be slightly steeper than our strong line 
abundance gradients with 
$\Delta(\log$(O/H))$/\Delta$R$_G\sim$$-$0.020 dex kpc$^{-1}$.  
Our oxygen abundance gradients are noticeably less steep than previous results.  We attribute this partially to
the possible atypically low abundance of the most distant HII region in previous data sets, M\"unch 1.  We 
calculate a lower oxygen abundance for this region than for any other region in our sample, and it is not 
clear if this is a true reflection of the average metallicity at its radial distance.  More deep spectroscopic data is 
needed for HII regions near M\"unch 1 to determine whether or not this region has an anomalously low metallicity
and whether M81 shows a broken abundance profile with a flat outer disk gradient.  
Our shallow gradient is corroborated by our reanalysis of previous data.  We combine our data with 
three published HII region data sets and recalculate oxygen abundances using a consistent strong line method for all 
data.  The reanalysis of these data show a flatter gradient for M81.  The shallow gradient may be flatter than some chemical 
evolution models predict for a galaxy of M81's Hubble type and mass.  This may be explained by radial redistribution
of gas and inefficient outer disk star formation caused by the interaction of M81 and its companions.

\acknowledgments
We would like to thank the anonymous reviewer for insightful comments that have helped to improve this paper.  In addition to
the new data presented here, we have made use of archived data from the NASA Galaxy Evolution Explorer. 
GALEX is operated for NASA by the California Institute of Technology under NASA contract NAS5-98034.  
R.A.M.W. and M.T.P acknowledge support from the Research Corporation for Science Advancement.  M.T.P. would also
like to acknowledge support from a New Mexico Space Grant Graduate Research Fellowship.

\bibliographystyle{mn2e}
\bibliography{biblio}

\begin{thebibliography}{98}
\expandafter\ifx\csname natexlab\endcsname\relax\def\natexlab#1{#1}\fi

\bibitem[{{Allen} {et~al}\mbox{.}(2008){Allen}, {Groves}, {Dopita},
  {Sutherland}, \& {Kewley}}]{Allen2008}
{Allen} M.~G., {Groves} B.~A., {Dopita} M.~A., {Sutherland} R.~S., {Kewley}
  L.~J., 2008, \apjs, 178, 20

\bibitem[{{Appleton}, {Davies} \& {Stephenson}(1981){Appleton}, {Davies}, \&
  {Stephenson}}]{Appleton1981}
{Appleton} P.~N., {Davies} R.~D., {Stephenson} R.~J., 1981, \mnras, 195, 327

\bibitem[{{Arp}(1965)}]{Arp1965}
{Arp} H., 1965, Science, 148, 363

\bibitem[{{Baldwin}, {Phillips} \& {Terlevich}(1981){Baldwin}, {Phillips}, \&
  {Terlevich}}]{Baldwin1981}
{Baldwin} J.~A., {Phillips} M.~M., {Terlevich} R., 1981, \pasp, 93, 5

\bibitem[{{Barker} {et~al}\mbox{.}(2009){Barker}, {Ferguson}, {Irwin},
  {Arimoto}, \& {Jablonka}}]{Barker2009}
{Barker} M.~K., {Ferguson} A.~M.~N., {Irwin} M., {Arimoto} N., {Jablonka} P.,
  2009, \aj, 138, 1469

\bibitem[{{Boissier} \& {Prantzos}(1999)}]{Boissier1999}
{Boissier} S., {Prantzos} N., 1999, \mnras, 307, 857

\bibitem[{{Bresolin}(2007)}]{B07}
{Bresolin} F., 2007, \apj, 656, 186

\bibitem[{{Bresolin}, {Garnett} \& {Kennicutt}(2004){Bresolin}, {Garnett}, \&
  {Kennicutt}}]{Bresolin2004}
{Bresolin} F., {Garnett} D.~R., {Kennicutt}, Jr. R.~C., 2004, \apj, 615, 228

\bibitem[{{Bresolin} {et~al}\mbox{.}(2009){Bresolin}, {Ryan-Weber},
  {Kennicutt}, \& {Goddard}}]{Bresolin2009}
{Bresolin} F., {Ryan-Weber} E., {Kennicutt} R.~C., {Goddard} Q., 2009, \apj,
  695, 580

\bibitem[{{Cardelli}, {Clayton} \& {Mathis}(1989){Cardelli}, {Clayton}, \&
  {Mathis}}]{Cardelli1989}
{Cardelli} J.~A., {Clayton} G.~C., {Mathis} J.~S., 1989, \apj, 345, 245

\bibitem[{{Chiappini}, {Matteucci} \& {Gratton}(1997){Chiappini}, {Matteucci},
  \& {Gratton}}]{Chiappini1997}
{Chiappini} C., {Matteucci} F., {Gratton} R., 1997, \apj, 477, 765

\bibitem[{{Chiappini}, {Matteucci} \& {Romano}(2001){Chiappini}, {Matteucci},
  \& {Romano}}]{Chiappini2001}
{Chiappini} C., {Matteucci} F., {Romano} D., 2001, \apj, 554, 1044

\bibitem[{{Chiappini}, {Romano} \& {Matteucci}(2003){Chiappini}, {Romano}, \&
  {Matteucci}}]{Chiappini2003}
{Chiappini} C., {Romano} D., {Matteucci} F., 2003, \mnras, 339, 63

\bibitem[{{Chiboucas}, {Karachentsev} \& {Tully}(2009){Chiboucas},
  {Karachentsev}, \& {Tully}}]{Chiboucas2009}
{Chiboucas} K., {Karachentsev} I.~D., {Tully} R.~B., 2009, \aj, 137, 3009

\bibitem[{{Chynoweth} {et~al}\mbox{.}(2008){Chynoweth}, {Langston}, {Yun},
  {Lockman}, {Rubin}, \& {Scoles}}]{Chynoweth2008}
{Chynoweth} K.~M., {Langston} G.~I., {Yun} M.~S., {Lockman} F.~J., {Rubin}
  K.~H.~R., {Scoles} S.~A., 2008, \aj, 135, 1983

\bibitem[{{Contini} {et~al}\mbox{.}(2002){Contini}, {Treyer}, {Sullivan}, \&
  {Ellis}}]{Contini2002}
{Contini} T., {Treyer} M.~A., {Sullivan} M., {Ellis} R.~S., 2002, \mnras, 330,
  75

\bibitem[{{Croxall} {et~al}\mbox{.}(2009){Croxall}, {van Zee}, {Lee},
  {Skillman}, {Lee}, {C{\^o}t{\'e}}, {Kennicutt}, \& {Miller}}]{Croxall2009}
{Croxall} K.~V., {van Zee} L., {Lee} H., {Skillman} E.~D., {Lee} J.~C.,
  {C{\^o}t{\'e}} S., {Kennicutt}, Jr. R.~C., {Miller} B.~W., 2009, \apj, 705,
  723

\bibitem[{{Davidge}(2009)}]{Davidge2009}
{Davidge} T.~J., 2009, \apj, 697, 1439

\bibitem[{{Davies} {et~al}\mbox{.}(2010){Davies}, {Wilson}, {Auld}, {Baes},
  {Barlow}, {Bendo}, {Bock}, {Boselli}, {Bradford}, {Buat}, {Castro-Rodriguez},
  {Chanial}, {Charlot}, {Ciesla}, {Clements}, {Cooray}, {Cormier}, {Cortese},
  {Dwek}, {Eales}, {Elbaz}, {Galametz}, {Galliano}, {Gear}, {Glenn}, {Gomez},
  {Griffin}, {Hony}, {Isaak}, {Levenson}, {Lu}, {Madden}, {O'Halloran},
  {Okumura}, {Oliver}, {Page}, {Panuzzo}, {Papageorgiou}, {Parkin},
  {Perez-Fournon}, {Pohlen}, {Rangwala}, {Rigby}, {Roussel}, {Rykala},
  {Sacchi}, {Sauvage}, {Schulz}, {Schirm}, {Smith}, {Spinoglio}, {Stevens},
  {Srinivasan}, {Symeonidis}, {Trichas}, {Vaccari}, {Vigroux}, {Wozniak},
  {Wright}, \& {Zeilinger}}]{Davies2010}
{Davies} J.~I. {et~al.}, 2010, \mnras, 409, 102

\bibitem[{{de Blok} \& {van der Hulst}(1998)}]{deBlok1998}
{de Blok} W.~J.~G., {van der Hulst} J.~M., 1998, \aap, 335, 421

\bibitem[{{de Mello} {et~al}\mbox{.}(2008){de Mello}, {Smith}, {Sabbi},
  {Gallagher}, {Mountain}, \& {Harbeck}}]{deMello2008}
{de Mello} D.~F., {Smith} L.~J., {Sabbi} E., {Gallagher} J.~S., {Mountain} M.,
  {Harbeck} D.~R., 2008, \aj, 135, 548

\bibitem[{{De Robertis}, {Dufour} \& {Hunt}(1987){De Robertis}, {Dufour}, \&
  {Hunt}}]{DeRobertis1987}
{De Robertis} M.~M., {Dufour} R.~J., {Hunt} R.~W., 1987, \jrasc, 81, 195

\bibitem[{{de Vaucouleurs} {et~al}\mbox{.}(1991){de Vaucouleurs}, {de
  Vaucouleurs}, {Corwin}, {Buta}, {Paturel}, \& {Fouque}}]{deVaucouleurs1991}
{de Vaucouleurs} G., {de Vaucouleurs} A., {Corwin}, Jr. H.~G., {Buta} R.~J.,
  {Paturel} G., {Fouque} P., 1991, {Third Reference Catalogue of Bright
  Galaxies}, {de Vaucouleurs, G., de Vaucouleurs, A., Corwin, H.~G., Jr., Buta,
  R.~J., Paturel, G., \& Fouque, P.}, ed.

\bibitem[{{Deharveng} {et~al}\mbox{.}(2000){Deharveng}, {Pe{\~n}a}, {Caplan},
  \& {Costero}}]{Deharveng2000}
{Deharveng} L., {Pe{\~n}a} M., {Caplan} J., {Costero} R., 2000, \mnras, 311,
  329

\bibitem[{{Durrell} {et~al}\mbox{.}(2004){Durrell}, {Decesar}, {Ciardullo},
  {Hurley-Keller}, \& {Feldmeier}}]{Durrell2004}
{Durrell} P.~R., {Decesar} M.~E., {Ciardullo} R., {Hurley-Keller} D.,
  {Feldmeier} J.~J., 2004, in IAU Symposium, Vol. 217, Recycling Intergalactic
  and Interstellar Matter, {P.-A.~Duc, J.~Braine, \& E.~Brinks}, ed., p.~90

\bibitem[{{Durrell}, {Sarajedini} \& {Chandar}(2010){Durrell}, {Sarajedini}, \&
  {Chandar}}]{Durrell2010}
{Durrell} P.~R., {Sarajedini} A., {Chandar} R., 2010, \apj, 718, 1118

\bibitem[{{Esteban} {et~al}\mbox{.}(2005){Esteban}, {Garc{\'{\i}}a-Rojas},
  {Peimbert}, {Peimbert}, {Ruiz}, {Rodr{\'{\i}}guez}, \&
  {Carigi}}]{Esteban2005}
{Esteban} C., {Garc{\'{\i}}a-Rojas} J., {Peimbert} M., {Peimbert} A., {Ruiz}
  M.~T., {Rodr{\'{\i}}guez} M., {Carigi} L., 2005, \apjl, 618, L95

\bibitem[{{Fabbiano}(1988)}]{Fabbiano1988}
{Fabbiano} G., 1988, \apj, 325, 544

\bibitem[{{Ferguson}, {Gallagher} \& {Wyse}(1998){Ferguson}, {Gallagher}, \&
  {Wyse}}]{Ferguson1998}
{Ferguson} A.~M.~N., {Gallagher} J.~S., {Wyse} R.~F.~G., 1998, \aj, 116, 673

\bibitem[{{Freedman} {et~al}\mbox{.}(2001){Freedman}, {Madore}, {Gibson},
  {Ferrarese}, {Kelson}, {Sakai}, {Mould}, {Kennicutt}, {Ford}, {Graham},
  {Huchra}, {Hughes}, {Illingworth}, {Macri}, \& {Stetson}}]{Freedman2001}
{Freedman} W.~L. {et~al.}, 2001, \apj, 553, 47

\bibitem[{{Garc{\'{\i}}a-Benito} {et~al}\mbox{.}(2010){Garc{\'{\i}}a-Benito},
  {D{\'{\i}}az}, {H{\"a}gele}, {P{\'e}rez-Montero}, {L{\'o}pez},
  {V{\'{\i}}lchez}, {P{\'e}rez}, {Terlevich}, {Terlevich}, \&
  {Rosa-Gonz{\'a}lez}}]{Garcia2010}
{Garc{\'{\i}}a-Benito} R. {et~al.}, 2010, \mnras, 408, 2234

\bibitem[{{Garnett}(1992)}]{Garnett1992}
{Garnett} D.~R., 1992, \aj, 103, 1330

\bibitem[{{Garnett} \& {Shields}(1987)}]{Garnett1987}
{Garnett} D.~R., {Shields} G.~A., 1987, \apj, 317, 82

\bibitem[{{Gil de Paz} {et~al}\mbox{.}(2007){Gil de Paz}, {Madore}, {Boissier},
  {Thilker}, {Bianchi}, {S{\'a}nchez Contreras}, {Barlow}, {Conrow}, {Forster},
  {Friedman}, {Martin}, {Morrissey}, {Neff}, {Rich}, {Schiminovich}, {Seibert},
  {Small}, {Donas}, {Heckman}, {Lee}, {Milliard}, {Szalay}, {Wyder}, \&
  {Yi}}]{GildePaz2007}
{Gil de Paz} A. {et~al.}, 2007, \apj, 661, 115

\bibitem[{{Goddard} {et~al}\mbox{.}(2011){Goddard}, {Bresolin}, {Kennicutt},
  {Ryan-Weber}, \& {Rosales-Ortega}}]{Goddard2011}
{Goddard} Q.~E., {Bresolin} F., {Kennicutt} R.~C., {Ryan-Weber} E.~V.,
  {Rosales-Ortega} F.~F., 2011, \mnras, 412, 1246

\bibitem[{{Gogarten} {et~al}\mbox{.}(2010){Gogarten}, {Dalcanton}, {Williams},
  {Ro{\v s}kar}, {Holtzman}, {Seth}, {Dolphin}, {Weisz}, {Cole}, {Debattista},
  {Gilbert}, {Olsen}, {Skillman}, {de Jong}, {Karachentsev}, \&
  {Quinn}}]{Gogarten2010}
{Gogarten} S.~M. {et~al.}, 2010, \apj, 712, 858

\bibitem[{{Gogarten} {et~al}\mbox{.}(2009){Gogarten}, {Dalcanton}, {Williams},
  {Seth}, {Dolphin}, {Weisz}, {Skillman}, {Holtzman}, {Cole}, {Girardi}, {de
  Jong}, {Karachentsev}, {Olsen}, \& {Rosema}}]{Gogarten2009}
{Gogarten} S.~M. {et~al.}, 2009, \apj, 691, 115

\bibitem[{{Greenawalt} {et~al}\mbox{.}(1998){Greenawalt}, {Walterbos},
  {Thilker}, \& {Hoopes}}]{Greenawalt1998}
{Greenawalt} B., {Walterbos} R.~A.~M., {Thilker} D., {Hoopes} C.~G., 1998,
  \apj, 506, 135

\bibitem[{{Gris{\'e}} {et~al}\mbox{.}(2011){Gris{\'e}}, {Kaaret}, {Pakull}, \&
  {Motch}}]{Grise2011}
{Gris{\'e}} F., {Kaaret} P., {Pakull} M.~W., {Motch} C., 2011, \apj, 734, 23

\bibitem[{{Hodge} \& {Kennicutt}(1983)}]{Hodge1983}
{Hodge} P.~W., {Kennicutt}, Jr. R.~C., 1983, \aj, 88, 296

\bibitem[{{Hoversten} {et~al}\mbox{.}(2011){Hoversten}, {Gronwall}, {Vanden
  Berk}, {Basu-Zych}, {Breeveld}, {Brown}, {Kuin}, {Page}, {Roming}, \&
  {Siegel}}]{Hoversten2011}
{Hoversten} E.~A. {et~al.}, 2011, \aj, 141, 205

\bibitem[{{Immler} \& {Wang}(2001)}]{Immler2001}
{Immler} S., {Wang} Q.~D., 2001, \apj, 554, 202

\bibitem[{{Karachentsev} {et~al}\mbox{.}(2004){Karachentsev}, {Karachentseva},
  {Huchtmeier}, \& {Makarov}}]{Karachentsev2004}
{Karachentsev} I.~D., {Karachentseva} V.~E., {Huchtmeier} W.~K., {Makarov}
  D.~I., 2004, \aj, 127, 2031

\bibitem[{{Kennicutt}, {Bresolin} \& {Garnett}(2003){Kennicutt}, {Bresolin}, \&
  {Garnett}}]{Kennicutt2003}
{Kennicutt}, Jr. R.~C., {Bresolin} F., {Garnett} D.~R., 2003, \apj, 591, 801

\bibitem[{{Kewley} \& {Dopita}(2002)}]{KD2002}
{Kewley} L.~J., {Dopita} M.~A., 2002, \apjs, 142, 35

\bibitem[{{Kewley} \& {Ellison}(2008)}]{Kewley2008}
{Kewley} L.~J., {Ellison} S.~L., 2008, \apj, 681, 1183

\bibitem[{{Kewley}, {Geller} \& {Barton}(2006){Kewley}, {Geller}, \&
  {Barton}}]{Kewley2006}
{Kewley} L.~J., {Geller} M.~J., {Barton} E.~J., 2006, \aj, 131, 2004

\bibitem[{{Kobulnicky} \& {Kewley}(2004)}]{KK2004}
{Kobulnicky} H.~A., {Kewley} L.~J., 2004, \apj, 617, 240

\bibitem[{{Kong} {et~al}\mbox{.}(2000){Kong}, {Zhou}, {Chen}, {Cheng}, {Jiang},
  {Zhu}, {Zheng}, {Mao}, {Shang}, {Fan}, {Byun}, {Chen}, {Chen}, {Deng},
  {Hester}, {Li}, {Lin}, {Su}, {Sun}, {Tsay}, {Windhorst}, {Wu}, {Xia}, {Xu},
  {Xue}, {Yan}, {Zheng}, \& {Zou}}]{Kong2000}
{Kong} X. {et~al.}, 2000, \aj, 119, 2745

\bibitem[{{Kudritzki} {et~al}\mbox{.}(2011){Kudritzki}, {Urbaneja}, {Gazak},
  {Bresolin}, {Przybilla}, {Gieren}, \& {Pietrzynski}}]{Kudritzki2011}
{Kudritzki} R.-P., {Urbaneja} M.~A., {Gazak} Z., {Bresolin} F., {Przybilla} N.,
  {Gieren} W., {Pietrzynski} G., 2011, ArXiv e-prints

\bibitem[{{Lin} {et~al}\mbox{.}(2003){Lin}, {Zhou}, {Burstein}, {Windhorst},
  {Chen}, {Chen}, {Jiang}, {Kong}, {Ma}, {Sun}, {Wu}, {Xue}, \&
  {Zhu}}]{Lin2003}
{Lin} W. {et~al.}, 2003, \aj, 126, 1286

\bibitem[{{Makarova} {et~al}\mbox{.}(2010){Makarova}, {Koleva}, {Makarov}, \&
  {Prugniel}}]{Makarova2010}
{Makarova} L., {Koleva} M., {Makarov} D., {Prugniel} P., 2010, \mnras, 406,
  1152

\bibitem[{{Makarova} {et~al}\mbox{.}(2002){Makarova}, {Grebel}, {Karachentsev},
  {Dolphin}, {Karachentseva}, {Sharina}, {Geisler}, {Guhathakurta}, {Hodge},
  {Sarajedini}, \& {Seitzer}}]{Makarova2002}
{Makarova} L.~N. {et~al.}, 2002, \aap, 396, 473

\bibitem[{{Marcon-Uchida}, {Matteucci} \& {Costa}(2010){Marcon-Uchida},
  {Matteucci}, \& {Costa}}]{Marcon2010}
{Marcon-Uchida} M.~M., {Matteucci} F., {Costa} R.~D.~D., 2010, \aap, 520, A35+

\bibitem[{{Matteucci} \& {Francois}(1989)}]{Matteucci1989}
{Matteucci} F., {Francois} P., 1989, \mnras, 239, 885

\bibitem[{{McGaugh}(1991)}]{McGaugh1991}
{McGaugh} S.~S., 1991, \apj, 380, 140

\bibitem[{{Miller}(1995)}]{Miller1995}
{Miller} B.~W., 1995, \apjl, 446, L75+

\bibitem[{{Miller} \& {Hodge}(1994)}]{Miller1994}
{Miller} B.~W., {Hodge} P., 1994, \apj, 427, 656

\bibitem[{{Moll{\'a}} \& {D{\'{\i}}az}(2005)}]{Molla2005}
{Moll{\'a}} M., {D{\'{\i}}az} A.~I., 2005, \mnras, 358, 521

\bibitem[{{Molla}, {Ferrini} \& {Diaz}(1997){Molla}, {Ferrini}, \&
  {Diaz}}]{Molla1997}
{Molla} M., {Ferrini} F., {Diaz} A.~I., 1997, \apj, 475, 519

\bibitem[{{Mouhcine} \& {Ibata}(2009)}]{Mouhcine2009}
{Mouhcine} M., {Ibata} R., 2009, \mnras, 399, 737

\bibitem[{{Moustakas} {et~al}\mbox{.}(2010){Moustakas}, {Kennicutt},
  {Tremonti}, {Dale}, {Smith}, \& {Calzetti}}]{Moustakas2010}
{Moustakas} J., {Kennicutt}, Jr. R.~C., {Tremonti} C.~A., {Dale} D.~A., {Smith}
  J., {Calzetti} D., 2010, \apjs, 190, 233

\bibitem[{{M{\"u}nch}(1959)}]{Munch1959}
{M{\"u}nch} G., 1959, \pasp, 71, 101

\bibitem[{{Osterbrock} \& {Ferland}(2006)}]{Osterbrock2006}
{Osterbrock} D.~E., {Ferland} G.~J., 2006, {Astrophysics of gaseous nebulae and
  active galactic nuclei}, {Osterbrock, D.~E.~\& Ferland, G.~J.}, ed.

\bibitem[{{Pagel} {et~al}\mbox{.}(1979){Pagel}, {Edmunds}, {Blackwell}, {Chun},
  \& {Smith}}]{Pagel1979}
{Pagel} B.~E.~J., {Edmunds} M.~G., {Blackwell} D.~E., {Chun} M.~S., {Smith} G.,
  1979, \mnras, 189, 95

\bibitem[{{Perez}, {Michel-Dansac} \& {Tissera}(2011){Perez}, {Michel-Dansac},
  \& {Tissera}}]{Perez2011}
{Perez} J., {Michel-Dansac} L., {Tissera} P.~B., 2011, \mnras, 417, 580

\bibitem[{{P{\'e}rez-Gonz{\'a}lez}
  {et~al}\mbox{.}(2006){P{\'e}rez-Gonz{\'a}lez}, {Kennicutt}, {Gordon},
  {Misselt}, {Gil de Paz}, {Engelbracht}, {Rieke}, {Bendo}, {Bianchi},
  {Boissier}, {Calzetti}, {Dale}, {Draine}, {Jarrett}, {Hollenbach}, \&
  {Prescott}}]{Perez2006}
{P{\'e}rez-Gonz{\'a}lez} P.~G. {et~al.}, 2006, \apj, 648, 987

\bibitem[{{Petit}, {Sivan} \& {Karachentsev}(1988){Petit}, {Sivan}, \&
  {Karachentsev}}]{PSK1988}
{Petit} H., {Sivan} J.-P., {Karachentsev} I.~D., 1988, \aaps, 74, 475

\bibitem[{{Pilyugin} \& {Thuan}(2005)}]{PT2005}
{Pilyugin} L.~S., {Thuan} T.~X., 2005, \apj, 631, 231

\bibitem[{{Prantzos} \& {Boissier}(2000)}]{Prantzos2000}
{Prantzos} N., {Boissier} S., 2000, \mnras, 313, 338

\bibitem[{{Press} {et~al}\mbox{.}(1992){Press}, {Teukolsky}, {Vetterling}, \&
  {Flannery}}]{Press1992}
{Press} W.~H., {Teukolsky} S.~A., {Vetterling} W.~T., {Flannery} B.~P., 1992,
  {Numerical recipes in C. The art of scientific computing}, {Press, W.~H.,
  Teukolsky, S.~A., Vetterling, W.~T., \& Flannery, B.~P. }, ed.

\bibitem[{{Renda} {et~al}\mbox{.}(2005){Renda}, {Kawata}, {Fenner}, \&
  {Gibson}}]{Renda2005}
{Renda} A., {Kawata} D., {Fenner} Y., {Gibson} B.~K., 2005, \mnras, 356, 1071

\bibitem[{{Rohlfs} \& {Kreitschmann}(1980)}]{Rohlfs1980}
{Rohlfs} K., {Kreitschmann} J., 1980, \aap, 87, 175

\bibitem[{{Rood} {et~al}\mbox{.}(2007){Rood}, {Quireza}, {Bania}, {Balser}, \&
  {Maciel}}]{Rood2007}
{Rood} R.~T., {Quireza} C., {Bania} T.~M., {Balser} D.~S., {Maciel} W.~J.,
  2007, in Astronomical Society of the Pacific Conference Series, Vol. 374,
  From Stars to Galaxies: Building the Pieces to Build Up the Universe,
  {A.~Vallenari, R.~Tantalo, L.~Portinari, \& A.~Moretti}, ed., pp. 169--+

\bibitem[{{Rudolph} {et~al}\mbox{.}(2006){Rudolph}, {Fich}, {Bell}, {Norsen},
  {Simpson}, {Haas}, \& {Erickson}}]{Rudolph2006}
{Rudolph} A.~L., {Fich} M., {Bell} G.~R., {Norsen} T., {Simpson} J.~P., {Haas}
  M.~R., {Erickson} E.~F., 2006, \apjs, 162, 346

\bibitem[{{Rupke}, {Kewley} \& {Barnes}(2010){Rupke}, {Kewley}, \&
  {Barnes}}]{Rupke2010a}
{Rupke} D.~S.~N., {Kewley} L.~J., {Barnes} J.~E., 2010, \apjl, 710, L156

\bibitem[{{Rupke}, {Kewley} \& {Chien}(2010){Rupke}, {Kewley}, \&
  {Chien}}]{Rupke2010b}
{Rupke} D.~S.~N., {Kewley} L.~J., {Chien} L.-H., 2010, \apj, 723, 1255

\bibitem[{{Sabbi} {et~al}\mbox{.}(2008){Sabbi}, {Gallagher}, {Smith}, {de
  Mello}, \& {Mountain}}]{Sabbi2008}
{Sabbi} E., {Gallagher} J.~S., {Smith} L.~J., {de Mello} D.~F., {Mountain} M.,
  2008, \apjl, 676, L113

\bibitem[{{Shaw} \& {Dufour}(1995)}]{Shaw1995}
{Shaw} R.~A., {Dufour} R.~J., 1995, \pasp, 107, 896

\bibitem[{{Skillman}, {Kennicutt} \& {Hodge}(1989){Skillman}, {Kennicutt}, \&
  {Hodge}}]{Skillman1989}
{Skillman} E.~D., {Kennicutt} R.~C., {Hodge} P.~W., 1989, \apj, 347, 875

\bibitem[{{Sollima} {et~al}\mbox{.}(2010){Sollima}, {Gil de Paz},
  {Martinez-Delgado}, {Gabany}, {Gallego-Laborda}, \& {Hallas}}]{Sollima2010}
{Sollima} A., {Gil de Paz} A., {Martinez-Delgado} D., {Gabany} R.~J.,
  {Gallego-Laborda} J.~J., {Hallas} T., 2010, \aap, 516, A83+

\bibitem[{{Stanghellini} {et~al}\mbox{.}(2010){Stanghellini}, {Magrini},
  {Villaver}, \& {Galli}}]{Stanghellini2010}
{Stanghellini} L., {Magrini} L., {Villaver} E., {Galli} D., 2010, \aap, 521,
  A3+

\bibitem[{{Sternberg}, {Hoffmann} \& {Pauldrach}(2003){Sternberg}, {Hoffmann},
  \& {Pauldrach}}]{Sternberg2003}
{Sternberg} A., {Hoffmann} T.~L., {Pauldrach} A.~W.~A., 2003, \apj, 599, 1333

\bibitem[{{Thilker} {et~al}\mbox{.}(2005){Thilker}, {Bianchi}, {Boissier}, {Gil
  de Paz}, {Madore}, {Martin}, {Meurer}, {Neff}, {Rich}, {Schiminovich},
  {Seibert}, {Wyder}, {Barlow}, {Byun}, {Donas}, {Forster}, {Friedman},
  {Heckman}, {Jelinsky}, {Lee}, {Malina}, {Milliard}, {Morrissey}, {Siegmund},
  {Small}, {Szalay}, \& {Welsh}}]{Thilker2005}
{Thilker} D.~A. {et~al.}, 2005, \apjl, 619, L79

\bibitem[{{Thilker} {et~al}\mbox{.}(2007){Thilker}, {Bianchi}, {Meurer}, {Gil
  de Paz}, {Boissier}, {Madore}, {Boselli}, {Ferguson}, {Mu{\~n}oz-Mateos},
  {Madsen}, {Hameed}, {Overzier}, {Forster}, {Friedman}, {Martin}, {Morrissey},
  {Neff}, {Schiminovich}, {Seibert}, {Small}, {Wyder}, {Donas}, {Heckman},
  {Lee}, {Milliard}, {Rich}, {Szalay}, {Welsh}, \& {Yi}}]{Thilker2007}
{Thilker} D.~A. {et~al.}, 2007, \apjs, 173, 538

\bibitem[{{Tody}(1993)}]{Tody1993}
{Tody} D., 1993, in Astronomical Society of the Pacific Conference Series,
  Vol.~52, Astronomical Data Analysis Software and Systems II, {R.~J.~Hanisch,
  R.~J.~V.~Brissenden, \& J.~Barnes}, ed., pp. 173--+

\bibitem[{{Vacca}, {Garmany} \& {Shull}(1996){Vacca}, {Garmany}, \&
  {Shull}}]{Vacca1996}
{Vacca} W.~D., {Garmany} C.~D., {Shull} J.~M., 1996, \apj, 460, 914

\bibitem[{{Vlaji{\'c}}, {Bland-Hawthorn} \& {Freeman}(2009){Vlaji{\'c}},
  {Bland-Hawthorn}, \& {Freeman}}]{Vlajic2009}
{Vlaji{\'c}} M., {Bland-Hawthorn} J., {Freeman} K.~C., 2009, \apj, 697, 361

\bibitem[{{Vlaji{\'c}}, {Bland-Hawthorn} \& {Freeman}(2011){Vlaji{\'c}},
  {Bland-Hawthorn}, \& {Freeman}}]{Vlajic2011}
{Vlaji{\'c}} M., {Bland-Hawthorn} J., {Freeman} K.~C., 2011, \apj, 732, 7

\bibitem[{{Walter} {et~al}\mbox{.}(2002){Walter}, {Weiss}, {Martin}, \&
  {Scoville}}]{Walter2002}
{Walter} F., {Weiss} A., {Martin} C., {Scoville} N., 2002, \aj, 123, 225

\bibitem[{{Wang}(2002)}]{Wang2002}
{Wang} Q.~D., 2002, \mnras, 332, 764

\bibitem[{{Weisz} {et~al}\mbox{.}(2008){Weisz}, {Skillman}, {Cannon},
  {Dolphin}, {Kennicutt}, {Lee}, \& {Walter}}]{Weisz2008}
{Weisz} D.~R., {Skillman} E.~D., {Cannon} J.~M., {Dolphin} A.~E., {Kennicutt},
  Jr. R.~C., {Lee} J., {Walter} F., 2008, \apj, 689, 160

\bibitem[{{Werk} {et~al}\mbox{.}(2011){Werk}, {Putman}, {Meurer}, \&
  {Santiago-Figueroa}}]{Werk2011}
{Werk} J.~K., {Putman} M.~E., {Meurer} G.~R., {Santiago-Figueroa} N., 2011,
  \apj, 735, 71

\bibitem[{{Werk} {et~al}\mbox{.}(2010){Werk}, {Putman}, {Meurer}, {Thilker},
  {Allen}, {Bland-Hawthorn}, {Kravtsov}, \& {Freeman}}]{Werk2010b}
{Werk} J.~K., {Putman} M.~E., {Meurer} G.~R., {Thilker} D.~A., {Allen} R.~J.,
  {Bland-Hawthorn} J., {Kravtsov} A., {Freeman} K., 2010, \apj, 715, 656

\bibitem[{{Yin} {et~al}\mbox{.}(2009){Yin}, {Hou}, {Prantzos}, {Boissier},
  {Chang}, {Shen}, \& {Zhang}}]{Yin2009}
{Yin} J., {Hou} J.~L., {Prantzos} N., {Boissier} S., {Chang} R.~X., {Shen}
  S.~Y., {Zhang} B., 2009, \aap, 505, 497

\bibitem[{{Yun}(1999)}]{Yun1999}
{Yun} M.~S., 1999, in IAU Symposium, Vol. 186, Galaxy Interactions at Low and
  High Redshift, {J.~E.~Barnes \& D.~B.~Sanders}, ed., pp. 81--+

\bibitem[{{Yun}, {Ho} \& {Lo}(1994){Yun}, {Ho}, \& {Lo}}]{Yun1994}
{Yun} M.~S., {Ho} P.~T.~P., {Lo} K.~Y., 1994, \nat, 372, 530

\bibitem[{{Zaritsky}, {Kennicutt} \& {Huchra}(1994){Zaritsky}, {Kennicutt}, \&
  {Huchra}}]{Zaritsky1994}
{Zaritsky} D., {Kennicutt}, Jr. R.~C., {Huchra} J.~P., 1994, \apj, 420, 87

\end{thebibliography}



{\footnotesize
\begin{longtable}{c|ccccccc}
  \caption{HII Region Sample} \\
  \hline \hline
  ID     &  aka &  R.A.    &    Dec.    &   R/R$_{25}$         &  log($L_{H\alpha}$)    & aperture &  spectra     \\
         &      &          &            &                      &                      & radius        &  exposure     \\
         &      & (J2000)   &  (J2000)   & ($R_{25}$=14.6 kpc)  &  (ergs s$^{-1}$)   &(arcsec)   &   (seconds)  \\
  \hline 
  \endhead
  
  1    &  \nodata  & 09:52:38.95  & +69:14:24.0  &  1.76 &   37.36  &  9.1    &\nodata        \\ 
  2    &  \nodata  & 09:53:06.42  & +69:15:10.5  &  1.49 &   37.63  &  12.2   &2$\times$1800  \\ 
  3    &  \nodata  & 09:53:10.31  & +69:15:36.4  &  1.46 &   37.48  &  12.2   &2$\times$1200  \\ 
  4    &  \nodata  & 09:53:26.23  & +69:16:47.5  &  1.35 &   37.21  &  8.1    &\nodata   \\ 
  5    &  \nodata  & 09:53:59.52  & +69:19:21.1  &  1.27 &   36.91  &  8.1    &\nodata   \\ 
  6    &  \nodata  & 09:54:11.08  & +69:11:27.0  &  0.86 &   37.19  &  8.1    &\nodata   \\ 
  7    &  \nodata  & 09:54:13.95  & +68:53:32.2  &  1.60 &   36.94  &  8.1    &2$\times$1800  \\ 
  8    &  \nodata  & 09:54:34.90  & +69:02:42.7  &  0.74 &   37.41  &  13.2   &\nodata   \\ 
  9    &  \nodata  & 09:54:36.66  & +69:17:00.5  &  1.02 &   36.63  &  4.1    &\nodata   \\ 
  10   &  \nodata  & 09:54:53.24  & +69:17:19.4  &  1.05 &   36.87  &  8.1    &\nodata   \\ 
  11   &  PSK149, HK543  & 09:55:08.30  & +68:56:16.8  &  0.86 &   37.37  &  7.1  &\nodata   \\ 
  12   &  PSK175, HK487  & 09:55:14.63  & +68:55:32.4  &  0.87 &   37.34  &  9.1  &\nodata   \\ 
  13   &  \nodata  & 09:55:26.35  & +68:53:10.7  &  1.00 &   37.38  &  16.2   &\nodata   \\ 
  14   &  \nodata  & 09:55:40.98  & +68:52:21.9  &  0.99 &   37.34  &  8.1   &2$\times$1800 \\ 
  15   &  \nodata  & 09:55:43.81  & +69:16:57.6  &  1.23 &   36.86  &  5.1   &\nodata   \\ 
  16   &  \nodata  & 09:55:44.19  & +68:51:58.1  &  1.01 &   37.77  &  12.2  &\nodata   \\ 
  17   &  \nodata  & 09:55:49.66  & +69:19:56.7  &  1.53 &   37.59  &  10.1  &2$\times$1200  \\ 
  18   &  \nodata  & 09:55:53.19  & +68:51:33.1  &  1.01 &   37.56  &  10.1  &\nodata   \\ 
  19   &  \nodata  & 09:55:58.59  & +68:51:05.8  &  1.03 &   37.51  &  11.2  &\nodata   \\ 
  20   &  \nodata  & 09:56:06.63  & +68:52:32.4  &  0.89 &   37.03  &  8.1   &\nodata   \\   
  21   &  M\"unch 1 & 09:56:18.50  & +68:49:43.2  &  1.11 &   38.18  & 14.2  & 2$\times$1200 \\   
  22   &  PSK487    & 09:56:31.64  & +69:06:21.9  &  0.81 &   37.31  & 8.1   & \nodata    \\   
  23   &  PSK489, HK007  & 09:56:35.88  & +69:05:50.7  &  0.83 &   37.69  &  15.2 &\nodata   \\  
  24   &  \nodata  & 09:56:41.83  & +68:50:00.4  &  1.10 &   37.21  &  8.1   &2$\times$1200  \\
  25   &  \nodata  & 09:56:48.48  & +68:51:28.9  &  1.03 &   37.15  &  8.1   &2$\times$1800  \\
  26   &  \nodata  & 09:56:50.06  & +69:21:54.4  &  2.18 &   37.36  &  8.1   &2$\times$1200  \\
  27   &  \nodata  & 09:57:06.32  & +69:22:25.7  &  2.37 &   37.05  &  8.1   &\nodata   \\
  28   &  KDG 61-9 & 09:57:07.66  & +68:35:54.0  &  2.18 &   37.88  &  12.2  &2$\times$1200 \\
  29   &  \nodata  & 09:57:15.05  & +69:16:48.0  &  2.01 &   36.95  &  7.1   &2$\times$1800 \\
  30   &  \nodata  & 09:57:23.07  & +69:19:42.0  &  2.31 &   36.71  &  6.1   &\nodata   \\
  31   &  \nodata  & 09:57:27.77  & +68:49:37.3  &  1.33 &   37.17  &  11.2  &\nodata   \\
  32   &  \nodata  & 09:57:30.22  & +69:16:01.9  &  2.10 &   36.78  &  8.1   &\nodata   \\
  33   &  \nodata  & 09:57:34.49  & +69:17:40.5  &  2.27 &   36.93  &  8.1   &2$\times$1200  \\
  34   &  \nodata  & 09:57:41.68  & +69:05:47.8  &  1.58 &   36.79  &  8.1   &\nodata   \\
  35   &  HoIX-MH9,MH10 & 09:57:53.32  & +69:03:50.7  &  1.63 &   38.12  &  18.3  &2$\times$1200  \\
  36   &  \nodata  & 09:57:54.57  & +68:53:50.1  &  1.45 &   37.18  &  10.1  &\nodata   \\
  37   &  \nodata  & 09:57:55.23  & +68:56:45.7  &  1.46 &   37.60  &  15.2  &2$\times$1200  \\
  38   &  \nodata  & 09:58:01.35  & +68:58:12.3  &  1.55 &   37.00  &  8.1   &\nodata   \\
  39   &  \nodata   & 09:58:41.83  & +69:18:17.8  &  3.02 &   36.67  &  7.1  &\nodata   \\
  40   &  \nodata  & 09:58:52.79  & +69:15:34.5  &  2.96 &   36.65  &  6.1   &\nodata   \\
  \hline                                                    
  disk1  & PSK97, HK652 & 09:54:56.74  & +69:08:45.0  &  0.43 &   38.92$^{\dag}$  & 4.85$^{\dag}$  & 2$\times$1200 \\
  disk2  &  \nodata& 09:54:59.89  & +69:14:12.7  &  0.81 &   38.00         & 8.1 &  2$\times$1200  \\
  disk3  & M\"unch 18 & 09:55:01.70  & +69:12:57.2  &  0.71 &   38.65      & 8.1 &  2$\times$1200  \\
  disk4  & PSK123, HK615  & 09:55:03.65  & +69:10:54.7  &  0.55 &   38.32$^{\dag}$  &  4.4$^{\dag}$ &2$\times$1200 \\
         & M\"unch 17     &              &              &       &                  &              &              \\     
  disk5  & PSK178, HK500 & 09:55:16.80  & +69:08:56.7  &  0.39 &   38.83$^{\dag}$  &  6.75$^{\dag}$ & 2$\times$1200 \\
  disk6  & PSK209, HK453 & 09:55:24.92  & +69:08:16.6  &  0.35 &   38.79$^{\dag}$  &  6.0$^{\dag}$  & 2$\times$1200  \\
  disk7  & PSK259, HK360 & 09:55:39.19  & +69:05:42.8  &  0.20 &   37.56$^{\dag}$  &  3.3$^{\dag}$  & 2$\times$1800  \\
  \hline
  
  \caption{Locations, galactocentric distances, H$\alpha$ luminosities (after [NII] correction), 
    and aperture radii for our HII region sample.
    We assume a distance to M81 of 3.63 Mpc \citep{Freedman2001}, a rotation angle of 157$^{\circ}$ and 
    inclination of 59$^{\circ}$ \citep{Kong2000}.  We assume the $R_{25}$ radius is 13.8 arcmin (14.6 kpc at 3.63 Mpc)
    \citep{deVaucouleurs1991}.
    The alternative IDs are from \citet{PSK1988}, \citet{Hodge1983}, \citet{Munch1959}, \citet{Miller1994} and 
    \citetalias{Croxall2009}.  The luminosities and radii marked with $^{\dag}$ are calculated from H$\alpha$ + [NII] 
    fluxes from \citet{Lin2003}.  The last column indicates whether or not we have spectroscopic data for the region. }
  \label{tab:sample}
  
\end{longtable}
}


\begin{deluxetable}{l|ccccccccccc}
  \center
  \setlength{\tabcolsep}{0.04in} 
  \rotate{}
  \tabletypesize{\scriptsize}
  \tablewidth{0.0pt}
  \tablecolumns{13}
  \tablecaption{Dereddened Line Fluxes and Errors}
  \tablehead{
    \colhead{Line} &
    \colhead{02}   &
    \colhead{03}   &
    \colhead{07}   &
    \colhead{14}   &
    \colhead{17}   &
    \colhead{21}   &
    \colhead{24}   &
    \colhead{25}   &
    \colhead{26}  &
    \colhead{28}   &
    \colhead{29}     }
  \startdata
      [OII]     $\lambda$3727       & 400$\pm$26     & 332$\pm$18    & 514$\pm$262    & 228$\pm$16      & 309$\pm$20     & 204$\pm$9      & 318$\pm$21     & 363$\pm$49    & 275$\pm$17        & 170$\pm$9      & 364$\pm$24    \\
      $$[NeIII]   $\lambda$3869       & \nodata        & 10.5$\pm$3.7  & \nodata        & \nodata         & \nodata        & 48.4$\pm$2.3   & 11.7$\pm$6.0   & \nodata       & 14.4$\pm$4.2    & 72.7$\pm$4.0   & \nodata       \\
      $$H$\delta$ $\lambda$4101       & 28$\pm$4       & 32$\pm$3      & 31$\pm$7       & 28$\pm$3        & 30$\pm$3       & 25$\pm$1       & 24$\pm$3       & 13$\pm$5      & 32$\pm$3        & 24$\pm$2       & 28$\pm$4      \\
      $$H$\gamma$ $\lambda$4340       & 49$\pm$4       & 48$\pm$3      & 64$\pm$8       & 46$\pm$3        & 51$\pm$4       & 46$\pm$2       & 47$\pm$4       & 45$\pm$7      & 46$\pm$3        & 45$\pm$2       & 45$\pm$4      \\
      $$[OIII]    $\lambda$4363       & \nodata        & \nodata       & \nodata        & \nodata         & \nodata        & 4.7$\pm$1.0    & \nodata        & \nodata       & \nodata         & 9.5$\pm$1.5    & \nodata       \\
      $$[OIII]    $\lambda$4959       & 33.1$\pm$2.8   & 49.1$\pm$2.8  & 18.6$\pm$5.3   & 79.3$\pm$4.1    & 53.6$\pm$3.4   & 166.2$\pm$7.2  & 68.7$\pm$4.5   & 11.5$\pm$6.8  & 87.3$\pm$4.6    & 241.4$\pm$10.8 & 8.2 $\pm$1.6  \\  
      $$[OIII]    $\lambda$5007       & 86.7$\pm$5.0   & 144.9$\pm$7.2 & 24.0$\pm$4.0   & 243.7$\pm$11.6  & 175.1$\pm$9.2  & 498.5$\pm$21.5 & 190.1$\pm$10.5 & 47.9$\pm$7.0  & 247.6$\pm$12.3  & 713.2$\pm$31.3 & 20.5$\pm$2.1  \\
      $$[NII]     $\lambda$5755       & \nodata        & \nodata       & \nodata        & \nodata         & \nodata        & 1.1$\pm$1.0    & \nodata        & \nodata       & \nodata         & \nodata        & \nodata       \\ 
      $$HeI       $\lambda$5876       & 6.9$\pm$3.3    & 11.5$\pm$1.2  & \nodata        & 17.2$\pm$1.9    & \nodata        & 12.7$\pm$0.6   & 13.5$\pm$2.3   & \nodata       & 5.3$\pm$1.1     & 11.9$\pm$0.9   & 6.9$\pm$1.1   \\
      $$[SIII]    $\lambda$6312       & \nodata        & \nodata       & \nodata        & \nodata         & \nodata        & 1.50$\pm$0.31  & \nodata        & \nodata       & \nodata         & \nodata        & \nodata       \\ 
      $$[OI]      $\lambda$6363       & \nodata        & \nodata       & 16.62$\pm$5.22 & 5.36$\pm$1.63   & \nodata        & 2.04$\pm$0.26  & \nodata        & \nodata       & 4.48$\pm$0.92   & 4.75$\pm$1.16  & 16.93$\pm$3.15 \\
      $$[NII]     $\lambda$6548       & 22.2$\pm$2.2   & 23.2$\pm$1.8  & 40.5$\pm$12.7  & 17.7$\pm$1.9    & 12.2$\pm$1.8   & 13.3$\pm$0.7   & 20.5$\pm$2.3   & 21.1$\pm$4.0  & 11.7$\pm$1.5    & 9.7$\pm$0.9    & 20.9$\pm$2.2   \\
      $$H$\alpha$ $\lambda$6563       & 301$\pm$15     & 292$\pm$14    & 305$\pm$29     & 292$\pm$14      & 304$\pm$16     & 303$\pm$13     & 290$\pm$16     & 264$\pm$26    & 295$\pm$15      & 283$\pm$13     & 310$\pm$16     \\
      $$[NII]     $\lambda$6583       & 75$\pm$4       & 71$\pm$4      & 83$\pm$9       & 51$\pm$3        & 29$\pm$3       & 40$\pm$2       & 58$\pm$4       & 85$\pm$10     & 29$\pm$2        & 25$\pm$1       & 45$\pm$3      \\
      $$HeI       $\lambda$6678       & \nodata        & 4.1$\pm$1.1   & 8.8$\pm$5.0    & 5.4$\pm$1.4     & 6.7$\pm$1.1    & 3.4$\pm$0.3    & 4.3$\pm$2.1    & \nodata       & 5.6$\pm$2.3     & 2.9$\pm$0.8    & 4.1$\pm$1.6   \\ 
      $$[SII]     $\lambda$6717       & 23.5$\pm$2.5   & 35.2$\pm$2.2  & 44.7$\pm$14.0  & 15.7$\pm$1.5    & 16.0$\pm$2.1   & 18.9$\pm$0.9   & 22.7$\pm$2.8   & 57.3$\pm$18.2 & 22.2$\pm$6.7    & 17.3$\pm$1.1   & 39.9$\pm$2.7  \\
      $$[SII]     $\lambda$6731       & 10.6$\pm$2.1   & 24.3$\pm$1.8  & 26.1$\pm$4.0   & 9.6$\pm$1.2     & 13.8$\pm$2.2   & 13.5$\pm$0.7   & 17.5$\pm$2.5   & 37.3$\pm$6.7  & 15.4$\pm$4.7    & 11.8$\pm$0.9   & 24.0$\pm$1.9  \\ 
      $$[ArIII]   $\lambda$7135       & \nodata        & 4.0$\pm$1.0   & \nodata        & 6.4$\pm$1.9     & 9.0$\pm$1.4    & 5.5$\pm$0.3    & 7.2$\pm$3.6    & \nodata       & 7.6$\pm$4.2     & 6.2$\pm$0.7    & \nodata       \\  
      $$c(H$\beta$)          & 0.38           & 0.21          & 0.22           & 0.17            & 0.20           & 0.24           & 0.07           & 0.17          & 0.11            & 0.12           & 0.28          \\
      $$H$\beta$ flux (E-15 )& 2.27           & 1.65          & 0.63           & 1.49            & 1.56           & 12.16          & 0.74           & 0.56          & 1.18            & 3.70           & 0.85          \\
      WR features            & \nodata        & \nodata       & \nodata        & \nodata         & \nodata        & yes            & \nodata        & \nodata       & \nodata         & yes            & \nodata       \\
      \hline
              
      \enddata
      \tablecomments{All fluxes are given relative to H$\beta$=100.}
      \label{tab:fluxes1}
\end{deluxetable} 


\begin{deluxetable}{l|cccccccccc}
  \center
  \setlength{\tabcolsep}{0.04in} 
  \rotate{}
  \tabletypesize{\scriptsize}
  \tablewidth{0.0pt}
  \tablecolumns{12}
  \tablecaption{Dereddened Line Fluxes and Errors (continued)}
  \tablehead{
    \colhead{Line} &
    \colhead{33   }   &
    \colhead{35   }   &
    \colhead{37   }   &
    \colhead{disk1}   &
    \colhead{disk2}   &
    \colhead{disk3}   &
    \colhead{disk4}   &
    \colhead{disk5}  &
    \colhead{disk6}  &
    \colhead{disk7}    }
  \startdata
      [OII]     $\lambda$3727       & 342$\pm$39     & 734$\pm$38      & 466$\pm$41     & 320$\pm$14      & 334$\pm$19     & 329$\pm$26      & 282$\pm$12     & 227$\pm$19    & 233$\pm$20    &  225$\pm$21    \\
      $$[NeIII]   $\lambda$3869       & 24.2$\pm$12.8  & 46.0$\pm$5.5    & \nodata        & 3.8$\pm$0.5     & \nodata        & 14.6$\pm$0.9    & \nodata        & 7.4$\pm$1.0   & 5.4$\pm$1.2   &  \nodata       \\
      $$H$\delta$ $\lambda$4101       & 36$\pm$9       & 43$\pm$6        & 46$\pm$10      & 25$\pm$1        & 23$\pm$3       & 26$\pm$1        & 26$\pm$1       & 27$\pm$2      & 25$\pm$2      &  31$\pm$3      \\
      $$H$\gamma$ $\lambda$4340       & 47$\pm$24      & 61$\pm$5        & 45$\pm$7       & 48$\pm$2        & 53$\pm$3       & 49$\pm$2        & 47$\pm$2       & 48$\pm$4      & 47$\pm$4      &  48$\pm$4      \\
      $$[OIII]    $\lambda$4363       & \nodata        & \nodata         & \nodata        & 0.15$\pm$0.05   & \nodata        & 1.4$\pm$0.4     & \nodata        & \nodata       & \nodata       &  \nodata       \\
      $$[OIII]    $\lambda$4959       & 71.5$\pm$7.5   & 54.7$\pm$3.4    & 51.7$\pm$10.8  & 25.7$\pm$1.1    & 21.2$\pm$1.5   & 85.9$\pm$3.6    & 26.8$\pm$1.2   & 32.8$\pm$2.8  & 25.4$\pm$2.2  &  7.6 $\pm$3.8  \\  
      $$[OIII]    $\lambda$5007       & 224$\pm$21.4   & 170.1$\pm$8.7   & 121.5$\pm$8.7  & 75.8$\pm$3.2    & 58.4$\pm$3.1   & 252.3$\pm$10.7  & 76.4$\pm$3.3   & 96.8$\pm$8.2  & 75.9$\pm$6.4  &  26.3$\pm$2.5  \\
      $$[NII]     $\lambda$5755       & \nodata        & \nodata         & \nodata        & 0.7$\pm$0.2     & \nodata        & 0.7$\pm$0.3     & 0.7$\pm$0.2    & 0.7$\pm$0.2   & \nodata       &  \nodata       \\ 
      $$HeI       $\lambda$5876       & 71.9$\pm$36.6  & 14.1$\pm$8.2    & 20.0$\pm$13.3  & 10.6$\pm$0.5    & 10.3$\pm$1.3   & 12.8$\pm$0.6    & 11.8$\pm$0.6   & 10.9$\pm$0.9  & 10.7$\pm$0.9  &  8.5$\pm$4.3   \\
      $$[SIII]    $\lambda$6312       & 9.21$\pm$4.69  & \nodata         & \nodata        & 1.09$\pm$0.11   & \nodata        & 1.68$\pm$0.14   & \nodata        & 1.07$\pm$0.17 & 1.16$\pm$0.26 &  2.04$\pm$0.64 \\ 
      $$[OI]      $\lambda$6363       & 14.09$\pm$7.17 & 29.98$\pm$3.01  & 24.73$\pm$5.40 & 1.50$\pm$0.13   & 6.15$\pm$0.84  & 1.83$\pm$0.22   & 0.59$\pm$0.16  & 1.04$\pm$0.22 & 1.33$\pm$0.30 &  2.61$\pm$0.82 \\
      $$[NII]     $\lambda$6548       & 31.0$\pm$9.8   & 30.6$\pm$2.5    & 26.0$\pm$5.2   & 36.7$\pm$1.6    & 38.1$\pm$2.2   & 25.9$\pm$1.1    & 35.2$\pm$1.5   & 32.7$\pm$2.8  & 33.8$\pm$2.9  &  53.1$\pm$4.7  \\
      $$H$\alpha$ $\lambda$6563       & 282$\pm$27     & 313$\pm$16      & 311$\pm$20     & 298$\pm$13      & 303$\pm$14     & 307$\pm$13      & 301$\pm$13     & 302$\pm$26    & 305$\pm$26    &  306$\pm$26    \\
      $$[NII]     $\lambda$6583       & 27$\pm$4       & 98$\pm$5        & 69$\pm$7       & 121$\pm$5       & 105$\pm$5      & 81$\pm$3        & 106$\pm$4      & 102$\pm$9     & 106$\pm$9     &  156$\pm$13    \\
      $$HeI       $\lambda$6678       & \nodata        & \nodata         & \nodata        & 2.8$\pm$0.2     & 4.1$\pm$1.2    & 3.6$\pm$0.2     & 3.7$\pm$0.3    & 3.0$\pm$0.3   & 3.0$\pm$0.3   &  4.4$\pm$1.4   \\ 
      $$[SII]     $\lambda$6717       & 29.7$\pm$5.3   & 130.4$\pm$6.8   & 34.3$\pm$5.2   & 44.8$\pm$1.9    & 51.6$\pm$2.7   & 27.9$\pm$1.2    & 26.7$\pm$1.2   & 37.5$\pm$3.2  & 37.6$\pm$3.2  &  58.9$\pm$5.1  \\
      $$[SII]     $\lambda$6731       & 24.5$\pm$7.7   & 83.9$\pm$4.6    & 26.0$\pm$5.4   & 31.7$\pm$1.3    & 37.1$\pm$2.1   & 19.8$\pm$0.9    & 17.8$\pm$0.8   & 26.2$\pm$2.2  & 26.6$\pm$2.3  &  42.8$\pm$4.2  \\ 
      $$[ArIII]   $\lambda$7135       & 19.8$\pm$6.0   & \nodata         & \nodata        & 4.2$\pm$0.2     & 0.7$\pm$0.8    & 6.1$\pm$0.3     & 5.8$\pm$0.3    & 4.1$\pm$0.4   &  3.6$\pm$0.4  &  4.5$\pm$1.3   \\  
      $$c(H$\beta$)          & 0.14           & 0.15               & 0.12           & 0.32            & 0.31           & 0.30            & 0.22           & 0.34          & 0.43          &  0.38          \\
      $$H$\beta$ flux (E-15 )& 0.66           & 1.70               & 0.49           & 71.94           & 3.68           & 29.07           & 15.92          & 38.89         & 46.28         &  5.66          \\
      WR features            & \nodata        & \nodata            & \nodata        & yes             & \nodata        & yes             & yes            & yes           & \nodata       & \nodata        \\
      \hline
      
      \enddata 
      \tablecomments{All fluxes are given relative to H$\beta$=100.}
      \label{tab:fluxes2}  
\end{deluxetable}         


\begin{deluxetable}{lrrrr}
  \center
  \tabletypesize{\footnotesize}
  \tablewidth{0.0pt}
  \tablecolumns{5}
  \tablecaption{Temperatures}
  \tablehead{
    \colhead{ID} &
    \multicolumn{2}{|c|}{Measured} &
    \multicolumn{2}{c}{Adopted} \\
    \hline
    \colhead{\null}  &
    \colhead{T[NII]}   &
    \colhead{T[OIII]}  & 
    \colhead{T[OII]}   &
    \colhead{T[OIII]}  \\
    \colhead{\null}  &
    \colhead{(K)}    &
    \colhead{(K)}    & 
    \colhead{(K)}    &
    \colhead{(K)}    
  }
  \startdata
  
  21           &  14000$^{+6000}_{-3300}$    &  11800$^{+600}_{-1200}$  &  14000$^{+6000}_{-3300}$    &  11800$^{+600}_{-1200}$      \\
  26           &  \nodata                 &  11000$^{+3000}_{-1200}$ &  10700$^{+2100}_{-800}$ &   11000$^{+3000}_{-1200}$       \\  
  28           &  \nodata                 &  12900$^{+900}_{-800}$  &  12000$^{+600}_{-500}$     &  12900$^{+900}_{-800}$       \\
  disk1        &   7500$^{+800}_{-600}$     &  7400$^{+700}_{-500}$   &   7500$^{+800}_{-600}$     &  7400$^{+700}_{-500}$        \\
  disk3        &   8500$^{+2000}_{-900}$     &  9600$^{+1000}_{-700}$  &   8500$^{+2000}_{-900}$     &  9600$^{+1000}_{-700}$        \\
  disk4        &   7800$^{+900}_{-600}$     &  \nodata              &  7800$^{+900}_{-600}$     &  6900$^{+1200}_{-900}$     \\
  disk5        &   7900$^{+1000}_{-600}$     &  \nodata             &  7900$^{+1000}_{-600}$     &  7000$^{+1400}_{-900}$     \\ 
  \hline
  \enddata
  
  \tablecomments{Measured and adopted temperatures for abundances determination.  
    The adopted temperatures were derived as explained in the text.} 
  \label{tab:temps}
\end{deluxetable}


\begin{deluxetable}{lccc}
  \center
  \tabletypesize{\footnotesize}
  \tablewidth{0.0pt}
  \tablecolumns{4}
  \tablecaption{T$_{e}$ Based Oxygen Abundances}
  \tablehead{
    \colhead{ID} &
    \colhead{0$^+$/H$^+$}   &
    \colhead{O$^{+2}$/H$^+$}   &
    \colhead{12+log(O/H)}  
  }
  \startdata
  
  21           &  (2.2$\pm$1.4)E$-$5   &  (1.0$\pm$0.5)E$-$4    &   8.09$\pm$0.24          \\
  26           &  (7.8$\pm$3.0)E$-$5   &  (6.4$\pm$3.1)E$-$4    &   8.15$\pm$0.16          \\
  28           &  (3.1$\pm$1.0)E$-$5   &  (1.1$\pm$0.3)E$-$4    &   8.15$\pm$0.11          \\
  disk1        &  (4.9$\pm$2.0)E$-$4   &  (8.8$\pm$2.8)E$-$5    &   8.67$\pm$0.18          \\
  disk3        &  (2.6$\pm$1.6)E$-$4   &  (1.0$\pm$0.3)E$-$4    &   8.76$\pm$0.19          \\
  disk4        &  (3.5$\pm$1.5)E$-$4   &  (1.2$\pm$0.6)E$-$4    &   8.56$\pm$0.26          \\
  disk5        &  (2.6$\pm$1.5)E$-$4   &  (1.4$\pm$0.6)E$-$4    &   8.60$\pm$0.22          \\ \hline
  \enddata
  
  \tablecomments{Oxygen ion and element abundances derived from temperatures and line ratios.} 
  \label{tab:ionic}
\end{deluxetable}


\vspace{-1.00truein}
\begin{deluxetable}{l|ccrrrrccccc}
  \center
  \rotate{}
  \tabletypesize{\scriptsize}
  \tablewidth{-0.2pt}
  \tablecolumns{11}
  \tablecaption{Strong Line and T$_e$ Derived Abundances}
  \tablehead{
    \colhead{ID}                &
    \colhead{R/R$_{25}$}         &
    \colhead{branch}            &
    \colhead{$R_{23}$ }          &   
    \colhead{$O_{32}$ }          &
    \colhead{$P$ }              &
    \colhead{log([NII]/[OII])}  &
    \colhead{12+log(O/H)} &
    \colhead{12+log(O/H)} &
    \colhead{12+log(O/H)} &
    \colhead{12+log(O/H)} &
    \colhead{12+log(O/H)} \\
    \colhead{} &
    \colhead{} &
    \colhead{} &
    \colhead{} &
    \colhead{} &
    \colhead{} &
    \colhead{} &
    \colhead{KK04} &
    \colhead{PT05} &
    \colhead{KD02} &
    \colhead{B07} &
    \colhead{$T_e$} \\
    \colhead{(1)}                 &
    \colhead{(2)}                 &
    \colhead{(3)}                 &
    \colhead{(4)}                 &   
    \colhead{(5)}                 &
    \colhead{(6)}                 &
    \colhead{(7)}                 &
    \colhead{(8)}                 &
    \colhead{(9)}                 &
    \colhead{(10)}                &
    \colhead{(11)}                &
    \colhead{(12)}}
  \startdata
  02     &    1.49   &   U   &  5.20$\pm$0.40  &  0.30$\pm$0.02  &  0.23$\pm$0.02  &   $-$0.73$\pm$0.04  &   8.79$\pm$0.23   &   8.08$\pm$0.21  &  8.77$\pm$0.10  &    8.31$\pm$0.20  & \nodata       \\
  03     &    1.46   &   U   &  5.26$\pm$0.33  &  0.58$\pm$0.04  &  0.37$\pm$0.02  &   $-$0.67$\pm$0.03  &   8.84$\pm$0.26   &   8.25$\pm$0.25  &  8.80$\pm$0.10  &    8.34$\pm$0.20  & \nodata       \\
  07     &    1.60   &   U   &  5.57$\pm$2.67  &  0.08$\pm$0.40  &  0.08$\pm$0.04  &   $-$0.79$\pm$0.23  &   8.73$\pm$0.63   &   7.78$\pm$0.56  &  8.73$\pm$0.15  &    8.27$\pm$0.22  & \nodata       \\
  14     &    0.99   &   U   &  5.51$\pm$0.34  &  1.41$\pm$0.11  &  0.59$\pm$0.03  &   $-$0.65$\pm$0.04  &   8.82$\pm$0.36   &   8.41$\pm$0.34  &  8.81$\pm$0.10  &    8.35$\pm$0.20  & \nodata       \\
  17     &    1.53   &   T   &  5.37$\pm$0.38  &  0.74$\pm$0.06  &  0.43$\pm$0.03  &   $-$1.02$\pm$0.05  &   8.50$\pm$0.26   &   8.03$\pm$0.25  &  8.58$\pm$0.10  &    8.12$\pm$0.20  & \nodata       \\
  21     &    1.11   &   U   &  8.69$\pm$0.45  &  3.26$\pm$0.18  &  0.76$\pm$0.03  &   $-$0.71$\pm$0.03  &   8.51$\pm$0.35   &   8.27$\pm$0.34  &  8.78$\pm$0.10  &    8.32$\pm$0.20  & 8.09$\pm$0.24 \\
  24     &    1.10   &   U   &  5.77$\pm$0.43  &  0.81$\pm$0.07  &  0.45$\pm$0.03  &   $-$0.74$\pm$0.04  &   8.78$\pm$0.33   &   8.28$\pm$0.31  &  8.76$\pm$0.10  &    8.30$\pm$0.20  & \nodata       \\
  25     &    1.03   &   U   &  4.23$\pm$0.68  &  0.16$\pm$0.03  &  0.14$\pm$0.03  &   $-$0.63$\pm$0.08  &   8.89$\pm$0.39   &   8.07$\pm$0.35  &  8.82$\pm$0.11  &    8.37$\pm$0.20  & \nodata       \\
  26     &    2.18   &   T   &  6.09$\pm$0.38  &  1.22$\pm$0.09  &  0.55$\pm$0.03  &   $-$0.97$\pm$0.04  &   8.47$\pm$0.29   &   8.10$\pm$0.28  &  8.61$\pm$0.10  &    8.15$\pm$0.20  &  8.15$\pm$0.16  \\
  28     &    2.18   &   T   & 11.24$\pm$0.61  &  5.63$\pm$0.36  &  0.85$\pm$0.04  &   $-$0.83$\pm$0.03  &   8.26$\pm$0.35   &   8.20$\pm$0.34  &  8.71$\pm$0.10  &   8.24$\pm$0.20  &  8.15$\pm$0.11 \\
  29     &    2.01   &   U   &  3.93$\pm$0.33  &  0.08$\pm$0.01  &  0.07$\pm$0.01  &   $-$0.91$\pm$0.04  &   8.92$\pm$0.16   &   7.99$\pm$0.14  &  8.66$\pm$0.10  &   8.19$\pm$0.20   &  \nodata       \\
  33     &    2.27   &   T   &  6.37$\pm$0.79  &  0.86$\pm$0.12  &  0.46$\pm$0.05  &   $-$1.10$\pm$0.08  &   8.49$\pm$0.48   &   8.09$\pm$0.46  &  8.51$\pm$0.11  &    8.06$\pm$0.20  &  \nodata       \\
  35     &    1.63   &   U   &  9.59$\pm$0.65  &  0.31$\pm$0.02  &  0.23$\pm$0.01  &   $-$0.88$\pm$0.03  &   8.31$\pm$0.20   &   7.61$\pm$0.19  &  8.68$\pm$0.10  &   8.21$\pm$0.20  &  \nodata       \\
  37     &    1.46   &   U   &  6.39$\pm$0.64  &  0.37$\pm$0.04  &  0.27$\pm$0.03  &   $-$0.83$\pm$0.06  &   8.70$\pm$0.36   &   8.01$\pm$0.33  &  8.71$\pm$0.10  &   8.24$\pm$0.20  &  \nodata       \\
  disk1  &    0.43   &   U   &  4.21$\pm$0.23  &  0.32$\pm$0.02  &  0.24$\pm$0.01  &   $-$0.42$\pm$0.03  &   8.90$\pm$0.18   &   8.21$\pm$0.17  &  8.93$\pm$0.10  & 8.48$\pm$0.20  &  8.67$\pm$0.18 \\
  disk2  &    0.81   &   U   &  4.13$\pm$0.27  &  0.24$\pm$0.02  &  0.19$\pm$0.01  &   $-$0.50$\pm$0.03  &   8.91$\pm$0.19   &   8.16$\pm$0.17  &  8.89$\pm$0.10  & 8.44$\pm$0.20  &  \nodata       \\
  disk3  &    0.71   &   U   &  6.67$\pm$0.40  &  1.03$\pm$0.09  &  0.51$\pm$0.03  &   $-$0.61$\pm$0.04  &   8.69$\pm$0.32   &   8.25$\pm$0.31  &  8.83$\pm$0.10  & 8.38$\pm$0.20 &   8.76$\pm$0.19 \\ 
  disk4  &    0.55   &   U   &  3.85$\pm$0.21  &  0.37$\pm$0.02  &  0.27$\pm$0.02  &   $-$0.43$\pm$0.03  &   8.94$\pm$0.19   &   8.29$\pm$0.18  &  8.92$\pm$0.10  & 8.48$\pm$0.20  &  8.56$\pm$0.26 \\
  disk5  &    0.39   &   U   &  3.57$\pm$0.37  &  0.57$\pm$0.06  &  0.36$\pm$0.03  &   $-$0.35$\pm$0.05  &   8.97$\pm$0.41   &   8.42$\pm$0.39  &  8.96$\pm$0.10  & 8.51$\pm$0.20  &  8.60$\pm$0.22 \\
  disk6  &    0.35   &   U   &  3.34$\pm$0.35  &  0.44$\pm$0.05  &  0.30$\pm$0.03  &   $-$0.34$\pm$0.05  &   8.99$\pm$0.36   &   8.39$\pm$0.34  &  8.96$\pm$0.10  & 8.52$\pm$0.20  &  \nodata       \\
  disk7  &    0.20   &   U   &  2.59$\pm$0.31  &  0.15$\pm$0.02  &  0.13$\pm$0.02  &   $-$0.16$\pm$0.06  &   9.07$\pm$0.30   &   8.27$\pm$0.27  &  9.04$\pm$0.10  & 8.60$\pm$0.20  &  \nodata       \\
  \hline
  \enddata
  
  \tablecomments{{\bf(1)} ID number. 
    {\bf(2)} Galactocentric distance assuming $R_{25}$=14.6 kpc.
    {\bf(3)} Adopted branch of the $R_{23}$ vs. (O/H) relation. $U$=upper, $T$=turnaround. 
    {\bf(4)} $R_{23}$ = {([OII]$\lambda$3727+[OIII]$\lambda\lambda$4959,5007)}/{H$\beta\lambda$4861.} 
    {\bf(5)} $O_{32}$ =  {[OIII]$\lambda\lambda$4959,5007}/{[OII]$\lambda$3727}.
    {\bf(6)} Value of excitation parameter $P$ = {[OIII]$\lambda\lambda$4959,5007}/({[OII]$\lambda$3727+[OIII]$\lambda\lambda$4959,5007}).
    {\bf(7)} Value of log([NII]$\lambda$6584/[OII]$\lambda$3727).
    {\bf(8)} 12+log(O/H) from theoretical $R_{23}$ based calculation using \citet{KK2004}. 
    {\bf(9)} 12+log(O/H) from empirical $R_{23}$ based calculation using \citet{PT2005}.
    {\bf(10)} 12+log(O/H) from theoretical log([NII]/[OII]) based calculation using \citet{KD2002}.
    {\bf(11)} 12+log(O/H) from empirical log([NII]/[OII]) based calculation using \citet{B07}.
    {\bf(12)} 12+log(O/H) derived from temperature sensitive lines.}
  \label{tab:R23abund}
\end{deluxetable}         


\begin{deluxetable}{lccccc}
  \center
  \tabletypesize{\footnotesize}
  \tablewidth{0.0pt}
  \tablecolumns{6}
  \tablecaption{Oxygen Abundance Gradients}
  \tablehead{
    \colhead{Method} &
    \colhead{N$_{regions}$} &
    \colhead{$\Delta$(dex kpc$^{-1}$)}   &
    \colhead{$\Delta$(dex/$R_d$)}    &
    \colhead{$\Delta$(dex/$R_{25}$)}    &
    \colhead{A(O)$_0$}     }
  \startdata
  \multicolumn{6}{c}{\bf{these data} }\\
  \hline  
  KK04       &  21      &  $-$0.014$\pm$0.006  & $-$0.040$\pm$0.017 & $-$0.204$\pm$0.088   &  9.01$\pm$0.12  \\
  PT05       &  21      &  $-$0.013$\pm$0.006   & $-$0.037$\pm$0.017 & $-$0.190$\pm$0.088   &  8.34$\pm$0.12   \\
  KD02       &  21      &  $-$0.013$\pm$0.002  & $-$0.037$\pm$0.006 & $-$0.190$\pm$0.029   &  9.02$\pm$0.05  \\
  B07          &  21     &   $-$0.014$\pm$0.005 & $-$0.040$\pm$0.014 & $-$0.204$\pm$0.073   &  8.58$\pm$0.10   \\
  T$_e$      &   7      &  $-$0.020$\pm$0.006    & $-$0.057$\pm$0.017 & $-$0.292$\pm$0.088  &  8.76$\pm$0.13    \\
  \hline
  \multicolumn{6}{c}{\bf{all four data sets} } \\
  \hline
  KK04    &  49     &  $-$0.008$\pm$0.005   & $-$0.023$\pm$0.014 & $-$0.117$\pm$0.073 &  8.98$\pm$0.06  \\
  PT05    &  49     &  $-$0.016$\pm$0.004   & $-$0.046$\pm$0.011 & $-$0.234$\pm$0.058 &  8.47$\pm$0.06  \\
  KD02    &  49     &  $-$0.016$\pm$0.002   & $-$0.046$\pm$0.006 & $-$0.234$\pm$0.029 &  9.10$\pm$0.03  \\
  B07      &   49    &   $-$0.017$\pm$0.004   & $-$0.048$\pm$0.011 &  $-$0.248$\pm$0.058 &  8.66$\pm$0.05 \\
  \hline
  \enddata
  
  \tablecomments{Coefficients of weighted least-squares fits to the oxygen abundance gradient for 
    12+log(O/H) = A(O)$_0$ + $\Delta(\log$(O/H)/R).  We assume the scale length of $R_d$= 2.7 arcmin (2.85 kpc)
  \citep{Barker2009} and an optical size of $R_{25}$= 14.6 kpc.} 
  \label{tab:grad}
\end{deluxetable}
\pagebreak


\clearpage

\clearpage

\clearpage

\clearpage

\clearpage

\clearpage

\clearpage

\clearpage

\clearpage

\clearpage

\end{document}